\documentclass[aps,prl,twocolumn,showpacs,preprintnumbers,amsmath,amssymb,superscriptaddress]{revtex4-2}

\usepackage{graphicx}
\usepackage{dcolumn}
\usepackage{bm}
\usepackage{hyperref}
\hypersetup{
	colorlinks,
	citecolor=red,
	filecolor=black,
	linkcolor=blue,
	urlcolor=blue
}
\usepackage{lipsum}
\usepackage{natbib}
\usepackage{soul}

\usepackage{dsfont}
\usepackage{tgtermes}

\newcommand{\be}{\begin{equation}}
\newcommand{\ee}{\end{equation}}
\newcommand{\btcb}[2]{\begin{tcolorbox}[arc=#1mm,boxsep=#2mm]}
\newcommand{\etcb}{\end{tcolorbox}}
\newcommand{\bd}{\textbf}

\newcommand{\bds}{\boldsymbol}
\newcommand{\derd}{\, \mathrm d}

\newcommand{\ham}{\mathcal{H}}
\newcommand{\hham}{\hat{\mathcal{H}}}

\newcommand{\iOmega}{\mathit{\Omega}}
\newcommand{\la}{\langle}

\newcommand{\ra}{\rangle}

\newcommand{\tit}{\textit}

\newcommand{\sgn}{\text{sgn}}

\newcommand{\tr}{\text{Tr}}

\newcommand{\re}{\text{Re}}
\newcommand{\im}{\text{Im}}
\newcommand{\wt}{\widetilde}

\newtheorem{propt}{Proposition}

\usepackage{graphicx}
\usepackage{dcolumn}
\usepackage{bm}

\begin{document}


\title{Ideal quantum geometry of the surface states of rhombohedral graphite and its effects on the surface superconductivity}
\author{Guodong Jiang}
\email{guodong.jiang@aalto.fi}
\affiliation{Department of Applied Physics, Aalto University School of Science, FI-00076 Aalto, Finland}
\author{Tero T. Heikkil{\"a}}
\affiliation{Department of Physics and Nanoscience Center, University of Jyv{\"a}skyl{\"a}, FI-40014 University of Jyv\"askyl\"a, Finland}
\author{P{\"a}ivi T{\"o}rm{\"a}}
\email{paivi.torma@aalto.fi}
\affiliation{Department of Applied Physics, Aalto University School of Science, FI-00076 Aalto, Finland}


\begin{abstract}
The interplay of quantum geometry and interactions determines the correlated state properties of flat bands. Here, we investigate the ideal quantum geometry (IQG), i.e., the property that the trace of quantum metric equals the Berry curvature, in the surface flat bands of rhombohedrally stacked graphene (RG) multilayers. We show that RG represents a class of semimetals with IQG, among which only RG has a nonvanishing IQG at the center of the surface bands. In the presence of long-range hoppings, a displacement field polarizes the electron density to one of the two surfaces, stabilizing the IQG and boosting transitions to correlated phases like superconductivity. Analyzing the superfluid stiffness of the superconducting state in a many-layer RG, we reveal a heavy-fermion picture where the ``heavy electrons" carry a nonzero supercurrent. This unusual behavior arises from the IQG of the surface states, which suggests a large fraction of supercurrent running on the surface of RG.
\end{abstract}
\maketitle
{\em Introduction}.---The quantum geometric tensor (QGT)~\cite{provost1980riemannian,resta2011insulating}, which describes the wavefunction change in parameter space, has shown connections with various quantum phenomena~\cite{paivi2023essay,yu2024quantum,liu2025quantum,gao2025quantum}. Its imaginary part (Berry curvature (BC, $\Omega$)~\cite{berry1984}) integrated over the Brillouin zone (BZ) gives the quantized Hall conductance of Hall fluids~\cite{tknn1982} and Chern bands~\cite{haldane1988}. Its real part (quantum metric (QM, $g$)), influences correlated states in narrow bands, e.g., it enables supercurrent in a flat-band superconductor~\cite{peotta2015superfluidity,torma2022tbgreview}. In the lowest Landau level (LLL), the inequality between the magnitude of BC and the trace of QM, $\tr g\geq |\Omega|$, saturates (i.e., becomes equality)~\cite{roy2014band,peotta2015superfluidity,ozawa2021relations}, leading to the Girvin-MacDonald-Platzman (GMP) algebra~\cite{girvin1986magneto} of density operators, thus stabilizing the fractional quantum Hall states. This trace condition, or the condition of ideal quantum geometry (IQG)~\cite{claassen2015position,wang2021exact,wang2022hierachy}, has also been suggested to stabilize the fractional Chern insulating states~\cite{parameswaran2012fractional,roy2014band,jackson2015geometric,neupert2011fractional,regnault2011fractional,sheng2011fractional,cai2023signatures,zeng2023thermodynamic,park2023observation,xu2023observation,lu2024fractional,xie2025tunable,lu2025extended,waters2025chern}. Intensive efforts have been focused on achieving IQG in insulating phases~\cite{claassen2015position,wang2021exact,wang2022hierachy,parameswaran2013fractional,liu2022recent,ledwith2022family,wang2022hierachy,ledwith2023vortexability}, while semimetals with IQG have received little attention. Only very recently~\cite{bernevig2025berry}, it was shown that the density operator of the surface states of multi-layer rhombohedrally stacked graphene (RG) satisfies the GMP algebra, thereby relating them to the LLL. The origin of this property for RG and its implications for other correlated phases than the fractional states, in particular superconductivity, are important open questions.

\begin{figure}[t!]
\centering
\includegraphics[width=0.48\textwidth]{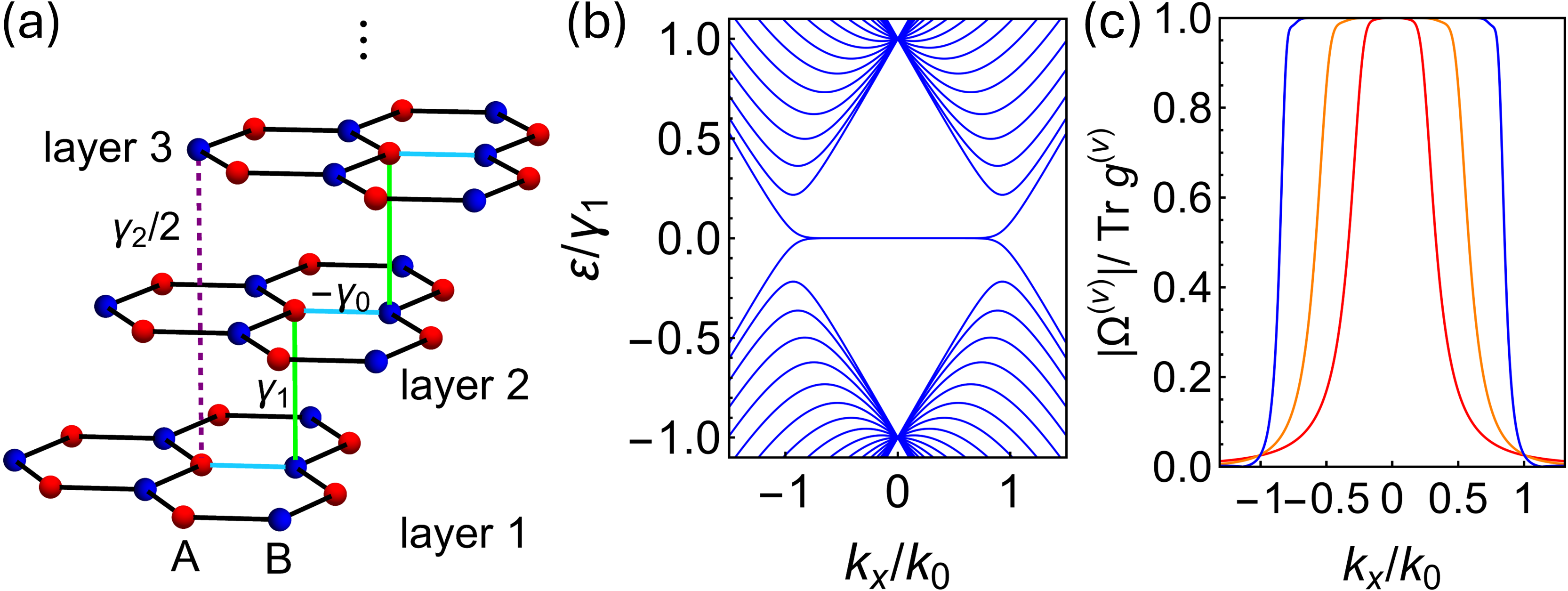}
\caption{(a) Lattice structure of multi-layer RG (3 layers are shown). The red or blue sphere represents the A/B sublattice. The light blue and green bonds (with hoppings $-\gamma_0$, $\gamma_1$) form the staircase unit cell. The dashed bond indicates the next-nearest interlayer hopping $\gamma_2/2$ (see SM Sec.~1 for additional hoppings). (b) Band structure of 20-layer RG at valley $K$ in the absence of a displacement field, with two degenerate drumhead-like surface bands at zero energy. (c) The trace condition holds, $\tr g=|\Omega|$, in the drumhead region, for 3 (red), 6 (orange) and 20 (blue)-layer RG with a surface potential $m=0.01\gamma_1$. (b)(c) are calculated from RG models with only $\gamma_0,\gamma_1$ hoppings.}
\label{fig:schematic}
\end{figure}

In this work, we establish the IQG of the surface bands of a broad class of semimetals, including RG, and investigate how it influences RG superconductivity. RG [Fig.~\ref{fig:schematic}(a)] is an example of ``nodal-line semimetals"~\cite{heikkila2011dimensional,heikkila2011flat,burkov2011topological,volovik2013topology,weng2015topological,yu2015topological,kim2015dirac,mullen2015line,xie2015new,hyart2018two} whose degeneracies in the bulk energy spectrum form lines in the 3D momentum space. When the nodal lines are projected to the surface BZ of an RG multilayer, they give rise to nearly dispersionless drumhead-like surface bands~\cite{heikkila2011dimensional,heikkila2011flat} [Fig.~\ref{fig:schematic}(b)], making RG a platform for studying superconductivity~\cite{kopnin2011high,kopnin2011surface,kopnin2013high,munoz2013tight,lothman2017universal,zhou2021superconductivity,han2025chiral,choi2024electric,morissette2025superconductivity,kumar2025superconductivity,nguyen2025hierarchy,yang2025,patterson2024} and other correlated phenomena~\cite{lee2013band,myhro2018large,shi2020electronic,hagymasi2022observation,zhou2021half,lee2022gate,zhou2024layer,bao2011stacking,lee2014competition,han2024correlated,arp2024intervalley,zhang2024correlated,zhang2024layer}. Our main results are: 1) We show that a class of semimetals exhibits IQG in their surface bands, with RG as the only case that has a nonvanishing IQG at the center of the drumhead region. 2) We describe the effects of the displacement field and long-range hopping effects on RG surface states and find that a displacement field stabilizes the IQG, and its surface polarization effect is significant to triggering the metal-superconductor transition in few-layer RG, e.g., in a trilayer~\cite{zhou2021superconductivity}. 3) We introduce a description of the superconductivity in many-layer RG (mRG) in terms of heavy and light fermions where the heavy fermions, counter-intuitively, carry supercurrent: the nonvanishing IQG provides a lower bound for the superfluid weight on the surface of RG.

{\em Ideal quantum geometry of the surface states of RG}.---We begin with an analysis of the QGT of RG surface bands and show that the trace condition is satisfied in the drumhead region. The drumhead region of RG at each valley has a radius $k_0=\gamma_1/v_f$ (we set $\hbar=1$), with $\gamma_1$ the nearest-neighbor interlayer hopping constant (see Fig.~\ref{fig:schematic}(a) for illustrations) and $v_f$ the Fermi velocity of single-layer graphene. This is quite a small area (0.3\%) compared to the surface BZ of RG, making its electron density tunable by gate voltage in a nearby electrode. The radius $k_0$ sets a natural momentum scale for RG surface bands, beyond which the surface states become bulk states. This implies that the decay length of the surface states, $\lambda$, varies with the in-plane momentum $\bd{k}$ \cite{koshino2010interlayer,heikkila2011dimensional}.

We first consider the $N$-layer RG model with the intra-layer hopping $-\gamma_0$ and interlayer hopping $\gamma_1$ only [Fig.~\ref{fig:schematic}(a)], in the limit of many layers, $N\rightarrow\infty$. Under these conditions, the two states localized at opposite surfaces have wavefunctions
\be\label{eq:twostates}
\begin{split}
&\psi^{(1)}_\bd{k}(z)=\frac{\sqrt{1-k^2}}{k}e^{\kappa(\bd{k})z}\begin{pmatrix}
    1 \\ 0
\end{pmatrix},\\
&\psi^{(2)}_\bd{k}(z)=\frac{\sqrt{1-k^2}}{k}e^{\kappa(\bd{k})^*(N+1-z)}\begin{pmatrix}
    0 \\ 1
\end{pmatrix}
\end{split}
\ee
at valley $K$, and those at valley $K'$ are related to them by time-reversal symmetry. Here, the two-component spinor is for A/B sublattice, $z$ is the layer index running from 1 to $N$ ($\rightarrow\infty$), $k=|\bd{k}|$ (in units of $k_0$), and function $\kappa(\bd{k})=\ln[-(k_x+ik_y)]$ has the meaning of the inverse of the decay length~\footnote{Function $\kappa(\bd{k})$ is understood as in the polar representation, with the gradient $\nabla\kappa(\bd{k})$ branch-independent. The singularity of $\kappa(\bd{k})$ at $\bd{k}=0$ becomes removable as we compute the QGT.}. The two states are degenerate in the entire drumhead region $k<1$ in the absence of a displacement field [Fig.~\ref{fig:schematic}(b)]. When a displacement field is applied perpendicular to the multilayer, the degeneracy is lifted and $\psi^{(1)}_\bd{k},\psi^{(2)}_\bd{k}$ remain as the eigenstates of the valence ($v$) and conduction ($c$) bands, respectively.

Then, QGT of the two states can be computed using its definition, $\mathcal{B}^{(j)}_{\mu\nu}(\bd{k})=\tr\{P^{(j)}_\bd{k}\partial_\mu P^{(j)}_\bd{k}\partial_\nu P^{(j)}_\bd{k}\}$ ($\mu$, $\nu$ denote $x$, $y$ and $\partial_\mu \equiv \partial/\partial k_\mu$), with $P^{(j)}_\bd{k}=|\psi^{(j)}_\bd{k}\ra\la\psi^{(j)}_\bd{k}|$ ($j=1,2$ or $v,c$) the band projection operator. We find (see SM Sec.~2.1)
\begin{align}\label{eq:qgtoftwobands}
\mathcal{B}^{(j)}_{\mu\nu}(\bd{k})=\frac{1}{(1-k^2)^2}\begin{pmatrix}
    1 & \pm i\\
    \mp i & 1
\end{pmatrix},
\end{align}
where the upper (lower) sign is for $\mathcal{B}^{(1)}$ ($\mathcal{B}^{(2)}$). Since $\mathcal{B}^{(j)}_{\mu\nu}\equiv g^{(j)}_{\mu\nu}-\frac{i}{2}\Omega^{(j)}_{\mu\nu}$, with $g_{\mu\nu}$ and $\Omega_{\mu\nu}$ the tensors of QM and BC, respectively, and $\Omega^{(j)}\equiv\Omega^{(j)}_{xy}$, it shows that the trace condition $\tr g^{(j)}=|\Omega^{(j)}|$ is satisfied in both bands. In addition, the prefactor in Eq.~\eqref{eq:qgtoftwobands}, $1/(1-k^2)^2$ is a slowly-varying function, so the IQG is approximately uniform near the center of the drumhead region. This uniform IQG makes the two bands analogous to a pair of LLLs with opposite magnetic fields, since the LLL has a constant IQG~\cite{roy2014band,peotta2015superfluidity,ozawa2021relations}. However, unlike LLL, which is a gapped state, RG is semi-metallic, with the drumhead region near charge neutrality only, so the QGT diverges as the surface states transit to bulk states at the rim of the drumhead region, $k=1$. This leads to a divergent total Berry phase in the drumhead region, which is consistent with the $N\rightarrow\infty$ limit we take, as the $N$-layer RG surface band carries a Berry phase of $N\pi$~\cite{koshino2009trigonal}.

\begin{figure}[t!]
\centering
\includegraphics[width=0.48\textwidth]{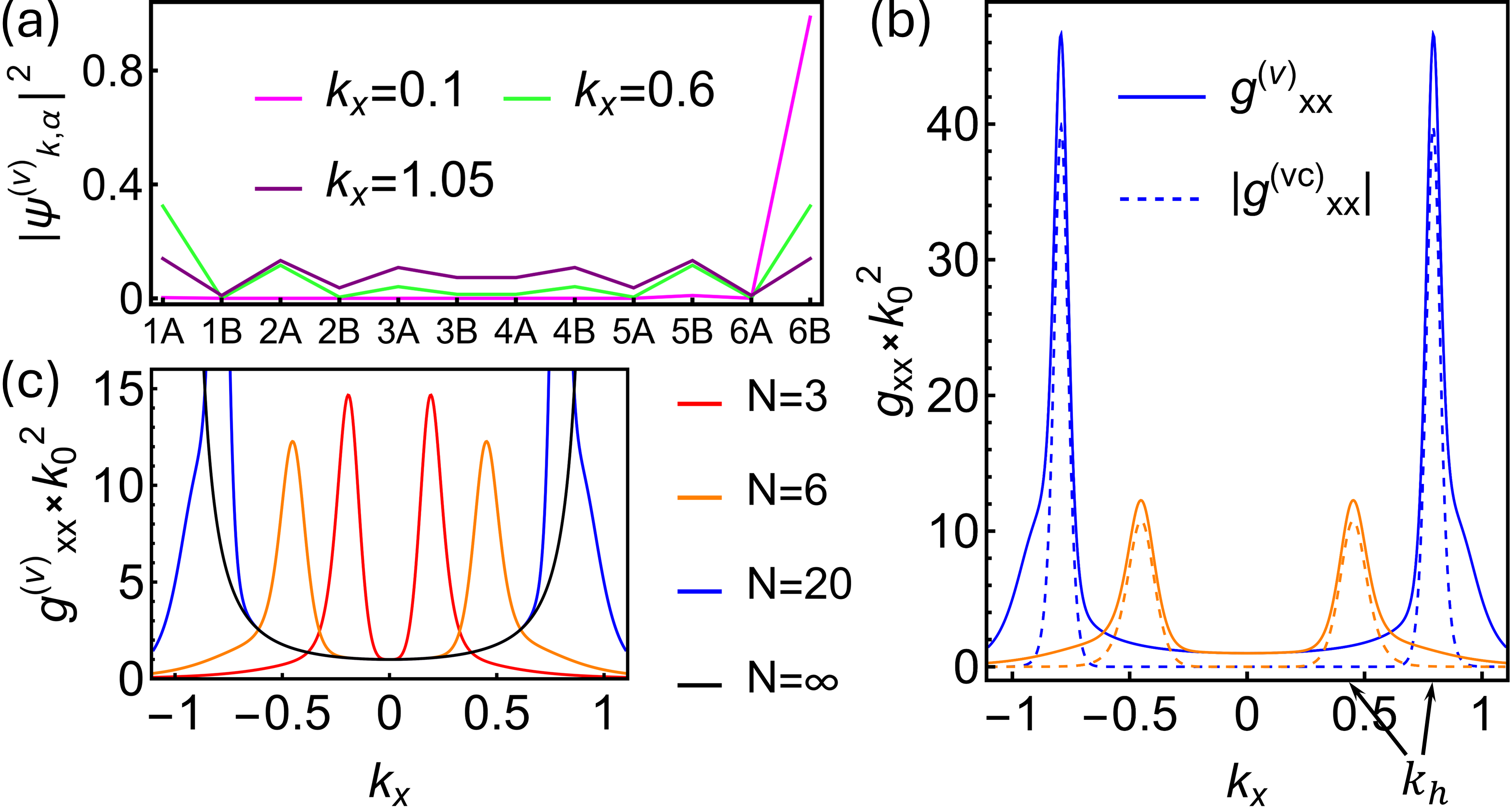}
\caption{RG model with $\gamma_0$, $\gamma_1$ hoppings and surface potential $m=0.01\gamma_1$ (to open a bandgap between the two bands). (a) Weight of orbital $\alpha$, $\rho^{(v)}_\alpha(\bd{k})=|\psi^{(v)}_{\bd{k},\alpha}|^2$ in the valence surface band of 6-layer RG at various $k_x$, with fixed $k_y=0$ at valley $K$. (b) Valence surface band QM $g^{(v)}_{xx}$ vs. the interband QM component, $|g^{(vc)}_{xx}|$, for 6 (orange) and 20 (blue) -layer RG. (c) $g^{(v)}_{xx}$ vs the multilayer thickness $N$. The black curve corresponds to the QM in the $N\rightarrow\infty$ limit in Eq.~\eqref{eq:qgtoftwobands}.}
\label{fig:qmplot}
\end{figure}

{\em Effect of finite thickness and surface hybridization}.---When $N$ is small, the drumhead region of flat bands shrinks to a smaller radius, $k_h$. It characterizes the momentum at which the decay length of the surface states gets comparable with the thickness, $\lambda=1/|\kappa|\sim N$. As a result, $k_h$ has an approximate relation to the thickness, $k_h=k_0e^{-\eta/N}$, with $\eta$ a constant of the order of unity~\footnote{This relation is obtained by setting the thickness equal to $\eta$ times the decay length, $N= \eta\lambda=-\eta/\ln k_h$, which is also equivalent to setting the dispersion function of $N$-layer RG equal to some small energy scale, $k_h^N=e^{-\eta}$. Since there is an arbitrariness in choosing $\eta$, $k_h$ cannot be defined precisely, but has a small span, which decreases with increasing $N$.}. In the shrunk drumhead region of these small-$N$-layer RG ($k<k_h$), the two surface states remain degenerate and are strongly localized. In the region $k_h<k<k_0$, the two surfaces hybridize due to the large decay length, lifting the degeneracy and forming the dispersive ``bulk-like" surface states, which are linear combinations of $\psi^{(1)}_\bd{k}$ and $\psi^{(2)}_\bd{k}$, as shown by the green curve in Fig.~\ref{fig:qmplot}(a).

The surface hybridization makes the QGT no longer ideal at $k>k_h$ [Fig.~\ref{fig:schematic}(c)]. In addition, it results in sharp peaks of QGT at $k=k_h$ [Fig.~\ref{fig:qmplot}(b)], in contrast to the nearly uniform QGT within the drumhead region. In Fig.~\ref{fig:qmplot}(b), we compare the QM of one surface band, e.g., the valence band, $g^{(v)}$, with the interband QM component between the two surface bands, $|g^{(vc)}|$. Here, the negative semidefinite $g^{(vc)}$ is defined as $g^{(vc)}_{\mu\nu}(\bd{k})=-\re\la\partial_\mu\psi^{(v)}_\bd{k}|\psi^{(c)}_\bd{k}\ra\la\psi^{(c)}_\bd{k}|\partial_\nu\psi^{(v)}_\bd{k}\ra$, which can be used to measure the wavefunction overlap between the two bands (see SM Sec.~3.2). We find that $|g^{(vc)}|$ constitutes a substantial portion of $g^{(v)}$, confirming that the QM peaks at $k=k_h$ are due to surface hybridization.

In Fig.~\ref{fig:qmplot}(c), we compare the surface state QM of different $N$-layer RG. Due to the thickness-independence of the surface state wavefunctions near the center, where they are subject to no surface hybridization, these QM curves converge to that in the $N\rightarrow\infty$ limit [Eq.~\eqref{eq:qgtoftwobands}] as $N$ increases. It shows that the IQG at the center of the drumhead region is a universal property independent of the multilayer thickness.

\begin{figure}[t!]
\centering
\includegraphics[width=0.48\textwidth]{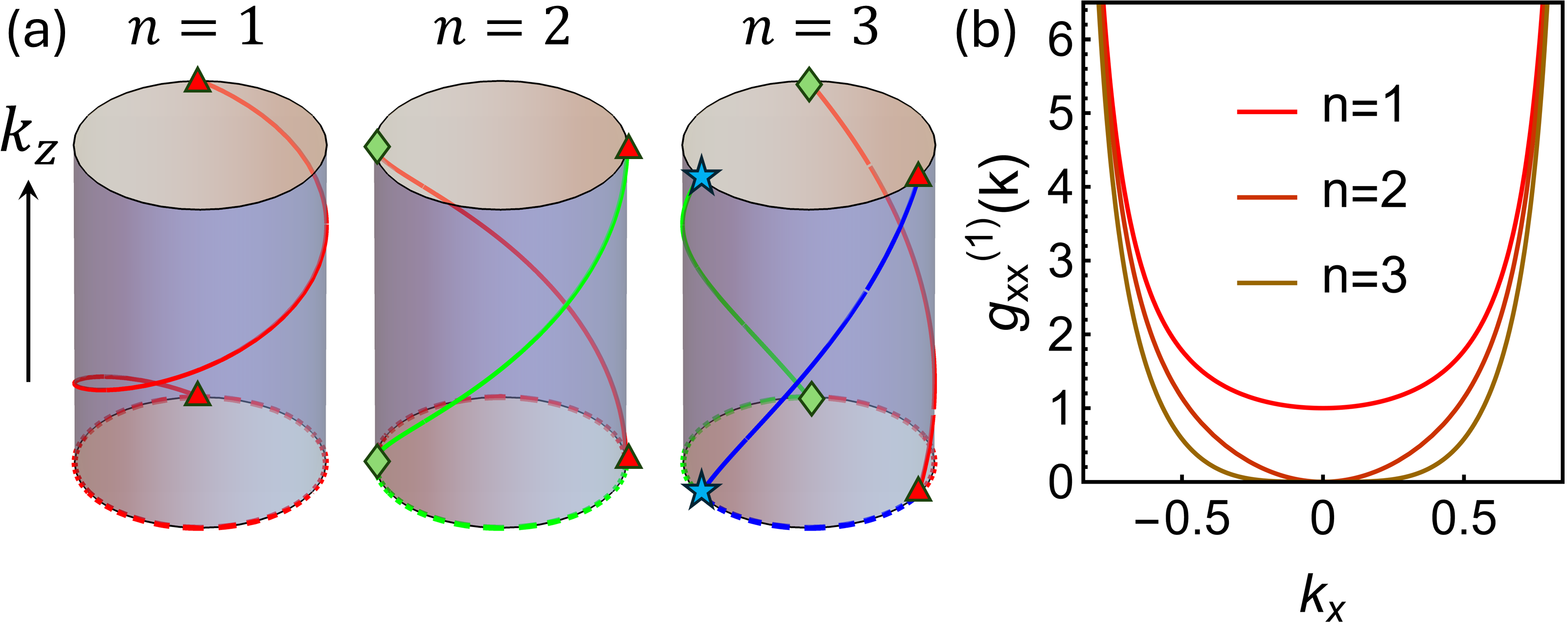}
\caption{Nodal lines and QM of the semimetal class Eq.~\eqref{eq:generalmodel} with IQG. (a) Visualization of nodal lines on a cylindrical surface of radius 1 and height $2\pi$ in momentum space, for models $n=1,2,3$ ($n=1$ corresponds to the RG model). The triangle, diamond, and star shapes indicate the same points in the 3D BZ. The dashed curves at the bottom indicate the nodal lines projected to the surface BZ. (b) Plots of the surface band QM, $g^{(1)}_{xx}(\bd{k})$ [from Eq.~\eqref{eq:qgtgeneral}] for models $n=1,2,3$.}
\label{fig:models}
\end{figure}

{\em Ideal quantum geometry in a class of semimetals}.---It is noteworthy that RG belongs to a class of semimetals having flat surface bands with IQG, but among them, only RG has a nearly uniform IQG in the drumhead region. This semimetal class has the bulk Hamiltonians in 3D momentum space
\begin{align}\label{eq:generalmodel}
\ham^{(n)}(k_x,k_y,k_z)=\begin{pmatrix}
0 & \text{c.c.}\\
(k_x+ik_y)^n+e^{ik_z} & 0
\end{pmatrix},
\end{align}
where c.c. refers to the complex conjugate entry that makes the Hamiltonian hermitian. Equation~\eqref{eq:generalmodel} describes 2D semimetals exhibiting $k^n$ dispersion and longer-range in-plane hoppings near the expansion point~\cite{zhu2018quadratic,gao2023topological}, and are coupled by the interlayer hopping $e^{ik_z}$ term. The exponent $n$ takes positive integer values, with $n=1$ corresponding to the RG case.

For the $n$th model above, energy degeneracies form $n$ nodal lines traversing from $k_z=0$ to $2\pi$, as shown in Fig.~\ref{fig:models}(a). When projected to the surface BZ in the $k_z$ direction, these nodal lines generate the complete boundary of a drumhead region defined by $k=\sqrt{k_x^2+k_y^2}<1$. Then, the surface state wavefunctions are similar to Eq.~\eqref{eq:twostates} but with replacements $k\rightarrow k^n$, $k_x+ik_y\rightarrow(k_x+ik_y)^n$, and the QGT is
\begin{align}\label{eq:qgtgeneral}
\mathcal{B}^{(j)}_{\mu\nu}(\bd{k})=\frac{n^2k^{2(n-1)}}{(1-k^{2n})^2}\begin{pmatrix}
    1 & \pm i\\
    \mp i & 1
\end{pmatrix}.
\end{align}
We note that the IQG $\propto k^{2(n-1)}$ near the center, which is nonvanishing for the RG case only [Fig.~\ref{fig:models}(b)]. The IQG of these models results from the holomorphic property of the surface state wavefunctions~\cite{claassen2015position,mera2021kahler,ozawa2021relations,wang2021exact}, which we explain in detail in SM Sec.~2.3. Therefore, Eq.~\eqref{eq:generalmodel} can be further generalized to a broader class of semimetals with IQG, if $(k_x+ik_y)^n$ is replaced with a polynomial of $k_x+ik_y$ (see SM Sec.~2.4).

{\em Effects of long-range hoppings and displacement field}.---In the energy scales relevant for correlated phenomena, the above discussed RG model is perturbed by long-range hoppings and the displacement field. We find that they compete with each other---the former tends to destroy the IQG, while the latter stabilizes it.

A general perturbation term $\hham'$ can be described by an effective Hamiltonian acting on the basis of Eq.~\eqref{eq:twostates}, $\la\psi^{(i)}_\bd{k}|\hham'|\psi^{(j)}_\bd{k}\ra$ ($i,j=1,2$), with the matrix elements computed in SM Sec.~3. For instance, the next-nearest-neighbor interlayer hopping term $\gamma_2$ couples different A/B sublattices [see Fig.~\ref{fig:schematic}(a)], leading to an off-diagonal matrix. Therefore, it mixes state $\psi^{(1)}_\bd{k}$ with $\psi^{(2)}_\bd{k}$ and destroys the IQG of both bands. Conversely, the displacement field term leads to a diagonal matrix $m(1-k^2)\sigma_z$, with $\pm m$ the potentials on the two surfaces and $\sigma_z$ the Pauli matrix, meaning a displacement field preserves $\psi^{(1)}_\bd{k}$, $\psi^{(2)}_\bd{k}$ as the eigenstates, thus preserving the IQG. We note that weak disorders can be viewed as random onsite potentials; therefore, their effect is similar to the displacement field, which we discuss briefly in SM Sec.~3.1.

The opposite roles of long-range hoppings and displacement field on the surface states can lead to intriguing phenomena. At zero displacement field, the large density of states (DOS) of a surface band is equally shared by the two surfaces due to the state mixing by the long-range hoppings. However, the displacement field effect will dominate if its magnitude exceeds a threshold value, restoring the two bands as $\psi^{(1)}_\bd{k}$ and $\psi^{(2)}_\bd{k}$, thereby polarizing the electron density of each band to one surface only~\cite{nissinen2021topological}. The effective Hamiltonians of the long-range hopping terms exponentially decay as $N$ increases (see SM Sec.~3.1), indicating that a large displacement field is particularly important for transitions to correlated states in few-layer RG. We use the superconducting state in rhombohedral trilayer graphene (RTG) to demonstrate this effect.

{\em Surface superconductivity}.---We assume an inter-valley pairing superconducting state in the valence surface band of RTG, with $s$-wave pairing symmetry arising from an effective onsite attractive interaction, $\hham_I=-U\sum_i\hat{n}_{i\uparrow}\hat{n}_{i\downarrow}$ ($U>0$), where $\hat{n}_{i\sigma}$ is the density operator for site $i$, spin $\sigma$. This state is consistent with the superconducting phase labeled by SC1 in recent experiments~\cite{zhou2021superconductivity}, for which we give a brief review in SM Sec.~5. Since SC1 lies in the weak coupling regime, the phase boundary of the metal-superconductor transition is highly sensitive to the DOS $D_0$, and the superconducting order parameter depends on $U$ and $D_0$ exponentially.

A large displacement field flattens the RTG surface bands and enhances the DOS, leading to the transition to SC1~\cite{zhou2021superconductivity,cea2022superconductivity,pantaleon2023review}. Within the RTG model including all the long-range hoppings and displacement field, we solve the zero-temperature mean-field gap equation $\Delta_{\alpha}=-U\la c_{\bd{R}\alpha\downarrow}c_{\bd{R}\alpha\uparrow}\ra$ for orbital $\alpha$ (which runs over 1A, 1B, ..., 3B for RTG), with $c_{\bd{R}\alpha\sigma}$ the annihilation operator for unit cell $\bd{R}$, orbital $\alpha$ and spin $\sigma$. The weak interaction $U$ guarantees that the only possibly nonzero order parameters are for the two surface orbitals, $\Delta_{1\text{A}}$ and $\Delta_{3\text{B}}$. We find that the larger one of the two (depending on the sign of $m$), e.g., $\Delta_{3\text{B}}$ depends on the surface potential $m$ via both the DOS $D_0(m)$ and the weight of orbital 3B in the band, $\rho^{(v)}_{3\text{B}}=|\psi^{(v)}_{3\text{B}}|^2$,
\begin{align}
\Delta_{3\text{B}}(m)=\frac{2\omega_c}{\rho^{(v)}_{3\text{B}}(m)}\exp\{-1/[2\rho^{(v)}_{3\text{B}}(m)D_0(m)U]\}
\end{align}
with $\omega_c$ an energy cutoff around the Fermi surface where the interaction is assumed attractive (see SM Sec.~5.2). As $m$ exceeds a threshold value, the valence band becomes polarized to the third layer, and $\rho^{(v)}_{3\text{B}}$ becomes doubled, contributing an additional factor of 2 to the coupling constant; therefore, the surface polarization aids the superconducting transition.

{\em Superfluid weight}.---The surface bands of mRG are barely affected by the long-range hoppings, and can remain dispersionless in the presence of a small displacement field (See SM Sec.~3.1), making them suitable for studies of flat-band superconductivity. Superfluid weight (stiffness), $D_s$, determines the existence of supercurrent and Meissner effect, and gives the penetration depth $\lambda_L\propto1/\sqrt{D_s}$~\cite{scalapino1992superfluid,scalapino1993insulator}. It contains a conventional contribution from the band dispersion, and a quantum geometric contribution that exists in multi-orbital systems and depends on the QGT of the band~\cite{peotta2015superfluidity,julku2016geometric,liang2017band,huhtinen2022revisiting,jiang2023pdw,jiang2024geometric,daido2024quantum}. The two contributions for a superconducting state of mRG doped to the valence band, which is similar to SC1, are $D_{s,\mu\nu}=\int\frac{\derd^2k}{(2\pi)^2}[f^\text{conv}_{\mu\nu}(\bd{k})+f^\text{geo}_{\mu\nu}(\bd{k})]$, with
\begin{align}
&f^\text{conv}_{\mu\nu}(\bd{k})=-\bigg(\frac{\xi_{v,\bd{k}}}{E_{v,\bd{k}}}+1\bigg)\partial_\mu\partial_\nu\xi_{v,\bd{k}},\\
f^\text{geo}_{\mu\nu}(\bd{k})&=\frac{1}{E_{v,\bd{k}}}\tr\{\partial_\mu P^{(v)}_\bd{k}\hat{\Delta}\partial_\nu P^{(v)}_\bd{k}\hat{\Delta}-\partial_\mu\partial_\nu P^{(v)}_\bd{k}\hat{\Delta}P^{(v)}_\bd{k}\hat{\Delta}\},\label{eq:integrands}
\end{align}
where $\xi_{v,\bd{k}}$ is the dispersion, $E_{v,\bd{k}}=\sqrt{\xi_{v,\bd{k}}^2+\Delta_{v,\bd{k}}^2}$ is the Bogoliubov quasiparticle energy, $\hat{\Delta}=\text{diag}(\Delta_\alpha)$ is the pairing gap in the orbital basis and $\Delta_{v,\bd{k}}=\la\psi^{(v)}_\bd{k}|\hat{\Delta}|\psi^{(v)}_\bd{k}\ra$ in the band basis~\cite{peotta2015superfluidity,jiang2023pdw} (see SM Sec.~4.3 for details). Here, the order parameter $\hat{\Delta}$ dresses the projection operator $P^{(v)}_\bd{k}$ to give the geometric integrand $f^\text{geo}_{\mu\nu}$. Expressions like Eq.~\eqref{eq:integrands} where interactions dress the QGT are also found for the susceptibilities of correlated phases in flat bands~\cite{han2024qgn,zhang2025identifying}.

\begin{figure}[b!]
\centering
\includegraphics[width=0.48\textwidth]{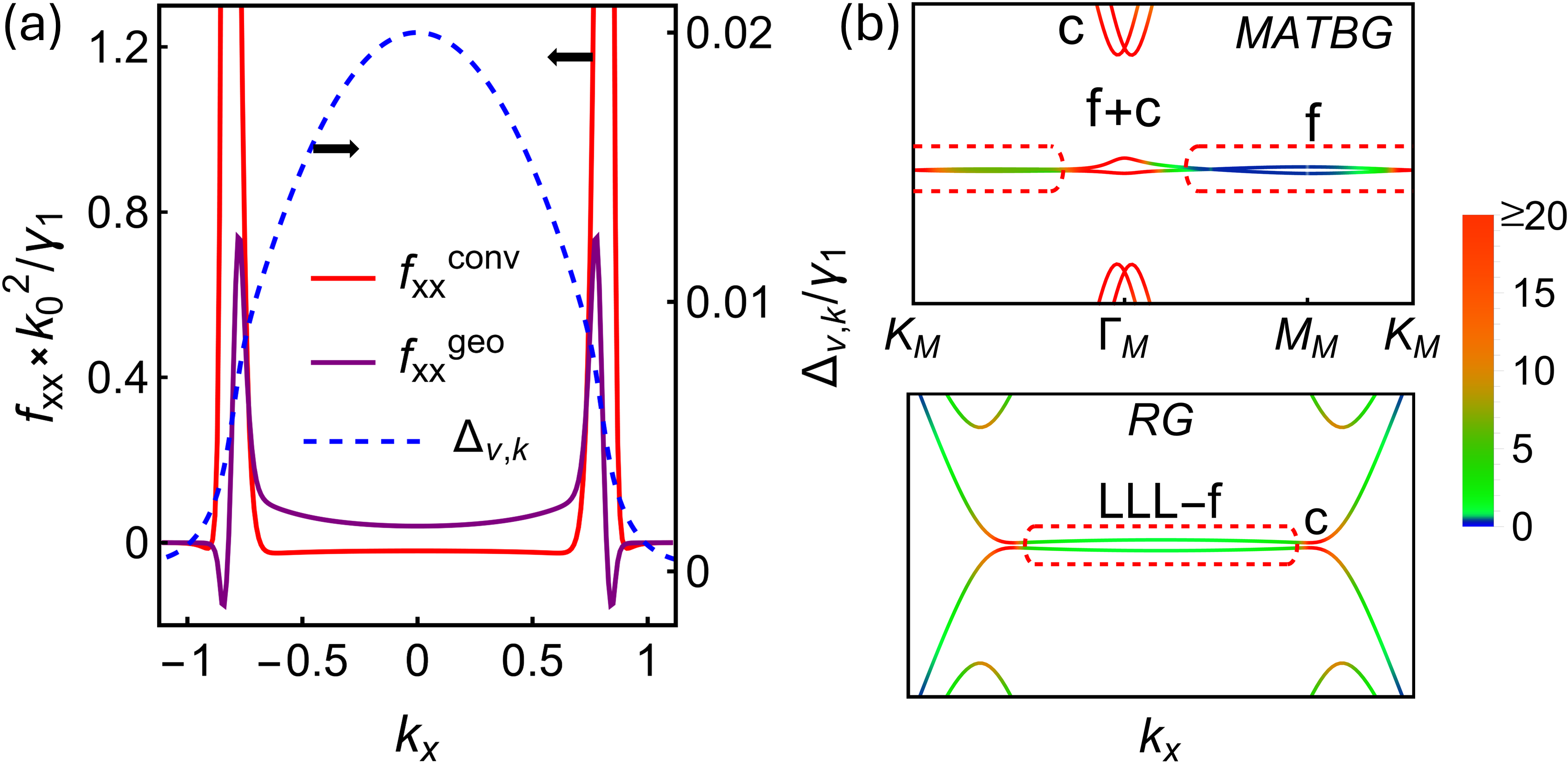}
\caption{(a) Superfluid weight integrands and order parameter $\Delta_{v,\bd{k}}$ for 20-layer RG doped to the valence band, using $\Delta_{20\text{B}}=0.02\gamma_1$, $m=-\mu=0.01\gamma_1$ ($\mu$ is the chemical potential) for RG model with $\gamma_0,\gamma_1$ hoppings. (b) Heavy-fermion pictures of MATBG vs RG, with the f-electrons enclosed in dashed contours. The band structures are calculated using a twist angle $\theta=1.05^\circ$ and interlayer hoppings $w_{AA}=\frac{0.9}{\sqrt{3}}$, $w_{AB}=\frac{1}{\sqrt{3}}$ for MATBG ($\Gamma_M,K_M,M_M$ are the high-symmetry points of the moir\'e BZ)~\cite{bernevig2021tbg}, and $m=0.01\gamma_1$ for 20-layer RG. Color of bands shows the QM scaled by the natural momentum units $k_\theta=\frac{8\pi}{3}\sin\frac{\theta}{2}$, and $k_0$ of the two systems, respectively.}
\label{fig:hfpicture}
\end{figure}

In Fig.~\ref{fig:hfpicture}(a), we first plot the two integrands for the 20-layer RG with solid curves. We consider a pairing matrix fully polarized to the 20th layer, $\hat{\Delta}=\text{diag}(0,...,0,\Delta_{20\text{B}})$, by the displacement field. The quantum geometric contribution becomes significant when $\Delta_{20\text{B}}$ is comparable or larger than the drumhead bandwidth, as chosen in the figure~\footnote{The order parameter used, $\Delta_{20\text{B}}=0.02\gamma_1$, is one order-of-magnitude larger than that is estimated for 20-layer RG using $U\sim40$ meV, which is extrapolated from SC1 of RTG.}. Both integrands peak at the rim of the drumhead region, as expected~\cite{kopnin2011surface}. The peaks of $f^\text{conv}_{\mu\nu}$ and $f^\text{geo}_{\mu\nu}$ are due to the strong dispersion and the nodes of $\Delta_{v,\bd{k}}$ (via the vanishing of denominator in Eq.~\eqref{eq:integrands}) at the rim, respectively. For mRG, $\Delta_{v,\bd{k}}$ approaches a bell-shape function $\Delta_{20\text{B}}(1-k^2)$~\cite{kopnin2011high}, as shown by the dashed curve in Fig.~\ref{fig:hfpicture}(a). What is unexpected is the nonzero, approximately uniform $f^\text{geo}_{\mu\nu}$ within the drumhead region, which resembles the QM there (see SM Sec.~4.4). These results suggest an unusual heavy-fermion description of the superconductivity in mRG, as we discuss below.

{\em Heavy-fermion picture and lower-bounded surface supercurrent}.---Heavy-fermion superconductors~\cite{steglich1979sc,sauls1994order,heffner1995heavy,pfleiderer2009superconducting} involve two species---the localized heavy electrons (``f") responsible for the Cooper pairing, and the itinerant light ones (``c") enabling phase coherence. An example of the usual heavy-fermion picture is the topological heavy-fermion model for the magic-angle twisted bilayer graphene (MATBG)~\cite{song2022magic,shi2022heavy,herzog2024topological}. In this model, the effective f-electrons refer to the molecular-like Wannier orbitals centered at the AA-stacking regions of MATBG, which constitute most of the flat bands, whereas c-electrons come from the high-energy conduction bands, which hybridize with f-electrons near the $\Gamma_M$ point only, see Fig.~\ref{fig:hfpicture}(b) (top panel) with the band structure calculated from the Bistritzer-MacDonald model~\cite{bistritzer2011moire} using parameters adopted from Ref.~\cite{bernevig2021tbg}. The localized f-electrons make the QGT in most regions of the MATBG flat bands negligible, as indicated by the deep blue color in Fig.~\ref{fig:hfpicture}(b), except for the hotspots at $\Gamma_M$ due to the hybridization with c-electrons, and at $K_M$, $K'_M$ (not shown) due to the f-band crossings. As a result, the superfluidity of MATBG~\cite{hu2019geometric,julku2020superfluid,xie2020topology,torma2022tbgreview} arises from these hotspots only.

In RG, the momentum scale $k_h$ classifies the electrons as dispersionless, strongly localized surface states (for $k<k_h$, $\lambda\ll N$) and dispersive, ``bulk-like" surface states (for $k\gtrsim k_h$, $\lambda\sim N$), which mimic the f and c-electrons in heavy-fermion compounds, respectively (Fig.~\ref{fig:hfpicture}(b), lower panel). The heavy-fermion picture of RG is intrinsically different from the usual ones like that of MATBG because its ``f-electrons" are localized in the $z$ direction only, but extended in the $x-y$ plane, as indicated by the nonvanishing IQG. In particular, the uniform IQG of its f-electrons is similar to the LLL; therefore, they can be dubbed as ``LLL-f-electrons".

Finally, we discuss the implications of the heavy-fermion picture in real space. The peaks of both integrands $f^\text{conv}_{\mu\nu}$ and $f^\text{geo}_{\mu\nu}$ at $k=k_h$ result from the ``bulk-like" surface states with $\lambda\sim N$, so they suggest a supercurrent flow penetrating into the bulk layers. On the contrary, the nearly uniform $f^\text{geo}_{\mu\nu}$ at $k<k_h$ results from these surface states with $\lambda\ll N$, implying a large fraction of the supercurrent running on the surface. The relative magnitude of the two integrands, and the magnitude of $f^\text{geo}_{\mu\nu}$ at the rim vs.~the center, depend on various parameters, which can be found in the additional figures in SM Sec.~4.5. When the drumhead region is at half filling, the surface fraction of superfluid weight has a lower-bound, $D^{\text{surf}}_{s,xx}\gtrsim \Delta_{N\text{B}}k_h^2/(2\pi)$ (see SM Sec.~4.4). The spatial profile of supercurrent affects the magnetic field screening inside RG, changing the profile of an external magnetic field near the RG surface, which can be probed by scanning magnetometry techniques.

{\em Discussion}.---Semimetals with IQG provide a pathway for realizing Chern bands with IQG, e.g., to engineer the bandgap near $k=k_h$ by forming an RG-hexagonal-boron-nitride heterostructure and tuning the twist angle between them. Importantly, the IQG of their surface states has negligible orbital-embedding dependence (see SM Sec.~2.3). This sheds light on the ongoing research~\cite{dong2024anomalous,herzog2024moire,dong2024theory,guo2024fractional,zhou2024fractional,huang2025fractional} on the fractional Chern insulating states observed in RG moir\'e systems~\cite{lu2024fractional,xie2025tunable,lu2025extended,waters2025chern}.

As the thickness of RG multilayers increases, the role of a large displacement field is weakened, and the superconducting state is anticipated to shift to the low-field regime, as was observed in recent experiments of 8-layer samples~\cite{kumar2025superconductivity}. The heavy-fermion picture established here also has consequences in phases competing with superconductivity, e.g., the ferromagnetic state in RG~\cite{lee2013band,myhro2018large,shi2020electronic,hagymasi2022observation,zhou2021half,lee2022gate,otani2010intrinsic,cuong2012magnetic,xu2012stacking,olsen2013ferro,awoga2023superconductivity}. Similar to the superfluid weight, the spin stiffness of a ferromagnetic state has a quantum geometric contribution that dominates in flat bands~\cite{wu2020quantum,kang2024quantum}; in ferromagentic mRG, this would affect the layer-dependent shape of magnetic textures and magnon eigenfunctions. 

\begin{acknowledgments}
{\em Acknowledgments.}---We thank Andrei Bernevig, Christophe De Beule, Guorui Chen, Anushree Datta, Aaron Dunbrack, Pertti Hakonen, Jonah Herzog-Arbeitman, Haoyu Hu, Jin-Xin Hu, Kry\v{s}tof Kol\'a\v{r}, Allan MacDonald, and Stevan Nadj-Perge for helpful discussions at different stages of this work and Sebastiano Peotta for his comments on the manuscript. This work was supported by the Research Council of Finland under project numbers 339313 and 354735, by Jane and Aatos Erkko Foundation, Keele Foundation, Magnus Ehrnrooth Foundation, and a collaboration between The Kavli Foundation, Klaus Tschira Stiftung, and Kevin Wells, as part of the SuperC collaboration, and by a grant from the Simons Foundation (SFI-MPS-NFS-00006741-12, P.T.) in the Simons Collaboration on New Frontiers in Superconductivity.
\end{acknowledgments}


%

\pagebreak
\onecolumngrid

\setcounter{figure}{0}
\setcounter{equation}{0}
\renewcommand\thefigure{S\arabic{figure}}
\renewcommand\theequation{S\arabic{equation}}

\section*{Supplementary material for ``Ideal quantum geometry of the surface states of rhombohedral graphite and its effects on the surface superconductivity"}
\section{1. Non-interacting Hamiltonian of rhombohedral graphite and rhombohedral multilayer graphene}
\begin{figure}[th]
	\centering	
	\includegraphics[height=3cm]{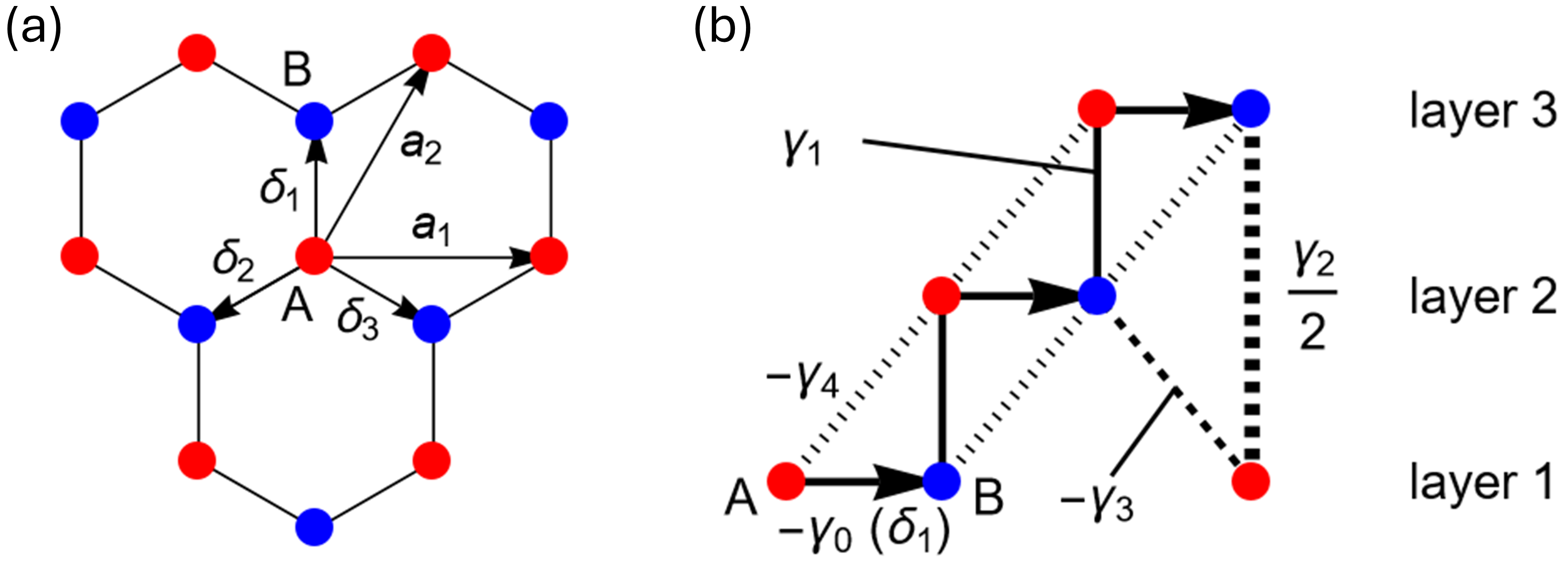}
	\caption{Summary of conventions for the RG multilayers. (a) $\bd{a}_1,\bd{a}_2$ are the Bravais lattice vectors and $\bds{\delta}_1,\bds{\delta}_2,\bds{\delta}_3$ are the three nearest-neighbor hopping vectors. (b) Side view of RG multilayers from $\bd{a}_1$ direction with hopping integrals $-\gamma_0$, $\gamma_1$, $\gamma_2/2$, $-\gamma_3$ and $-\gamma_4$.}
	\label{fig:conventions}
\end{figure}

The Bravais lattice vectors for a single-layer graphene are chosen as $\bd{a}_1=(1,0)a_0$, $\bd{a}_2=(\frac{1}{2},\frac{\sqrt{3}}{2})a_0$, and the three nearest-neighbor hopping vectors from sublattice A to B are $\bds{\delta}_1=(0,\frac{\sqrt{3}}{3})a_0$, $\bds{\delta}_2=(-\frac{1}{2},-\frac{\sqrt{3}}{6})a_0$, $\bds{\delta}_3=(\frac{1}{2},-\frac{\sqrt{3}}{6})a_0$ [Fig.~\ref{fig:conventions}(a)]. Atom A at $(0,0)$ and B at $\bds{\delta}_1$ are grouped as a unit cell. Dirac points of the two valleys, $K, K'$, are chosen as $K=\frac{2\pi}{a_0}(\frac{2}{3},0)$, $K'=\frac{2\pi}{a_0}(-\frac{2}{3},0)$. For multilayer RG, the $(n+1)$-th layer can be viewed as the $n$th layer shifted by vector $\bds{\delta}_1$ in the basal plane, therefore, the unit cell is chosen as the staircase structure consisting of horizontal ($\bds{\delta}_1$) and vertical ($c\hat{z}$) bonds, as shown in Fig.~\ref{fig:conventions}(b). The tight-binding Hamiltonian in real space for each spin $\sigma=\uparrow,\downarrow$, including all hoppings, reads
\be
\hham=-\gamma_0\sum_{\la i,j\ra\sim\gamma_0} c_i^\dagger c_j+\gamma_1\sum_{\la i,j\ra\sim\gamma_1} c_i^\dagger c_j+\frac{\gamma_2}{2}\sum_{\la i,j\ra\sim\gamma_2} c_i^\dagger c_j-\gamma_3\sum_{\la i,j\ra\sim\gamma_3} c_i^\dagger c_j-\gamma_4\sum_{\la i,j\ra\sim\gamma_4}c_i^\dagger c_j,
\ee
where $i,j$ generically label all the carbon sites. Symbol $\la i,j\ra\sim\gamma_s\,(s=0,1,...,4)$ refers to the pairs of sites $i,j$ that are linked by the hopping $\gamma_s$. The sum includes both the $\{i,j\}$ and the $\{j,i\}$ terms of the pair to make the Hamiltonian hermitian. The site label $i$ can also be replaced with the tuple $\bd{R}\alpha$, where the in-plane vector $\bd{R}$ labels the unit cell, and $\alpha$ runs over the $2N$ orbital indices in a unit cell, 1A, 1B, ..., $N$A, $N$B. We define $\bd{R}$ to be the position of atom A in the 1st layer (atom 1A).

To obtain the continuum model Hamiltonian, we perform the Fourier transformation of field operators with a local gauge choice, $c_{\bd{k}\alpha}\equiv\frac{1}{\sqrt{N_c}}\sum_\bd{R}e^{-i\bd{k}\cdot(\bd{R}+\bd{r}_\alpha)}c_{\bd{R}\alpha}$ ($\alpha=$1A, 1B,..., $N$A, $N$B), where $\bd{r}_{z\text{A}}=(z-1)\bds{\delta}_1$, $\bd{r}_{z\text{B}}=z\bds{\delta}_1$ ($z$ is the layer index) are the positions of the orbitals in the unit cell measured from 1A. Under these conventions and gauge choices, the intra-layer hopping term gives the Dirac effective Hamiltonian
\be\label{eq:hgamma0}
\hham^{(\gamma_0)}=\sum_{\bd{k}}\sum_{z=1}^N\psi_{\bd{k},z}^\dagger h^{(\xi)}_\bd{k}\psi_{\bd{k},z},
\ee
where $h^{(\xi)}_\bd{k}=\frac{\sqrt{3}}{2}\gamma_0a_0(\xi\sigma_xk_x+\sigma_y k_y)$, with $\xi=\pm$ for valley $K$ and $K'$, respectively, and $\bd{k}$ is the in-plane momentum measured from $K$ or $K'$. The two-component spinor $\psi_{\bd{k},z}\equiv(c_{\bd{k},z\text{A}},c_{\bd{k},z\text{B}})^T$.

\tit{Throughout the paper, we assume valley degeneracy}. The Hamiltonian at valley $K'$ is related to that at $K$ by time-reversal symmetry (TRS). The Hamiltonian terms associated with $\gamma_1-\gamma_4$ hoppings near $K$ are
\be\label{eq:hgamma1to4}
\hham^{(\gamma_1)}=\sum_{\bd{k}}\sum_{z=1}^{N-1}\psi_{\bd{k},z+1}^\dagger\gamma_1\sigma_+\psi_{\bd{k},z}+h.c.,
\ee
\be
\hham^{(\gamma_2)}=\sum_{\bd{k}}\sum_{z=1}^{N-2}\psi_{\bd{k},z+2}^\dagger\frac{\gamma_2}{2}\sigma_-\psi_{\bd{k},z}+h.c.,
\ee
\be
\hham^{(\gamma_3)}=\sum_{\bd{k}}\sum_{z=1}^{N-1}\psi_{\bd{k},z+1}^\dagger\big[\frac{\sqrt{3}}{2}\gamma_3a_0(k_x-ik_y)\sigma_-\big]\psi_{\bd{k},z}+h.c.,
\ee
\be
\hham^{(\gamma_4)}=\sum_{\bd{k}}\sum_{z=1}^{N-1}\psi_{\bd{k},z+1}^\dagger\big[\frac{\sqrt{3}}{2}\gamma_4a_0(k_x+ik_y)\sigma_0\big]\psi_{\bd{k},z}+h.c.
\ee
where $\sigma_\pm=\sigma_x\pm i\sigma_y$. Besides these terms, there is also the displacement field term. Without knowing the variation of the potential with layers, we make a metallic assumption that the potential only drops between the surfaces and their adjacent layers. Then, the displacement field term reads
\be\label{eq:hm}
\hham^{(m)}=\sum_{\bd{k}}m[\psi_{\bd{k},1}^\dagger \psi_{\bd{k},1}-\psi_{\bd{k},N}^\dagger \psi_{\bd{k},N}].
\ee

The Fermi velocity of single-layer graphene is $v_f=\frac{\sqrt{3}}{2}\gamma_0a_0$ and we define $v_3=\frac{\sqrt{3}}{2}\gamma_3a_0$, $v_4=\frac{\sqrt{3}}{2}\gamma_4a_0$. Choosing $\gamma_1$ and $k_0=\gamma_1/v_f$ as the units of energy and momentum, respectively, the Hamiltonians above can be written in terms of dimensionless parameters $\wt{\gamma}_2=\gamma_2/\gamma_1$, $\wt{m}=m/\gamma_1$, $\wt{v}_3=v_3/v_f=\gamma_3/\gamma_0$ and $\wt{v}_4=v_4/v_f=\gamma_4/\gamma_0$. Without loss of generality, we use the parameters of trilayer RG for the other multilayers. These are adopted from Ref.~\cite{zhou2021half}: $\gamma_0=3.1$ eV, $\gamma_1=0.38$ eV, $\gamma_2=-0.015$ eV, $\gamma_3=-0.29$ eV, $\gamma_4=-0.141$ eV. Then, the dimensionless parameters are $\wt{\gamma}_2=-0.039$, $\wt{v}_3=-0.094$, $\wt{v}_4=-0.045$.
\section{2. Ideal quantum geometry (IQG) of the semi-metallic surface states}
\subsection{2.1 IQG of RG continuum model calculation}
\label{sec:qgcontinuum}
We present the details of the QGT calculation for the RG continuum model. The surface eigenstates can be solved from the bulk Hamiltonian of RG subject to the open boundary condition \cite{heikkila2011dimensional}. The bulk Hamiltonian in the 3D momentum space is
\be\label{eq:h3d}
\ham(k_x,k_y,k_z)=\begin{pmatrix}
    0 & f(k_x,k_y,k_z)^*\\
    f(k_x,k_y,k_z) & 0
\end{pmatrix},
\ee
where $f(k_x,k_y,k_z)=-\gamma_0\sum_{i=1}^3e^{i\bd{k}\cdot\bds{\delta}_i}+\gamma_1e^{ick_z}$, with $\bd{k}=(k_x,k_y)$ the in-plane momentum, $\bds{\delta}_i$ ($i=1,2,3$) the three in-plane nearest neighbor hopping vectors, and $c$ the interlayer distance. As above, we choose $\gamma_1$ and $k_0=\gamma_1/v_f$ as the units. We also rescale the $z$ axis by taking $c=a_0=1$. Then, expansion near the $K$ point gives $f(k_x,k_y,k_z)\approx k_x+ik_y+e^{ik_z}$. Replacing $k_z\rightarrow-i\partial_z$ and making the ansatz wavefunction $e^{\kappa z}\phi_\bd{k}$, , the Hamiltonian leads to an eigen-equation for the surface states:
\be
\begin{pmatrix}
    0 & k_x-ik_y+e^{-\kappa}\\
    k_x+ik_y+e^{\kappa} & 0
\end{pmatrix}\phi_\bd{k}=\varepsilon\phi_\bd{k}
\ee
with two degenerate solutions of $\phi_{i,\bd{k}}$ and $\kappa_i$ $(i=1,2)$ for each energy $\varepsilon$. The final eigenstate is a linear combination
\be
\psi_{\bd{k},z}=c_1e^{\kappa_1 z}\phi_{1,\bd{k}}+c_2e^{\kappa_2 z}\phi_{2,\bd{k}},
\ee
with coefficients $c_1,c_2$ determined from the boundary condition $\psi_{\bd{k},z=0,\text{B}}=\psi_{\bd{k},z=N+1,\text{A}}=0$.

Exact solutions for finite $N$ lead to transcendental equations for $c_1,c_2$, which also determines the dispersion $\varepsilon=\pm k^N$ of the surface states. To avoid mathematical complexity, we take the limit $N\rightarrow \infty$. Then, we obtain the normalized eigenstates
\be\label{eq:surfacestates}
\psi^{(1)}_{\bd{k},z}=\frac{\sqrt{1-k^2}}{k}e^{\kappa(\bd{k})z}\begin{pmatrix}
    1 \\ 0
\end{pmatrix},\,\,\,\psi^{(2)}_{\bd{k},z}=\frac{\sqrt{1-k^2}}{k}e^{\kappa(\bd{k})^*(N+1-z)}\begin{pmatrix}
    0 \\ 1
\end{pmatrix},
\ee
with $\kappa(\bd{k})=\ln[-(k_x+ik_y)]$. QGT can be directly calculated in the orbital basis, e.g., for state $\psi^{(1)}_\bd{k}$, using $P^{(1)}_\bd{k} = |\psi^{(1)}_\bd{k}\rangle \langle \psi^{(1)}_\bd{k}|$
\be
\la z\text{A}|P^{(1)}_\bd{k}|z'\text{A}\ra=\frac{1-k^2}{k^2}e^{\kappa z}e^{\kappa^*z'},
\ee
\be
\la z\text{A}|\partial_\mu P^{(1)}_\bd{k}|z'\text{A}\ra=e^{\kappa z}e^{\kappa^*z'}\bigg[\partial_\mu\bigg(\frac{1-k^2}{k^2}\bigg)+\frac{1-k^2}{k^2}z\partial_\mu\kappa +\frac{1-k^2}{k^2}z'\partial_\mu\kappa^*\bigg],
\ee
where $\kappa(\bd{k})=\ln[-(k_x+ik_y)]$. The QGT $\mathcal{B}^{(1)}_{\mu\nu}$ can be expressed as
\be\label{eq:r1}
\begin{split}
\mathcal{B}^{(1)}_{\mu\nu}(\bd{k})=&\frac{1-k^2}{k^2}\partial_\mu\bigg(\frac{1-k^2}{k^2}\bigg)\partial_\nu\bigg(\frac{1-k^2}{k^2}\bigg)\la1\ra^3\\
&+\bigg(\frac{1-k^2}{k^2}\bigg)^2\bigg[\partial_\mu \bigg(\frac{1-k^2}{k^2}\bigg)\partial_\nu\kappa+\partial_\mu\bigg(\frac{1-k^2}{k^2}\bigg)\partial_\nu\kappa^*+\partial_\mu\kappa\partial_\nu\bigg(\frac{1-k^2}{k^2}\bigg)+\partial_\mu\kappa^*\partial_\nu\bigg(\frac{1-k^2}{k^2}\bigg)\bigg]\la1\ra^2\la z\ra\\
&+\bigg(\frac{1-k^2}{k^2}\bigg)^3(\partial_\mu\kappa\partial_\nu\kappa+\partial_\mu\kappa\partial_\nu\kappa^*+\partial_\mu\kappa^*\partial_\nu\kappa^*)\la1\ra\la z\ra^2+\bigg(\frac{1-k^2}{k^2}\bigg)^3\partial_\mu\kappa^*\partial_\nu\kappa\la1\ra^2\la z^2\ra,
\end{split}
\ee
where we have defined the moments
\be
\begin{split}
&\la1\ra\equiv \sum_{z=1}^\infty e^{(\kappa+\kappa^*)z}=\frac{k^2}{1-k^2},\\
&\la z\ra\equiv \sum_{z=1}^\infty ze^{(\kappa+\kappa^*)z}=\frac{k^2}{(1-k^2)^2},\\
&\la z^2\ra\equiv \sum_{z=1}^\infty z^2e^{(\kappa+\kappa^*)z}=\frac{k^2(1+k^2)}{(1-k^2)^3}.
\end{split}
\ee
Then Eq.~\eqref{eq:r1} gives the QGT presented in the MS. We want to comment on the origin of the nonzero QGT at the center of the drumhead region. From the procedures above one can see it arises from the gradient of both $\kappa(\bd{k})$ and $(1-k^2)/k^2$, but the $\bd{k}$-dependence of $(1-k^2)/k^2$ comes from that of $\kappa(\bd{k})$.

\subsection{2.2 Comparison with the QGT of RG two-orbital effective model}
\begin{figure}[ht]
	\centering	
	\includegraphics[height=3.5cm]{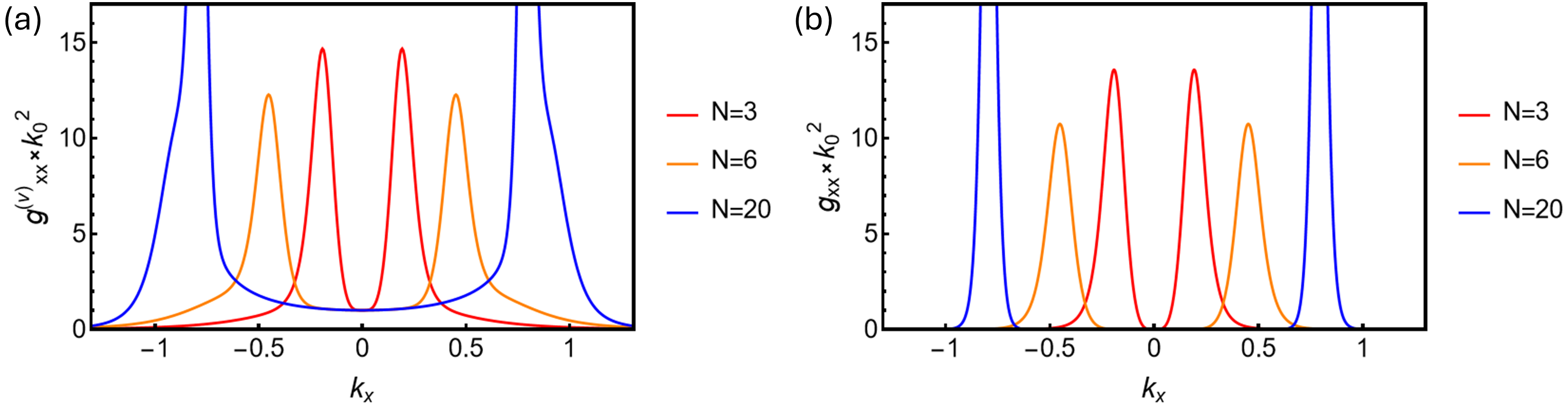}
	\caption{Contrasting the valence band QM of (a) the continuum model, $g^{(v)}_{xx}$ with (b) the two-band effective model, $\tilde{g}^{(v)}_{xx}$ along $k_y=0$, for $N=3,6,20$ at $m=0.01\gamma_1$.}
	\label{fig:qmxxm001}
\end{figure}

We show that the frequently adopted two-orbital effective Hamiltonian of RG does not contain the IQG information, since it neglects the momentum-dependent decaying property of the surface states. The two-orbital effective Hamiltonian of $N$-layer RG is derived by projecting the continuum model to orbitals 1A and $N$B~\cite{koshino2009trigonal,mccann2006landau,min2008chiral,guinea2006electronic,koshino2009trigonal}. The two-orbital Hamiltonian including $\gamma_0,\gamma_1$ and surface potential $m$ (in the units of $\gamma_1$) reads
\be\label{eq:twobandeff}
\ham_\text{eff}(\bd{k})=(-1)^{N-1}\begin{pmatrix}
m & \pi^{*N}\\
\pi^N & -m
\end{pmatrix},
\ee
with $\pi=\xi k_x+ik_y$ ($\xi=\pm$ for $K$ or $K'$). Quantum metric (QM) of this two-orbital model at valley $K$ is found to be \cite{hu2023effect,jiang2025superfluid}
\be\label{eq:qmtwobandeff}
\tilde{g}_{\mu\nu}^{(v)}(\bd{k})=\frac{N^2k^{2N-2}}{4(k^{2N}+m^2)^2}\bigg(m^2\delta_{\mu\nu}+\frac{k^2\delta_{\mu\nu}-k_\mu k_\nu}{k^{2-2N}}\bigg).
\ee
In Fig.~\ref{fig:qmxxm001}, we compare the valence band QM calculated from (a) the continuum model, $g_{\mu\nu}^{(v)}(\bd{k})$ (same as Fig.~2(c) in the manuscript (MS)), with (b) that from the the two-orbital model, $\tilde{g}_{\mu\nu}^{(v)}(\bd{k})$ at surface potential $m=0.01\gamma_1$. Since the two-orbital model ignores the layer degree of freedom, $\tilde{g}_{\mu\nu}^{(v)}(\bd{k})$ vanishes at the center of the drumhead region; it also loses some details near the rim of the drumhead region as it is unable to describe the bandgap closing effect between the surface and bulk bands. Nevertheless, it contains the A/B sublattice degree of freedom, and thus captures most of the features at $k=k_h$ due to surface hybridization.

\begin{figure}[ht]
	\centering	
	\includegraphics[height=3.5cm]{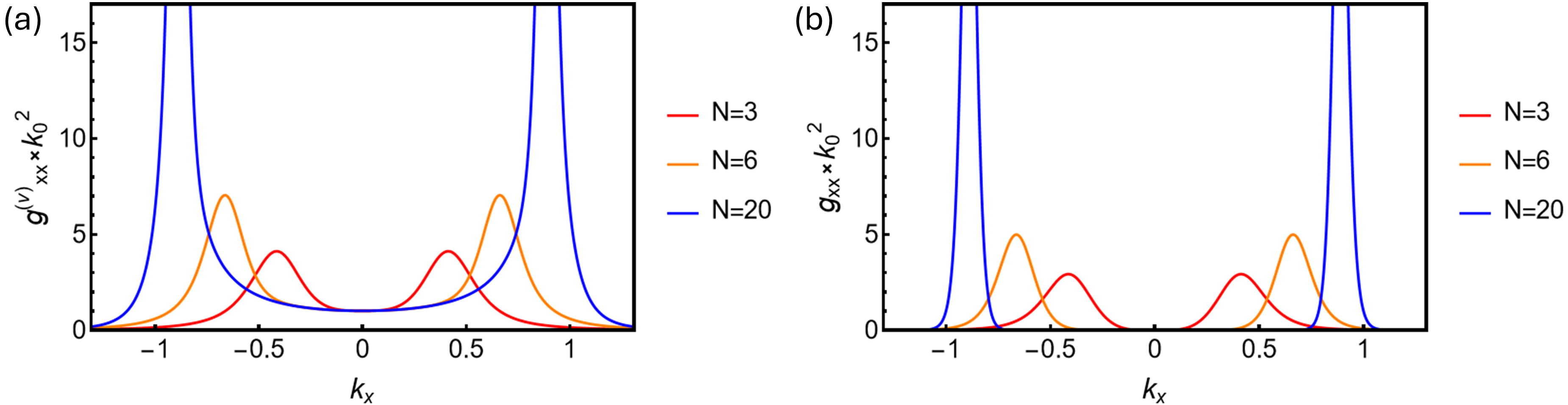}
	\caption{Same as Fig.~\ref{fig:qmxxm001}, but with a larger surface potential $m=0.1\gamma_1$.}
	\label{fig:qmxxm01}
\end{figure}

In Fig.~\ref{fig:qmxxm01} we plot these QMs for a larger surface potential $m=0.1\gamma_1$. Firstly, we notice that the QM at the center does not depend on $m$. Secondly, a larger $m$ leads to a larger bandgap between the two surface bands, making the QM peaks at $k_h$ less singular \cite{jiang2025superfluid}. Lastly, the peak position (rim of the drumhead), $k_h$, increases slightly with increasing $m$. For large $m$, we compute $k_h$ by finding the maxima of Eq.~\eqref{eq:qmtwobandeff}, giving
\be\label{eq:khwithh}
k_h=\bigg(\frac{N-1}{N+1}\bigg)^{1/N}m^{1/N}\approx m^{1/N}.
\ee
Eq.~\eqref{eq:khwithh} is only valid for large $m$, since the two-orbital model does not work near $k=0$.

\subsection{2.3 Origin of IQG and its orbital-embedding-dependence}
In this section, we explain the origin of IQG found in the surface states of RG, which lays the foundation for generalizing it to other semimetal models. We show the orbital-embedding-dependence~\cite{simon2020contrasting} of IQG, but we explain why the dependence is weak for the surface states of semimetals. \tit{Throughout this work, orbital-embedding refers to the $U(1)\times U(1)\times...\times U(1)$ gauge freedom of the Bloch Hamiltonian $\ham(\bd{k})$,} which is explained explicitly in Prop.~\ref{prop:prop2} below.

We first show an equivalent condition of IQG:
\begin{propt}\label{prop:prop1}
The quantum geometry is ideal if and only if one of the two quantities, $\tr\{P\frac{\partial P}{\partial\pi}\frac{\partial P}{\partial\pi^*}\}$ and $\tr\{P\frac{\partial P}{\partial\pi^*}\frac{\partial P}{\partial\pi}\}$, vanishes, where $\pi=k_x+ik_y$, $\pi^*=k_x-ik_y$, and $P$ is the projection operator to a single band or a set of bands.
\end{propt}
\tit{Proof:} A similar proof is given in Ref.~\cite{claassen2015position}. Note that both quantities are real, and they may not be equal. We replace the variables $k_x,k_y\rightarrow \pi,\pi^*$ and treat the projection operator as a function of $\pi,\pi^*$: $P_\bd{k}\rightarrow P_{\pi,\pi^*}$. After the substitution, we get
\be
\begin{split}
\tr g=&\mathcal{B}_{xx}+\mathcal{B}_{yy}=2\tr\big\{P\frac{\partial P}{\partial\pi}\frac{\partial P}{\partial\pi^*}+P\frac{\partial P}{\partial\pi^*}\frac{\partial P}{\partial\pi}\big\},\\[5pt]
\Omega=&-2\im\mathcal{B}_{xy}=2\tr\big\{P\frac{\partial P}{\partial\pi}\frac{\partial P}{\partial\pi^*}-P\frac{\partial P}{\partial\pi^*}\frac{\partial P}{\partial\pi}\big\}.
\end{split}
\ee
Both $\tr\{P\frac{\partial P}{\partial\pi}\frac{\partial P}{\partial\pi^*}\}$ and $\tr\{P\frac{\partial P}{\partial\pi^*}\frac{\partial P}{\partial\pi}\}$ are non-negative, e.g., for the single-band projection case,
\be\label{eq:singleband}
\tr\big\{P\frac{\partial P}{\partial\pi}\frac{\partial P}{\partial\pi^*}\big\}=\la\partial_\pi\psi|(1-|\psi\ra\la\psi|)|\partial_{\pi^*}\psi\ra=\la\partial_\pi\psi|Q|\partial_{\pi^*}\psi\ra\geq 0,
\ee
since $Q=1-P$ is a positive-semidefinite operator. Here, we use the convention that $\la\partial_\pi\psi|$ is the dual state of $|\partial_{\pi^*}\psi\ra$. Similarly, $\tr\{P\frac{\partial P}{\partial\pi^*}\frac{\partial P}{\partial\pi}\}=\la\partial_{\pi^*}\psi|Q|\partial_{\pi}\psi\ra\geq 0$. Therefore, it is easy to establish $\tr g\geq|\Omega|$, since $a+b\geq|a-b|$ for positive $a,b$, and the inequality saturates when $a=0$ or $b=0$.

Next, we show that the IQG is an orbital-embedding-dependent property with the proposition below.
\begin{propt}\label{prop:prop2}
The condition $\tr\{P\frac{\partial P}{\partial\pi}\frac{\partial P}{\partial\pi^*}\}=0$ or $\tr\{P\frac{\partial P}{\partial\pi^*}\frac{\partial P}{\partial\pi}\}=0$ is orbital-embedding-dependent.
\end{propt}
\tit{Proof:} As the orbital embedding changes, the Bloch Hamiltonian matrix transforms as $\ham(\bd{k})\rightarrow U^\dagger(\bd{k})\ham(\bd{k})U(\bd{k})$ and the state transforms as $|\psi_\bd{k}\ra\rightarrow U^\dagger(\bd{k})|\psi_\bd{k}\ra$, where the diagonal unitary matrix $U(\bd{k})=\text{diag}\{e^{i\bd{k}\cdot\delta\bd{r}_\alpha}\}$ has the dimension equal to the number of orbitals in a unit cell, with $\delta\bd{r}_\alpha=(x_\alpha,y_\alpha)$ the position shift of orbital $\alpha$. Since we have changed variables $k_x,k_y\rightarrow \pi,\pi^*$, we also introduce complex coordinates $X_\alpha=x_\alpha+iy_\alpha$, $X^*_\alpha=x_\alpha-iy_\alpha$, then, $U(\bd{k})=U_{\pi,\pi^*}=\text{diag}\{e^{(i/2)(\pi X^*_\alpha+\pi^* X_\alpha)}\}$. Using these notations, we find that under an orbital embedding transformation $P\rightarrow\wt{P}=U^\dagger PU$,
\be\label{eq:newtrace}
\tr\big\{\wt{P}\frac{\partial \wt{P}}{\partial\pi}\frac{\partial \wt{P}}{\partial\pi^*}\big\}=\tr\big\{P\frac{\partial P}{\partial\pi}\frac{\partial P}{\partial\pi^*}\big\}+\tr\big\{\frac{1}{4}(PX^*X-PX^*PX)-\frac{i}{2}P\frac{\partial P}{\partial \pi}X+\frac{i}{2}\frac{\partial P}{\partial \pi^*}PX^*\big\}.
\ee
with $X=\text{diag}\{X_\alpha\}$. On the r.h.s. of Eq.~\eqref{eq:newtrace}, the first term vanishes by assumption. If $X$ is proportional to the identity matrix, the second term also vanishes by the $U(1)$ gauge-invariance of QGT; otherwise, the second term is generally nonzero. As an example, if we shift the position of the first orbital only, i.e., $X=\text{diag}\{X_1,0,...,0\}$, then the second term of Eq.~\eqref{eq:newtrace}, when written in the orbital basis, becomes
\be
\frac{1}{4}(P_{11}-P_{11}^2)|X_1|^2-\frac{i}{2}\sum_\alpha(P_{1\alpha}\frac{\partial P_{\alpha 1}}{\partial\pi}X_1-\frac{\partial P_{1\alpha }}{\partial\pi^*}P_{\alpha 1}X_1^*)
\ee
which is positive when $|X_1|$ is large, since the coefficient $P_{11}-P_{11}^2=|\psi_1|^2-|\psi_1|^4>0$ (assuming $|\psi_1|<1$). Therefore, the general orbital-embedding-dependence is proved (QED).

A full analysis of Eq.~\eqref{eq:newtrace} is given right after Prop.~\ref{prop:prop4} below. Here, a remark is that \tit{even though IQG is an orbital-embedding-dependent property, it is still of importance if IQG is satisfied under a special physical orbital embedding,} e.g., for the RG case when we choose the local embedding where the carbon atoms are located.

From Prop.~\ref{prop:prop1}, we immediately get another equivalent condition:
\begin{propt}\label{prop:prop3}
The quantum geometry of a single band $\psi$ is ideal if and only if it is a function of $\pi$ or $\pi^*$ only, besides a normalization factor, i.e., $|\psi\ra=\mathcal{N}(\pi,\pi^*)|\phi(\pi)\ra$ or $|\psi\ra=\mathcal{N}(\pi,\pi^*)|\phi(\pi^*)\ra$. In particular, in the orbital basis, the state components $\psi_\alpha(\pi,\pi^*)=\mathcal{N}(\pi,\pi^*)\phi_\alpha(\pi)$, $\forall\,\alpha$ or $\mathcal{N}(\pi,\pi^*)\phi_\alpha(\pi^*)$, $\forall\,\alpha$.
\end{propt}
Similar statements have been shown or discussed in Ref.~\cite{claassen2015position,mera2021kahler,ozawa2021relations,wang2021exact}. We have it adapted to the orbital basis here. This result is obvious if we start from Eq.~\eqref{eq:singleband} and notice that $\la\partial_\pi\psi|Q|\partial_{\pi^*}\psi\ra=0$ means $\partial_{\pi^*}|\psi\ra\propto |\psi\ra$. Writing $\psi_\alpha(\pi,\pi^*)=\mathcal{N}(\pi,\pi^*)\phi_\alpha(\pi,\pi^*)$, it implies that in $\partial_{\pi^*}\psi_\alpha(\pi,\pi^*)$, the linear operator $\partial_{\pi^*}$ acts on the factor $\mathcal{N}(\pi,\pi^*)$ only, so $\phi_\alpha=\phi_\alpha(\pi)$.

This proposition explains why the surface states of RG, Eq.~\eqref{eq:surfacestates}, have an IQG. It also implies that the IQG is an orbital-embedding-dependent property, since an orbital embedding transformation $U_{\pi,\pi^*}$ generally makes $\phi_\alpha$ depend on both $\pi$ and $\pi^*$. However, for the surface states of RG, the following result is expected:
\begin{propt}\label{prop:prop4}
IQG in the drumhead region of RG surface bands has a weak orbital-embedding-dependence; at the center of the drumhead region, the IQG is orbital-embedding-independent.
\end{propt}
We sketch the proof by delineating a few useful results. We first note that when the first term in Eq.~\eqref{eq:newtrace} vanishes (in the case of IQG), the terms linear in $X,X^*$ in its second term must vanish also, i.e.,
\be
\tr\big\{P\frac{\partial P}{\partial \pi}\frac{\partial P}{\partial \pi^*}\big\}=0\,\,\,\Longrightarrow\,\,\,\la\alpha|P\frac{\partial P}{\partial\pi}|\alpha\ra=0\,\,\,\text{and}\,\,\,\la\alpha|\frac{\partial P}{\partial\pi^*}P|\alpha\ra=0,\,\,\,\forall\,\alpha,
\ee
which can be easily verified using Prop.~\ref{prop:prop3}. This means that when $P$ has IQG, Eq.~\eqref{eq:newtrace} is a positive-semidefinite quadratic form:
\be\label{eq:quadraticform}
\begin{split}
&\tr\big\{\wt{P}\frac{\partial \wt{P}}{\partial\pi}\frac{\partial \wt{P}}{\partial\pi^*}\big\}=\tr\big\{\frac{1}{4}(PX^*X-PX^*PX)\}=\frac{1}{4}\sum_{\alpha\beta}X_\alpha^*\mathcal{M}_{\alpha\beta}X_\beta,\\
&\text{with}\,\,\,\mathcal{M}_{\alpha\beta}=P_{\alpha\alpha}\delta_{\alpha\beta}-P_{\alpha\beta}P_{\beta\alpha}=|\psi_\alpha|^2\delta_{\alpha\beta}-|\psi_\alpha|^2|\psi_\beta|^2.
\end{split}
\ee
$P_{\alpha\beta}=\psi_\alpha\psi_\beta^*$ are the matrix elements of $P$ in the orbital basis. The positive-semidefinite $\mathcal{M}_{\alpha\beta}$ ($\mathcal{M}_{\alpha\beta}$ satisfies $\sum_\beta \mathcal{M}_{\alpha\beta}=0$ whose properties can be found in Ref.~\cite{jiang2024geometric}) ensures that $\tr\{\wt{P}\frac{\partial \wt{P}}{\partial\pi}\frac{\partial \wt{P}}{\partial\pi^*}\}\geq0$, and has $\text{Rank}(\mathcal{M}_{\alpha\beta})=\wt{N}_{orb}-1$, with $\wt{N}_{orb}$ the number of orbitals that have nonzero weight $|\psi_\alpha|^2$ in the state $\psi$.

In general, $\mathcal{M}_{\alpha\beta}$ measures how fast the IQG is destroyed by moving the orbital positions around the IQG embedding, which can be analyzed using the surface state $\psi^{(1)}$ from Eq.~\eqref{eq:surfacestates} as an example. For $N$-layer RG, near the rim of the drumhead region ($k\lesssim1$), the decay length $\lambda\rightarrow\infty$ and the surface state $\psi^{(1)}$ is uniformly distributed over the $\wt{N}_{orb}=N$ orbitals: 1A, 2A, ..., $N$A. Then, the nonzero entries of $\mathcal{M}_{\alpha\beta}$ form a $N\times N$ block,
\be
\begin{pmatrix}
\frac{1}{N}-\frac{1}{N^2} & -\frac{1}{N^2} & -\frac{1}{N^2} &..\\
-\frac{1}{N^2} & \frac{1}{N}-\frac{1}{N^2} & -\frac{1}{N^2} & ..\\
-\frac{1}{N^2} & -\frac{1}{N^2} & \frac{1}{N}-\frac{1}{N^2} & ..\\
.. & .. & .. & ..
\end{pmatrix}
\ee
which has one zero eigenvalue and $N-1$ degenerate eigenvalues of $1/N$. This means that when the orbital embedding changes by the size of the lattice constant of graphene, $|X_\alpha|\sim a_0$, $\tr\big\{P^{(1)}\frac{\partial P^{(1)}}{\partial\pi}\frac{\partial P^{(1)}}{\partial\pi^*}\big\}$ deviates from 0 by an amount $\sim(1/N)(k_0a_0)^2$, whereas the other quantity $\tr\big\{P^{(1)}\frac{\partial P^{(1)}}{\partial\pi^*}\frac{\partial P^{(1)}}{\partial\pi}\big\}=\frac{1}{2}\tr g^{(1)}=1/(1-k^2)^2\gg 1$ at the IQG embedding. As a result, under such an orbital embedding transformation, the ratio $|\Omega^{(1)}|/\tr g^{(1)}$ is decreased by
\be
\delta\bigg(\frac{|\Omega^{(1)}|}{\tr g^{(1)}}\bigg)\sim \frac{\delta\tr\{P^{(1)}\frac{\partial P^{(1)}}{\partial\pi}\frac{\partial P^{(1)}}{\partial\pi^*}\}}{\tr\{P^{(1)}\frac{\partial P^{(1)}}{\partial\pi^*}\frac{\partial P^{(1)}}{\partial\pi}\}|_{\text{IQG}}}\ll \frac{1}{N}(k_0a_0)^2\ll 1,
\ee
which means the effect on IQG is negligible. Here, the small deviation $\delta(|\Omega^{(1)}|/\tr g^{(1)})$ comes from both factors $1/N$ and the small value of $k_0a_0$ in RG.

On the other hand, near the center of the drumhead region, the surface states are strongly localized; therefore, the nonzero eigenvalues of $\mathcal{M}_{\alpha\beta}$ are all much smaller than $1/N$, meaning the orbital-embedding dependence is even weaker than at the rim. At the center of the drumhead region ($k=0$), the surface state $\psi^{(1)}$ is completely localized at orbital 1A, then $\wt{N}_{orb}=1$ and $\mathcal{M}_{\alpha\beta}$ becomes a zero matrix. Therefore, the IQG at $k=0$ is orbital-embedding-independent; this is consistent with having only the global $U(1)$ gauge freedom in the case of one orbital and QG being independent of the global gauge.

\subsection{2.4 Generalized semimetal models}
Here, we provide more details about the generalized semimetal models introduced in the MS. We focus on the most general case of Eq.\eqref{eq:h3d}, when $f(k_x,k_y,k_z)=p_n(\pi)+e^{ik_z}$, where
\be
p_n(\pi)=a_n\pi^n+...+a_1\pi+a_0
\ee
is a polynomial of $\pi=k_x+ik_y$ of degree $n$, with complex coefficients $a_n,...,a_0$. Then, the Hamiltonian Eq.\eqref{eq:h3d} can be understood as the 2D effective Hamiltonian $h(\bd{k})=\sigma_x\re p_n(\pi)+\sigma_y\im p_n(\pi)$ coupled by interlayer hoppings along the $z$ direction.

The gapless points of the model are guaranteed to form nodal lines, which can be solved from
\be\label{eq:polyeq}
p_n(\pi)+e^{ik_z}=0.
\ee
This polynomial equation has $n$ complex roots, which are labeled by $s$, and each nodal line is represented by one root as a function of $k_z$, $\pi^{(s)}(k_z)$. We make the assumption that the coefficient $a_0=0$, i.e., $p_n(0)=0$, which means that the 2D effective Hamiltonian $h(\bd{k})$ is gapless at $\bd{k}=0$. This implies that none of these nodal lines passes the origin $\pi=0$ in the $k_x-k_y$ plane, so winding numbers about $\pi=0$ can be defined. To define the winding number, we note that the polynomial equation is invariant if $k_z$ is increased by $2\pi$, restricting that one root evaluated at $k_z=2\pi$ must be identical to another root (or itself) at $k_z=0$, $\pi^{(s)}(2\pi)=\pi^{(s')}(0)$. This allows us to divide these $n$ roots into several groups (labeled by $i$):
\be
\{\pi^{(s_1)},\pi^{(s_2)},...,\pi^{(s_{n_1})}\},\,\,\,\{\pi^{(s_{n_1+1})},\pi^{(s_{n_1+2})},...,\pi^{(s_{n_1+n_2})}\},\,\,\,...
\ee
In each group, the roots are ordered by the successive relation $\pi^{(s_j)}(2\pi)=\pi^{(s_{j+1})}(0)$, while the last element is related to the first one in the same manner, e.g., in the first group, $\pi^{(s_{n_1})}(2\pi)=\pi^{(s_1)}(0)$. These relations indicate that when projected to the $k_x-k_y$ plane, these nodal lines in the same group form a loop. Then, a winding number can be defined for each loop,
\be
w_i=\frac{1}{2\pi}\sum_{s\in \text{group }i}\int_0^{2\pi}\derd k_z\frac{\derd\varphi^{(s)}(k_z)}{\derd k_z},
\ee
where $\varphi^{(s)}(k_z)$ is the azimuthal angle of $\pi^{(s)}(k_z)$. If $w_i=0$, then these nodal lines in group $i$ can be adiabatically shrunk into a point by other perturbations, without opening a gap for the 2D semimetal $h(\bd{k})$.

Following the same procedure described at the beginning of Sec.~2.1, it is easy to show that each loop corresponds to the boundary of a drumhead region (these drumheads may not have a round shape). The boundary condition decides which side of the loop the surface states live on. The Hamiltonian Eq.~\eqref{eq:h3d} with $f(k_x,k_y,k_z)=p_n(\pi)+e^{ik_z}$ tells us that sublattice B of layer $z$ is coupled with A of layer $z+1$ (see Fig.~\ref{fig:conventions}(b) for the RG case), so the boundary condition is still $\psi_\text{B}(z=0)=\psi_\text{A}(z=N+1)=0$. Then, two surface bands are formed on the side determined by $|p_n(\pi)|<1$, with normalized wavefunctions
\be
\psi^{(1)}_\bd{k}(z)=\frac{\sqrt{1-|p_n(\pi)|^2}}{|p_n(\pi)|}e^{\kappa(\bd{k})z}\begin{pmatrix}
1 \\ 0
\end{pmatrix},\,\,\,\psi^{(2)}_\bd{k}(z)=\frac{\sqrt{1-|p_n(\pi)|^2}}{|p_n(\pi)|}e^{\kappa(\bd{k})^*(N+1-z)}\begin{pmatrix}
0 \\ 1
\end{pmatrix},
\ee
where $\kappa(\bd{k})=\ln[-p_n(\pi)]$. Prop.~\ref{prop:prop3} and~\ref{prop:prop4} dictate that these surface bands have IQG with negligible orbital-embedding-dependence.

\section{3. Perturbation effects on the RG surface states}
\subsection{3.1 Effects of long-range hoppings, displacement field and disorder}
We treat the extra long-range hopping and displacement field terms using the degenerate perturbation method and compute their effective Hamiltonian matrices. These two-by-two matrices are given by $\ham'_{ij}(\bd{k})=\la\psi^{(i)}_\bd{k}|\hham'|\psi^{(j)}_\bd{k}\ra$ ($i,j=1,2$), where $\hham'$ refers to $\hham^{(\gamma_2)}$, $\hham^{(\gamma_3)}$, $\hham^{(\gamma_4)}$ and $\hham^{(m)}$. Using their expressions in Eq.~\eqref{eq:hgamma1to4} -~\eqref{eq:hm}, and the wavefunctions $\psi^{(1,2)}_\bd{k}$ given in Eq.~\eqref{eq:surfacestates}, it is straightforward to obtain
\be\label{eq:matrix1}
\ham^{(\gamma_2)}_{ij}(\bd{k})=(N-2)\frac{\wt{\gamma}_2}{2}\frac{1-k^2}{k^2}\begin{pmatrix}
    0 & c.c.\\
    (-\pi)^{N-1} & 0
\end{pmatrix};
\ee
\be
\ham^{(\gamma_3)}_{ij}(\bd{k})=-(N-1)\wt{v}_3(1-k^2)\begin{pmatrix}
	0 & c.c.\\
	(-\pi)^{N-1} & 0
\end{pmatrix};
\ee
\be\label{eq:hg4}
\ham^{(\gamma_4)}_{ij}(\bd{k})=-2\wt{v}_4k^2(1-k^{2N-2})\begin{pmatrix}
    1 & 0\\
    0 & 1
\end{pmatrix}\approx -2\wt{v}_4k^2\sigma_0;
\ee
\be\label{eq:hmeff}
\ham^{(m)}_{ij}(\bd{k})=\begin{pmatrix}
    m(1-k^2) & 0\\
    0 & -m(1-k^2)
\end{pmatrix},
\ee
where $\pi=\xi k_x+ik_y$. Note that choosing a different potential distribution function for the displacement field, e.g., a linear potential distribution $V(z)=m(1-2\frac{z-1}{N-1})$ only gives a small correction to Eq.~\eqref{eq:hmeff}, so makes no qualitative difference.

\begin{figure}[h!]
\centering	
\includegraphics[height=5.5cm]{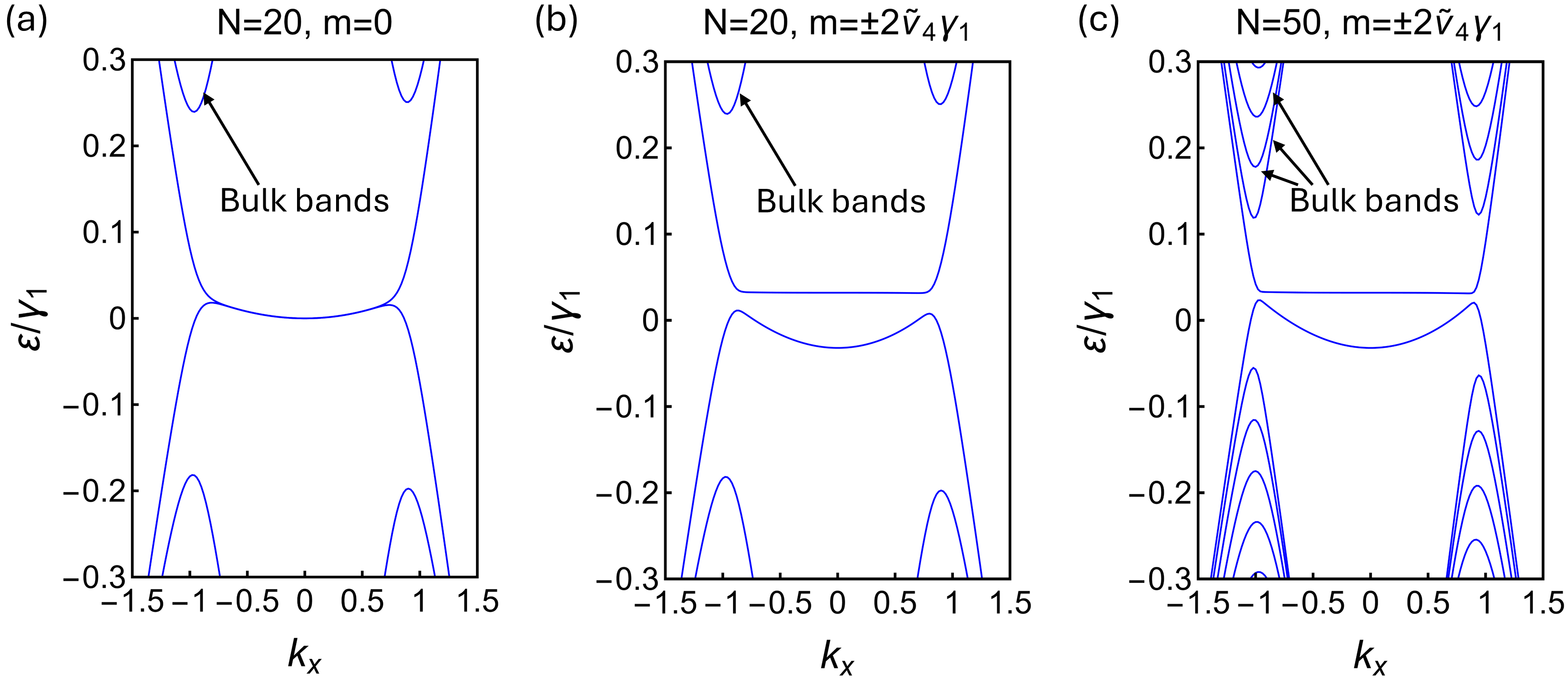}
\caption{Low-energy band structure calculated from the tight-binding method for (a) 20-layer RG without displacement field; (b)(c) 20 and 50-layer RG with surface potential $m=\pm2\wt{v}_4\gamma_1$. Hopping parameters for both 20 and 50-layer RG are adopted from the density-functional theory data in Ref.~\cite{kopnin2013high} for 20-layer RG: $\gamma_0=3.21$ eV, $\gamma_1=0.43$ eV, $\gamma_2=0$ eV, $\gamma_3=-0.12$ eV, $\gamma_4=-0.05$ eV. These band structures are consistent with the analysis based on the degenerate perturbation method, Eq.~\eqref{eq:matrix1} -~\eqref{eq:hmeff}.}
\label{fig:flatband}
\end{figure}

These effective Hamiltonians indicate the roles of each term on the surface states: (i) $\gamma_2$ and $\gamma_3$ terms both mix the two basis states $\psi^{(1)}_\bd{k}$ and $\psi^{(2)}_\bd{k}$, and split the two bands, but their effects exponentially decay as $N$ increases; (ii) $\gamma_4$ term does not affect the eigenstates or split the bands, but only gives the same quadratic ``bending" to both bands [see Fig.~\ref{fig:flatband}(a)], and this effect is almost $N$-independent; (iii) the displacement field term $m$ gives opposite quadratic ``bending" to the two bands, splitting the bands and preserving $\psi^{(1)}_\bd{k}$ and $\psi^{(2)}_\bd{k}$ as eigenstates, and this effect is also $N$-independent. When $N$ is large, effects of $\gamma_2,\gamma_3$ are negligible, while $\gamma_4$ and $m$ together can make one surface band much more dispersionless than the other. Comparing Eq.~\eqref{eq:hg4} with~\eqref{eq:hmeff}, we find that at special surface potential values, $m=\pm2\wt{v}_4$ (in units of $\gamma_1$), the conduction band becomes extremely flat, as shown in Fig.~\ref{fig:flatband}(b) and (c).

\begin{figure}[h!]
	\centering	
	\includegraphics[height=4cm]{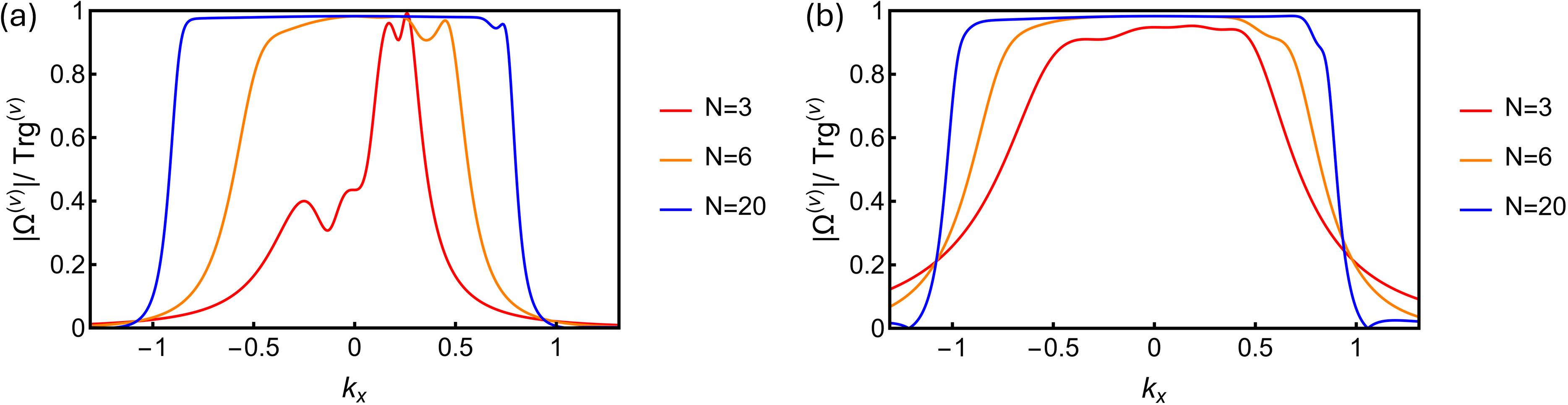}
	\caption{(a) Ratio $|\Omega^{(v)}|/\tr g^{(v)}$ for the valence surface band with $\gamma_2-\gamma_4$ terms (parameters used for 6-layer RG are the same as for RTG, given at the end of Sec.~1) and $m=0.01\gamma_1$. (b) Same as (a), but with $m=0.1\gamma_1$.}
	\label{fig:ratiovsm}
\end{figure}

The effects of different terms on IQG are illustrated in Fig.~\ref{fig:ratiovsm}. Fig.~\ref{fig:ratiovsm}(a) shows the ratio of $|\Omega^{(v)}|/\tr g^{(v)}$ when the long-range hoppings $\gamma_2-\gamma_4$ are present, where the trace condition in few-layer RG is strongly violated due to the state mixing effect by the $\gamma_2$ and $\gamma_3$ terms (c.f. Fig. 1(c) in the MS), whereas the violation is negligible for many-layer RG ($N=20$). However, when a large displacement field is applied [Fig.~\ref{fig:ratiovsm}(b)], the trace condition is restored, even for few-layer RG. According to the $k_h$-$m$ relation Eq.~\eqref{eq:khwithh}, the region where the trace condition is satisfied also expands as a displacement field is applied.

Finally, we briefly discuss the effect of disorder on the RG surface states and IQG. Disorder can be treated as random on-site potentials ($V_i$, where $i$ labels the lattice site), whose effect on the normal state is similar to the displacement field. For weak disorder ($|V_i|$ smaller than the gap between surface and bulk bands, see Fig.~\ref{fig:flatband}), the effective Hamiltonian matrix involves the two surface bands only, which is a diagonal two-by-two matrix. If the potentials $V_i$ have the same average on the two surfaces, then the matrix is proportional to the 2-dim identity matrix $\sigma_0$, which does not affect the surface states or QGT; however, if $V_i$ has different averages on the two surfaces, $\la V_i\ra_\text{top}\neq\la V_i\ra_\text{bottom}$ (e.g., due to different types of impurities), then the matrix contains a $\sigma_z$ term. Then, similar to the displacement field, this kind of disorder can even help preserve the surface states and IQG of both bands.

Strong disorder whose strength exceeds the gap between the surface and bulk bands enables transitions between them. As a result, the surface state properties (including IQG) are no longer protected. For correlated phases, the disorder strength must also be compared with the correlated gap, e.g., the superconducting gap of the superconducting state~\cite{bouadim2011,seibold2012,samanta2022}. When $|V_i|$ is smaller than the superconducting gap, the qualitative properties of RG superconductivity and superfluidity are expected to persist.

\subsection{3.2 Effect of surface hybridization}
In this section, we first provide an explanation for the surface hybridization origin of the QM peaks at $k=k_h$ for the small-$N$-layer RG. We then derive the two-by-two effective Hamiltonian for the surface hybridization, which is similar to the matrices $\ham^{(\gamma_{2,3,4})}_{ij}$ and $\ham^{(m)}_{ij}$ for the extra hopping terms in the previous section.

QGT of a surface band, e.g., the valence band, can be decomposed into the interband QGT components:
\be\label{eq:kubo}
\mathcal{B}^{(v)}_{\mu\nu}(\bd{k})=\sum_{j\neq v}\la\partial_\mu\psi^{(v)}_\bd{k}|\psi^{(j)}_\bd{k}\ra\la\psi^{(j)}_\bd{k}|\partial_\nu\psi^{(v)}_\bd{k}\ra=\sum_{j\neq v}\frac{\la\psi^{(v)}_\bd{k}|\partial_\mu\ham(\bd{k})|\psi^{(j)}_\bd{k}\ra\la\psi^{(j)}_\bd{k}|\partial_\nu\ham(\bd{k})|\psi^{(v)}_\bd{k}\ra}{(\varepsilon_{v,\bd{k}}-\varepsilon_{j\bd{k}})^2}.
\ee
Here, for the first equality, $j$ is summed over all bands except the valence band; we use the Kubo formula to get the second equality. Eq.~\eqref{eq:kubo} shows that the peaks of QGT can result from the small band gaps $\varepsilon_{v,\bd{k}}-\varepsilon_{j\bd{k}}$. In our case, the smallest bandgap to the valence band is from the conduction surface band, which is the reason why we compare $g^{(v)}$ with the interband component $g^{(vc)}$ in Fig. 2(b) of the MS. In addition, when the two surface states have a small decay length, $\lambda\ll N$,
\be
g^{(vc)}_{\mu\nu}(\bd{k})=g^{(12)}_{\mu\nu}(\bd{k})=-\re\la\partial_\mu\psi^{(1)}_\bd{k}|\psi^{(2)}_\bd{k}\ra\la \psi^{(2)}_\bd{k}|\partial_\nu\psi^{(1)}_\bd{k}\ra
\ee
vanishes, since $\psi^{(1)}_\bd{k}$ and $\psi^{(2)}_\bd{k}$ have no overlap. Therefore, $g^{(vc)}_{\mu\nu}(\bd{k})$ can only exhibit peaks when the decay length becomes large, $\lambda\sim N$, which confirms that the peaks of $g^{(v)}_{\mu\nu}$ result from the surface hybridization.

Next, we derive the two-by-two effective Hamiltonian for the surface hybridization. For finite-$N$-layer RG, one can still choose the same wavefunctions $\psi^{(1,2)}_\bd{k}$ from Eq.~\eqref{eq:surfacestates} as the basis, but have to truncate the Hilbert space by restricting $z$ running from 1 to a finite $N$. This means now $\psi^{(1,2)}_\bd{k}$ are only approximately normalized. Then, the effective Hamiltonian matrix is
\be
\ham^{(\text{hybd})}_{ij}(\bd{k})=\la \psi^{(i)}_\bd{k}|\hham^{(\gamma_0)}+\hham^{(\gamma_1)}|\psi^{(j)}_\bd{k}\ra,
\ee
with $\hham^{(\gamma_0)}$ and $\hham^{(\gamma_1)}$ given in Eq.~\eqref{eq:hgamma0} and ~\eqref{eq:hgamma1to4}. Evaluating the matrix elements, we find
\be\label{eq:hhybd}
\ham^{(\text{hybd})}_{ij}(\bd{k})=-\frac{1-k^2}{k^2}\begin{pmatrix}
0 & c.c.\\
(-\pi)^{N+2} & 0
\end{pmatrix}.
\ee
Note that this effective Hamiltonian is intrinsically different from Eq.~\eqref{eq:twobandeff}, since its basis states are $\psi^{(1,2)}_{\bd{k}}$, whereas the basis of Eq.~\eqref{eq:twobandeff} are orbital 1A and $N$B. Effective Hamiltonian $\ham^{(\text{hybd})}_{ij}$ becomes important when its matrix elements exceed some measure, i.e., a small value $e^{-\eta}$ ($\eta>0$). As long as $k$ is not too close to 1, we obtain
\be
\bigg|-\frac{1-k^2}{k^2}(-\pi)^{N+2}\bigg|\sim e^{-\eta}\,\,\,\rightarrow\,\,\,k\sim e^{-\eta/N},
\ee
which reproduces the definition of $k_h$ given in the MS. 

The hybridization term $\ham^{(\text{hybd})}_{ij}$ can also be considered together with the displacement field term, $\ham^{(m)}_{ij}$. Combining Eq.~\eqref{eq:hmeff} and ~\eqref{eq:hhybd}, we get (c.f. Eq.~\eqref{eq:twobandeff})
\be
\ham^{(m)}_{ij}(\bd{k})+\ham^{(\text{hybd})}_{ij}(\bd{k})=(1-k^2)\begin{pmatrix}
m & c.c.\\
\frac{(-\pi)^{N+2}}{k^2} & -m
\end{pmatrix}.
\ee
In this case, the surface potential $m$ provides a natural measure for the surface hybridization strength, e.g., it defines $k_h$ such that
\be
\bigg|\frac{(-\pi)^{N+2}}{k^2}\bigg|\sim m\,\,\,\rightarrow\,\,\,k_h\sim m^{1/N},
\ee
which reproduces the $k_h$-$m$ relation, Eq.~\eqref{eq:khwithh}.

\section{4. Self-consistency equations and superfluid weight calculation}
This section provides the general formalism for calculating the superconducting gap and superfluid weight.
\subsection{4.1 Self-consistent gap equation and electron number equation}
The onsite attractive density-density interaction for the RG is
\be
\hham_I=-U\sum_{\bd{R},\alpha}\hat{n}_{\bd{R}\alpha\uparrow}\hat{n}_{\bd{R}\alpha\downarrow}=-\frac{U}{N_c}\sum_{\bd{k}\bd{k}'\bd{q},\alpha}c_{\bd{k}+\bd{q},\alpha\uparrow}^\dagger c_{-\bd{k}+\bd{q},\alpha\downarrow}^\dagger c_{-\bd{k}'+\bd{q},\alpha\downarrow}c_{\bd{k}'+\bd{q},\alpha\uparrow}
\ee
with $U>0$. After mean-field decoupling, it leads to an intra-orbital pairing matrix which is diagonal in the orbital basis $\hat{\Delta}=\text{diag}\{\Delta_\alpha\}$ ($\alpha$ running over the $2N$ orbitals) and
\be\label{eq:gap0}
\Delta_\alpha\equiv-U\la c_{\bd{R}\alpha\downarrow}c_{\bd{R}\alpha\uparrow}\ra=-\frac{2U}{N_c}\sum_{\bd{k}\in\Lambda_K}\la c_{-\bd{k}\alpha\downarrow}c_{\bd{k}\alpha\uparrow}\ra,
\ee
where the sum over $\bd{k}$ is over a cutoff region $\Lambda_K$ near $K$ and the factor 2 counts the valley degeneracy. For an interaction $U \ll \gamma_1$, it is a valid approximation that the only nonzero order parameters are $\Delta_{1\text{A}}$ and $\Delta_{N\text{B}}$. Therefore, $\hat{\Delta}=\text{diag}\{\Delta_{1\text{A}},0,0,..,0,\Delta_{N\text{B}}\}$.

The Bogoliubov-de-Gennes (BdG) equation
\be
\ham_\text{BdG}(\bd{k})\phi_i(\bd{k})=E_{i,\bd{k}}\phi_i(\bd{k})
\ee
can be diagonalized numerically. Here $i$ labels the $4N$ Bogoliubov bands and
\be\label{eq:hbdg}
\ham_\text{BdG}(\bd{k})=\begin{pmatrix}
    \ham^\uparrow(\bd{k})-\mu & \hat{\Delta}\\
    \hat{\Delta} & -[\ham^{\downarrow T}(-\bd{k})-\mu]
\end{pmatrix}.
\ee
By TRS, $\ham^\uparrow(\bd{k})=\ham^{\downarrow T}(-\bd{k})\equiv \ham(\bd{k})$ with $\ham(\bd{k})$ the $2N$-dim noninteracting Hamiltonian of $N$-layer RG. Denote the $i$th eigenstate of $\ham_\text{BdG}(\bd{k})$ by $\phi_i$, then Eq.~\eqref{eq:gap0} becomes
\be\label{eq:gapnumerical}
\Delta_\alpha=-\frac{2U}{N_c}\sum_{\bd{k}\in\Lambda_K}\sum_{i=1}^{4N}\phi_{i,\alpha\downarrow}(\bd{k})^*\phi_{i,\alpha\uparrow}(\bd{k})n_i(\bd{k}),
\ee
where $n_i(\bd{k})=\frac{1}{1+\text{exp}(\beta E_{i,\bd{k}})}$ is the Fermi-Dirac function.

Similarly, the electron number equation in terms of eigenstates $\phi_i$ is
\be\label{eq:neeq0}
\begin{split}
N_e=&2\sum_{\bd{k}\in\Lambda_K}\sum_\alpha(\la c_{\bd{k}\alpha\uparrow}^\dagger c_{\bd{k}\alpha\uparrow}\ra+\la c_{-\bd{k}\alpha\downarrow}^\dagger c_{-\bd{k}\alpha\downarrow}\ra)\\
=&2\sum_{\bd{k}\in\Lambda_K}\sum_\alpha\sum_{i=1}^{4N}\big\{n_i(\bd{k})|\phi_{i,\alpha\uparrow}(\bd{k})|^2+[1-n_i(\bd{k})]|\phi_{i,\alpha\downarrow}(\bd{k})|^2\big\}.
\end{split}
\ee
The total electron number in the cutoff region for RG at charge neutrality is
\be\label{eq:neeq2}
N_{e,0}=2\sum_{\bd{k}\in\Lambda_K}\sum_\alpha(\frac{1}{2}+\frac{1}{2})=2\sum_{\bd{k}\in\Lambda_K}\sum_\alpha\sum_{i=1}^{4N}|\phi_{i,\alpha\downarrow}(\bd{k})|^2.
\ee
Then, the electron number measured from charge neutrality is
\be\label{eq:enumbernumerical}
\delta N_e\equiv N_e-N_{e,0}=2\sum_{\bd{k}\in\Lambda_K}\sum_\alpha\sum_{i=1}^{4N}n_i(\bd{k})\big[|\phi_{i,\alpha\uparrow}(\bd{k})|^2-|\phi_{i,\alpha\downarrow}(\bd{k})|^2\big].
\ee
The electron density $n_e\equiv\delta N_e/S$ ($S$ is the sample area) is in the units of $\text{cm}^{-2}$ [note that the momentum is in the units of $k_0=\gamma_1/(\hbar v_f)=0.39\text{ eV}/(6.58\times 10^{-16}\text{ eV}\cdot \text{s}\times 10^6\text{ m/s})=5.9\times 10^6\text{ cm}^{-1}$].

When the two surface bands of the RG are gapped by surface potential $m$, and the superconducting state is doped to only the valence band (requiring $m>\Delta_\alpha$), the single-band projection becomes valid. Therefore, the summation over the $4N$ Bogoliubov bands in Eq.~\eqref{eq:gapnumerical} and \ref{eq:enumbernumerical} can be replaced with summation over only the quasiparticle and quasihole bands associated with the valence band, $v+$ and $v-$, i.e. $\sum_{i=1}^{4N}\rightarrow\sum_{i=v+,v-}$.

\subsection{4.2 General superfluid weight formula}
We use the grand potential formalism \cite{peotta2015superfluidity} to compute the superfluid weight numerically. The superfluid weight tensor is the second total derivative of the free energy density with half of the Cooper pair center of mass momentum $\bd{q}$, $D_{s,\mu\nu}=\frac{1}{S}\frac{\derd^2F(\bd{q})}{\derd q_\mu \derd q_\nu}\big|_{\bd{q}=0}$ (we set $e=\hbar=1$; for convenience we set the sample area $S=1$ hereafter). Under TRS, using the relation between free energy $F$ and grand potential $\iOmega$, $F(\bd{q})=\iOmega(\bd{q},\hat{\Delta}_\bd{q},\hat{\Delta}_\bd{q}^*,\mu_\bd{q})+\mu_\bd{q}N_e$ where the $\bd{q}$-dependence of $\hat{\Delta}_\bd{q},\mu_\bd{q}$ arise from self-consistency equations (the electron number $N_e$ is fixed for different $\bd{q}$), the superfluid weight can be expressed as \cite{peotta2015superfluidity,huhtinen2022revisiting}:
\be\label{eq:dstotal}
D_{s,\mu\nu}=\bigg[\frac{\partial^2\iOmega}{\partial q_\mu\partial q_\nu}-\sum_{\alpha,\beta}\frac{\derd\Delta^I_{\bd{q},\alpha}}{\derd q_\mu} \frac{\partial^2\iOmega}{\partial \Delta^I_{\bd{q},\alpha}\partial \Delta^I_{\bd{q},\beta}}\frac{\derd\Delta^I_{\bd{q},\beta}}{\derd q_\nu}\bigg]\bigg|_{\bd{q}=0}.
\ee
Here, the second term of Eq.~\eqref{eq:dstotal} makes $D_{s,\mu\nu}$ gauge-invariant with respect to the orbital embedding $\bd{r}_\alpha$ of different orbitals $\alpha$ ($\hat{\Delta}_\bd{q}=\text{diag}\{\Delta_{\bd{q},\alpha}\}$ denotes the pairing matrix of the finite-$\bd{q}$ supercurrent state). Here $\Delta^I_{\bd{q},\alpha}$ is the imaginary part of $\Delta_{\bd{q},\alpha}$, which is 0 at $\bd{q}=0$ since all $\Delta_\alpha$ are real by virtue of TRS. TRS also implies that $\frac{\derd\Delta_{\bd{q},\alpha}}{\derd q_\mu}|_{\bd{q}=0}=i\frac{\derd\Delta^I_{\bd{q},\alpha}}{\derd q_\mu}|_{\bd{q}=0}=-\frac{\derd\Delta_{\bd{q},\alpha}^*}{\derd q_\mu}|_{\bd{q}=0}$. Taking into account a prefactor $1/(\mathcal{A}k_0^2)=54.9$, where $\mathcal{A}=5.24\times10^{-16}$cm$^2$ is the unit cell area coming from $1/S=1/(N_c\mathcal{A})$, the calculated $D_s$ has the dimension of energy. In other words, the final superfluid weight is the calculated value multiplied by a factor of $54.9e^2/\hbar^2$. Here, we compute the first term of Eq.~\eqref{eq:dstotal}; the second term is calculated in Sec.~4.5.

The BdG Hamiltonian for the supercurrent state is
\be\label{eq:hbdgkq}
\ham_\text{BdG}(\bd{k},\bd{q})=\begin{pmatrix}
    \ham^\uparrow(\bd{k}+\bd{q})-\mu_\bd{q}I_{2N} & \hat{\Delta}_\bd{q}\\
    \hat{\Delta}_\bd{q} & -[\ham^{\downarrow\,T}(-\bd{k}+\bd{q})-\mu_\bd{q}I_{2N}]
\end{pmatrix},
\ee
and the second-quantized total mean-field Hamiltonian is
\be\label{eq:hmf}
\hham_\text{MF}(\bd{q})=\sum_{\bd{k}\in\text{BZ}}\big\{\bd{C}_{\bd{k},\bd{q}}^\dagger\ham_\text{BdG}(\bd{k},\bd{q})\bd{C}_{\bd{k},\bd{q}}+\tr [\ham^\downarrow(-\bd{k}+\bd{q})-\mu_\bd{q}I_{2N}]\big\}+\frac{N_c}{U}\sum_\alpha|\Delta_{\bd{q},\alpha}|^2,
\ee
where $\bd{C}_{\bd{k},\bd{q}}=(c_{\bd{k}+\bd{q},1\text{A},\uparrow},..,c_{-\bd{k}+\bd{q},1\text{A},\downarrow}^\dagger,..)^T$, $I_{2N}$ is the $2N$-dim identity matrix and $\bd{k}$ runs over the entire Brillouin zone (BZ). Let $\hat{E}_\bd{k}(\bd{q})$ denote the diagonal matrix of the spectrum of $\ham_\text{BdG}(\bd{k},\bd{q})$ by diagonalization, then at temperature $T=1/\beta$ \cite{peotta2015superfluidity},
\be\label{eq:grandgeneral1}
\iOmega(\bd{q},\hat{\Delta}_\bd{q},\hat{\Delta}_\bd{q}^*,\mu_\bd{q})=-\frac{1}{\beta}\ln\tr\{e^{-\beta\hham_\text{MF}(\bd{q})}\}=-\frac{1}{2\beta}\sum_{\bd{k}\in\text{BZ}}\tr\ln[2+2\cosh\beta\hat{E}_\bd{k}(\bd{q})]+\frac{N_c}{U}\sum_\alpha|\Delta_{\bd{q},\alpha}|^2+\text{const.}
\ee
When computing the first term of Eq.~\eqref{eq:dstotal}, we treat $\hat{\Delta}_\bd{q}$ and $\mu_\bd{q}$ as $\bd{q}$-independent in Eq.~\eqref{eq:grandgeneral1}.

The Fermi surfaces of RG at valleys $K$ and $K'$ are small compared to the entire Brillouin zone. Therefore, it is necessary to convert Eq.~\eqref{eq:grandgeneral1} into a summation that converges in a finite cutoff regions at the two valleys. Note that the conduction (valence) surface band has a particle (hole)-like Fermi surface, i.e., as $\bd{k}$ is far from the cutoff regions, asymptotically
\be\label{eq:asymptotic}
E_{c\pm,\bd{k}}(\bd{q})\rightarrow \pm\xi_{c,\bd{k}\pm\bd{q}},\,\,\,E_{v\pm,\bd{k}}(\bd{q})\rightarrow\mp\xi_{v,\bd{k}\mp\bd{q}},
\ee
where $\xi_{c(v),\bd{k}}=\varepsilon_{c(v),\bd{k}}-\mu$, with $\varepsilon_{c(v),\bd{k}}$ the electronic band dispersion of the conduction (valence) band. Here $E_{c(v)\pm,\bd{k}}$ are the quasiparticle (+) and quasihole (-) band spectra associated with $c$ and $v$ bands in the diagonal entries of $\hat{E}_\bd{k}(\bd{q})$. We then transform the grand potential $\iOmega$ by adding a $\bd{q}$-independent boundary term: $\iOmega(\bd{q},\hat{\Delta}_\bd{q},\hat{\Delta}_\bd{q}^*,\mu_\bd{q})+\iOmega_0\rightarrow \iOmega(\bd{q},\hat{\Delta}_\bd{q},\hat{\Delta}_\bd{q}^*,\mu_\bd{q})$ (this process corresponds to ``subtracting the supercurrent in the normal state" in literature \cite{kopnin2008bcs,kopnin2011surface}). Specifically, we break each integral term (due to the trace of the diagonal matrix $\hat{E}_\bd{k}(\bd{q})$) of Eq.~\eqref{eq:grandgeneral1}, which integrates $\bd{k}$ over the entire BZ, into a sum of integrals inside and outside the cutoff region. e.g., for the $E_{c+,\bd{k}}(\bd{q})$ term:
\be
\begin{split}
&\sum_{\bd{k}\in\text{BZ}}\ln[2+2\cosh\beta E_{c+,\bd{k}}(\bd{q})]\\
=&\sum_{\bd{k}\in\Lambda_K}\ln[2+2\cosh\beta E_{c+,\bd{k}}(\bd{q})]+\sum_{\bd{k}\in\Lambda_{K'}}\ln[2+2\cosh\beta E_{c+,\bd{k}}(\bd{q})]+\sum_{\bd{k}\notin\Lambda_K\text{ or }\Lambda_{K'}}\ln[2+2\cosh\beta E_{c+,\bd{k}}(\bd{q})]\\
=&2\sum_{\bd{k}\in\Lambda_K}\ln[2+2\cosh\beta E_{c+,\bd{k}}(\bd{q})]+\sum_{\bd{k}\notin\Lambda_K\text{ or }\Lambda_{K'}}\ln[2+2\cosh\beta E_{c+,\bd{k}}(\bd{q})]\\
=&2\sum_{\bd{k}\in\Lambda_K}\ln[2+2\cosh\beta E_{c+,\bd{k}}(\bd{q})]+\sum_{\bd{k}\notin\Lambda_K\text{ or }\Lambda_{K'}}\ln[2+2\cosh\beta \xi_{c,\bd{k}+\bd{q}}]\\
=&2\sum_{\bd{k}\in\Lambda_K}\ln\frac{2+2\cosh\beta E_{c+,\bd{k}}(\bd{q})}{2+2\cosh\beta \xi_{c,\bd{k}+\bd{q}}}+\sum_{\bd{k}\in\text{BZ}}\ln[2+2\cosh\beta \xi_{c,\bd{k}+\bd{q}}].
\end{split}
\ee
Here $\Lambda_{K'}$ refers to the cutoff region at valley $K'$ which is of the same size as $\Lambda_K$; to get the second line, we used valley degeneracy; to get the third line, we used the asymptotic relation Eq.~\eqref{eq:asymptotic}; to get the last line, we used partial integral. The last term in the last line above is $\bd{q}$-independent as it integrates over the entire BZ, so it contributes to the boundary term $\iOmega_0$. We then obtain the following transformed grand potential which integrates over the region $\Lambda_K$ only:
\be\label{eq:grandgeneral2}
\begin{split}
\iOmega(\bd{q},\hat{\Delta}_\bd{q},\hat{\Delta}_\bd{q}^*,\mu_\bd{q})=-\frac{1}{\beta}\sum_{\bd{k}\in\Lambda_K}&\bigg[\ln\frac{1+\cosh\beta E_{c+,\bd{k}}(\bd{q})}{1+\cosh\beta\xi_{c,\bd{k}+\bd{q}}}+\ln\frac{1+\cosh\beta E_{c-,\bd{k}}(\bd{q})}{1+\cosh\beta\xi_{c,\bd{k}-\bd{q}}}\\
&+\ln\frac{1+\cosh\beta E_{v+,\bd{k}}(\bd{q})}{1+\cosh\beta\xi_{v,\bd{k}-\bd{q}}}+\ln\frac{1+\cosh\beta E_{v-,\bd{k}}(\bd{q})}{1+\cosh\beta\xi_{v,\bd{k}+\bd{q}}}\bigg].
\end{split}
\ee
At $T=0$ it can be replaced with a simple integral
\be\label{eq:grandgeneral2t0}
\iOmega(\bd{q},\hat{\Delta}_\bd{q},\hat{\Delta}_\bd{q}^*,\mu_\bd{q})=2\sum_{\bd{k}\in\Lambda_K}\big\{\big[E_{c-,\bd{k}}(\bd{q})+\xi_{c,\bd{k}-\bd{q}}\big]+\big[E_{v-,\bd{k}}(\bd{q})-\xi_{v,\bd{k}+\bd{q}}\big]\big\}.
\ee
If in the superconducting state, e.g., only the valence band is doped, i.e., the chemical potential is at the valence band and the superconducting gap is smaller than the bandgaps between the valence band and other bands, then only the last two terms in Eq.~\eqref{eq:grandgeneral2} and \eqref{eq:grandgeneral2t0} are nonzero.

\subsection{4.3 Separation into the conventional and geometric contributions}
We separate the first term of Eq.~\eqref{eq:dstotal} into the conventional and geometric contributions, $\frac{\partial^2\iOmega}{\partial q_\mu\partial q_\nu}\big|_{\bd{q}=0}=D_{s,\mu\nu}^{\text{conv}}+D_{s,\mu\nu}^{\text{geo}}$. When the valence surface band is doped, single-band projection leads to the BdG Hamiltonian in the band basis
\be\label{eq:hbdgkqband}
\ham_\text{BdG}(\bd{k},\bd{q})=\begin{pmatrix}
    \xi_{v,\bd{k}+\bd{q}} & \Delta_{v,\bd{k}}(\bd{q})\\
    \Delta_{v,\bd{k}}(\bd{q})^* & -\xi_{v,\bd{k}-\bd{q}}
\end{pmatrix},
\ee
where $\Delta_{v,\bd{k}}(\bd{q})=\la \psi^{(v)}_{\bd{k}+\bd{q}}|\hat{\Delta}|\psi^{(v)}_{\bd{k}-\bd{q}}\ra$ (here $\hat{\Delta}$ and $\mu$ are taken as $\bd{q}$-independent). The Bogoliubov spectra of Eq.~\eqref{eq:hbdgkqband} are
\be\label{eq:bogoliubovq}
E_{v\pm,\bd{k}}(\bd{q})=\frac{1}{2}\big[(\xi_{v,\bd{k}+\bd{q}}-\xi_{v,\bd{k}-\bd{q}})\pm \sqrt{(\xi_{v,\bd{k}+\bd{q}}+\xi_{v,\bd{k}-\bd{q}})^2+4|\Delta_{v,\bd{k}}(\bd{q})|^2}\big].
\ee
Inserting this into Eq.~\eqref{eq:grandgeneral2}, the second derivative $\frac{\partial^2\iOmega}{\partial q_\mu\partial q_\nu}$ (the first term of Eq.~\eqref{eq:dstotal}) can be separated into two contributions
\be\label{eq:dsfinitet}
\begin{split}
&D_{s,\mu\nu}^{\text{conv}}=-2\sum_{\bd{k}\in\Lambda_K}\bigg[\bigg(\frac{\xi_{v,\bd{k}}}{E_{v,\bd{k}}}\tanh\frac{\beta E_{v,\bd{k}}}{2}-\tanh\frac{\beta\xi_{v,\bd{k}}}{2}\bigg)\partial_\mu\partial_\nu\xi_{v,\bd{k}}+\frac{\beta}{2}\bigg(\text{sech}^2\frac{\beta E_{v,\bd{k}}}{2}-\text{sech}^2\frac{\beta \xi_{v,\bd{k}}}{2}\bigg)\partial_\mu\xi_{v,\bd{k}}\partial_\nu\xi_{v,\bd{k}}\bigg],\\
&D_{s,\mu\nu}^{\text{geo}}=-2\sum_{\bd{k}\in\Lambda_K}\frac{1}{2E_{v,\bd{k}}}\tanh\frac{\beta E_{v,\bd{k}}}{2}\partial_{q_\mu}\partial_{q_\nu}|\Delta_{v,\bd{k}}(\bd{q})|^2\big|_{\bd{q}=0},
\end{split}
\ee
where $E_{v,\bd{k}}=E_{v+,\bd{k}}(0)\equiv\sqrt{\xi_{v,\bd{k}}^2+\Delta_{v,\bd{k}}(0)^2}$ is the quasiparticle energy. At zero temperature, using $\lim_{\beta\rightarrow+\infty}\text{sech}^2\frac{\beta \xi_{v,\bd{k}}}{2}=\frac{4}{\beta}\delta(\xi_{v,\bd{k}})$, they reduce to
\be\label{eq:dst0}
\begin{split}
&D_{s,\mu\nu}^{\text{conv}}(T=0)=-2\sum_{\bd{k}\in\Lambda_K}\bigg\{\bigg[\frac{\xi_{v,\bd{k}}}{E_{v,\bd{k}}}-\sgn(\xi_{v,\bd{k}})\bigg]\partial_\mu\partial_\nu\xi_{v,\bd{k}}-2\delta(\xi_{v,\bd{k}})\partial_\mu \xi_{v,\bd{k}}\partial_\nu\xi_{v,\bd{k}}\bigg\} \equiv 2\sum_{\bd{k}\in\Lambda_K} f_{\mu\nu}^\text{conv},\\
&D_{s,\mu\nu}^{\text{geo}}(T=0)=-2\sum_{\bd{k}\in\Lambda_K}\frac{1}{2E_{v,\bd{k}}}\partial_{q_\mu}\partial_{q_\nu}|\Delta_{v,\bd{k}}(\bd{q})|^2\big|_{\bd{q}=0} \equiv 2\sum_{\bd{k}\in\Lambda_K} f_{\mu\nu}^\text{geo}.
\end{split}
\ee
Here the $\sgn(\xi_{v,\bd{k}})$ and $\delta(\xi_{v,\bd{k}})$ terms combine to give a boundary term, making $D_{s,\mu\nu}^{\text{conv}}(T=0)$ consistent with the usual integral formula over the entire Brillouin zone with the integrand $-\frac{\xi_{v,\bd{k}}}{E_{v,\bd{k}}}\partial_\mu\partial_\nu\xi_{v,\bd{k}}$. An equivalent and more convenient form is
\be\label{eq:dst0usual}
D_{s,\mu\nu}^{\text{conv}}(T=0)=-2\sum_{\bd{k}\in\Lambda_K}\bigg(\frac{\xi_{v,\bd{k}}}{E_{v,\bd{k}}}+1\bigg)\partial_\mu\partial_\nu\xi_{v,\bd{k}}.
\ee
Attention is needed as the integrand $f_{\mu\nu}^\text{conv}$ depends on the choice of the boundary term $\iOmega_0$. A physically meaningful integrand satisfies $f_{\mu\nu}^\text{conv}(\bd{k})\propto \partial_\mu\xi_{v,\bd{k}}$ and vanishes outside the cutoff region. The term $f_{\mu\nu}^\text{geo}$ has the following expression~\cite{jiang2023pdw}:
\be\label{eq:integrand}
\begin{split}
-\frac{1}{2E_{v,\bd{k}}}\partial_{q_\mu}\partial_{q_\nu}|\Delta_{v,\bd{k}}(\bd{q})|^2\big|_{\bd{q}=0}=&-\frac{1}{2E_{v,\bd{k}}}\partial_{q_\mu}\partial_{q_\nu}\tr\big\{P^{(v)}_{\bd{k}+\bd{q}}\hat{\Delta}P^{(v)}_{\bd{k}-\bd{q}}\hat{\Delta}\big\}\big|_{\bd{q}=0}\\
=&-\frac{1}{E_{v,\bd{k}}}\tr\big\{\partial_\mu\partial_\nu P^{(v)}_\bd{k}\hat{\Delta}P^{(v)}_\bd{k}\hat{\Delta}-\partial_\mu P^{(v)}_\bd{k}\hat{\Delta}\partial_\nu P^{(v)}_\bd{k}\hat{\Delta}\big\}.
\end{split}
\ee
\subsection{4.4 Analysis of the geometric integrand}
In this section, we analyze the integrand Eq.~\eqref{eq:integrand}: we show that (i) due to the decaying feature of the surface states and the surface-polarized pairing matrix $\hat{\Delta}=\text{diag}\{0,...,0,\Delta_{N\text{B}}\}$, Eq.~\eqref{eq:integrand} at the center of the drumhead region is nonvanishing and approximately uniform; (ii) after including the $\bd{k}$-dependence of the order parameter $\Delta_{v,\bd{k}}$, the integrand exhibits a $1/(1-k^2)^2$ behavior near the center, which is similar to the QM. These analyses provide the basis for obtaining the tight lower bound of the surface superfluid weight, $D_{s,xx}^\text{surf}\gtrsim \Delta_{N\text{B}}k_h^2/(2\pi)$ given in the MS.

For our pairing matrix $\hat{\Delta}=\text{diag}\{0,...,0,\Delta_{N\text{B}}\}$, using the projection operator in the orbital basis $P_{\alpha\beta}=\psi_\alpha\psi_\beta^*$, Eq.~\eqref{eq:integrand} can be rewritten in terms of the density of the surface orbital $N$B in the band, $\rho^{(v)}_{N\text{B}}(\bd{k})\equiv P^{(v)}_{N\text{B},N\text{B}}(\bd{k})$,
\be\label{eq:dddelta}
-\partial_{q_\mu}\partial_{q_\nu}|\Delta_{v,\bd{k}}(\bd{q})|^2\big|_{\bd{q}=0}=-2\Delta_{N\text{B}}^2\big[\rho^{(v)}_{N\text{B}}\partial_\mu\partial_\nu \rho^{(v)}_{N\text{B}}-\partial_\mu \rho^{(v)}_{N\text{B}}\partial_\nu \rho^{(v)}_{N\text{B}}\big].
\ee
To establish the connection to the QM, we first notice that when the pairing is uniform, i.e., when $\hat{\Delta}$ is proportional to the identity matrix, $\hat{\Delta}=\Delta_0\hat{\mathcal{I}}$, Eq.~\eqref{eq:integrand} reduces to the QM, as expected. This can be seen from
\be\label{eq:upccase}
\tr\{\partial_\mu P\partial_\nu P-P\partial_\mu\partial_\nu P\}=\tr\{\partial_\mu P\partial_\nu P-\partial_\mu(P\partial_\nu P)+\partial_\mu P\partial_\nu P\}=2\tr\{\partial_\mu P\partial_\nu P\}=4g_{\mu\nu},
\ee
since $\tr\{\partial_\mu(P\partial_\nu P)\}=\frac{1}{2}\tr\partial_\mu\partial_\nu P=0$ and $g_{\mu\nu}=\re \tr\{P\partial_\mu P\partial_\nu P\}=\frac{1}{2}\tr\{\partial_\mu P\partial_\nu P\}$.

Using the valence band wavefunction $\psi^{(2)}$ from Eq.~\eqref{eq:surfacestates}, we expand Eq.~\eqref{eq:upccase} into the orbital basis,
\be\label{eq:centerapprox}
\begin{split}
&\tr\{\partial_\mu P^{(v)}\partial_\nu P^{(v)}-P^{(v)}\partial_\mu\partial_\nu P^{(v)}\}=\sum_{\alpha\beta}(\partial_\mu P^{(v)}_{\alpha\beta}\partial_\nu P^{(v)}_{\beta\alpha}-P^{(v)}_{\alpha\beta}\partial_\mu\partial_\nu P^{(v)}_{\beta\alpha})\\
=&\sum_{z,z'=1}^N\partial_\mu\bigg[\frac{1-k^2}{k^2}e^{\kappa^*(N+1-z)+\kappa(N+1-z')}\bigg]\partial_\nu\bigg[\frac{1-k^2}{k^2}e^{\kappa^*(N+1-z')+\kappa(N+1-z)}\bigg]\\
&-\sum_{z,z'=1}^N\bigg[\frac{1-k^2}{k^2}e^{\kappa^*(N+1-z)+\kappa(N+1-z')}\bigg]\partial_\mu\partial_\nu\bigg[\frac{1-k^2}{k^2}e^{\kappa^*(N+1-z')+\kappa(N+1-z)}\bigg].
\end{split}
\ee
Near the center of the drumhead region, the diagonal components of Eq.~\eqref{eq:centerapprox} are $2-(-2)=4$, which is consistent with the QM $g^{(v)}_{xx}=g^{(v)}_{yy}=1$ that is computed from Eq.~\eqref{eq:r1}. Notice that the surface orbital term, $z=z'=N$ in Eq.~\eqref{eq:centerapprox}, is sufficient for calculating Eq.~\eqref{eq:dddelta}, and it evaluates to exactly half of the total of Eq.~\eqref{eq:centerapprox}, i.e., at $\bd{k}=0$,
\be
\partial_\mu \rho^{(v)}_{N\text{B}}\partial_\nu \rho^{(v)}_{N\text{B}}-\rho^{(v)}_{N\text{B}}\partial_\mu\partial_\nu \rho^{(v)}_{N\text{B}}\big|_{\bd{k}=0}=0-(-2)=2=2g_{\mu\nu}^{(v)}(0),
\ee
leading to $f^{\text{geo}}_{\mu\nu}(\bd{k})\simeq 2\Delta_{N\text{B}}^2/E_{v,\bd{k}}$, which explains why the geometric integrand is approximately uniform near the center.

Next, using Eq.~\eqref{eq:dddelta} and the density $\rho^{(v)}_{N\text{B}}(\bd{k})=1-k^2$ of state $\psi^{(2)}$, the surface orbital term $z=z'=N$ in the entire drumhead region for a many-layer RG ($N\rightarrow\infty$) can also be obtained analytically:
\be\label{eq:geo}
\partial_\mu \rho^{(v)}_{N\text{B}}\partial_\nu \rho^{(v)}_{N\text{B}}-\rho^{(v)}_{N\text{B}}\partial_\mu\partial_\nu \rho^{(v)}_{N\text{B}}=2(1-k^2)\delta_{\mu\nu}+4k_\mu k_\nu.
\ee
When the chemical potential is at the edge of the flat valence band, the quasiparticle energy $E_{v,\bd{k}}=\Delta_{v,\bd{k}}=\Delta_{N\text{B}}(1-k^2)$. Then the geometric integrand in the entire drumhead region has the expression
\be\label{eq:geofull}
f_{\mu\nu}^\text{geo}(\bd{k})=\frac{4\Delta_{N\text{B}}^2[(1-k^2)\delta_{\mu\nu}+2k_\mu k_\nu]}{2\Delta_{N\text{B}}(1-k^2)}=2\Delta_{N\text{B}}\bigg(\delta_{\mu\nu}+\frac{2}{1-k^2}k_\mu k_\nu\bigg).
\ee
At $k_y=0$ and small $k_x$,
\be
f_{xx}^\text{geo}(k_x,k_y=0)=2\Delta_{N\text{B}}\frac{1+k_x^2}{1-k_x^2}\simeq 2\Delta_{N\text{B}}\frac{1}{(1-k_x^2)^2},
\ee
is indeed a slowly varying function, which has a similar $\bd{k}$-dependence as $g^{(v)}_{xx}$. An alternative way to analyze $f_{\mu\nu}^\text{geo}(\bd{k})$ is to compute its eigenvalues and eigenvectors. The two eigenvalues are $d_1(\bd{k})=2\Delta_{N\text{B}}$ and $d_2(\bd{k})=2\Delta_{N\text{B}}\frac{1+k^2}{1-k^2}$, with eigenvectors $\bd{v}_1=(-k_y,k_x)$ and $\bd{v}_2=(k_x,k_y)$, respectively. At the center of the drumhead region, the supercurrent response is isotropic ($d_1=d_2$); as it moves away from the center, the local supercurrent response by a state with momentum $\bd{k}$ gets polarized to vector potential $\bd{A}$ that is parallel to $\bd{k}$ as $d_2>d_1$ (however, the total supercurrent after the $\bd{k}$-space integral is not polarized to any particular direction), with the eigenvalue $d_2\propto \frac{1+k^2}{1-k^2}\simeq \frac{1}{(1-k^2)^2}$.

The nonvanishing uniform $f^\text{geo}_{\mu\nu}$ implies a lower-bounded surface superfluid weight per valley,
\be
D_{s,xx}^\text{surf}=\sum_{k<k_h}f_{xx}^\text{geo}(\bd{k})\geq \sum_{k<k_h}f_{xx}^\text{geo}(0)=\int_{k<k_h}\frac{\derd^2 k}{(2\pi)^2}2\Delta_{N\text{B}}g^{(v)}_{xx}(0)=\frac{1}{2\pi}\Delta_{N\text{B}}k_h^2,
\ee
where $k_h$ can be set to $k_0=1$ as $N\rightarrow\infty$, and an additional prefactor of $54.9e^2/\hbar^2$ also needs to be included (see Sec.~4.2). Here, it is interesting to note the similarity of this result to the isolated flat band isotropic superfluid weight under uniform pairing and half-filling of the band, which is $4\Delta\int_{BZ}\frac{\derd^2 k}{(2\pi)^2} g_{xx}(k)$~\cite{peotta2015superfluidity}.
\subsection{4.5 Superfluid weight integrands vs order parameter, doping, and displacement field}
Here, we provide a parameter-dependent study of the superfluid weight integrands, $f_{\mu\nu}^\text{conv}$ and $f_{\mu\nu}^\text{geo}$, complementary to Fig.~4(a) in the MS. In the MS, we use the RG model with only $\gamma_0,\gamma_1$ hoppings. In real materials, other hoppings need to be included, which can lead to dispersion of the bands. However, as we discuss in Sec.~3 (Fig.~\ref{fig:flatband}), one surface band can be made dispersionless by tuning the displacement field. For this reason, one can still use the model with $\gamma_0,\gamma_1$ hoppings as a platform to study flat-band superconductivity, if we assume the superconducting order parameter can exceed or be of the order of the drumhead bandwidth.

\begin{figure}[hb]
	\centering	
	\includegraphics[height=4.5cm]{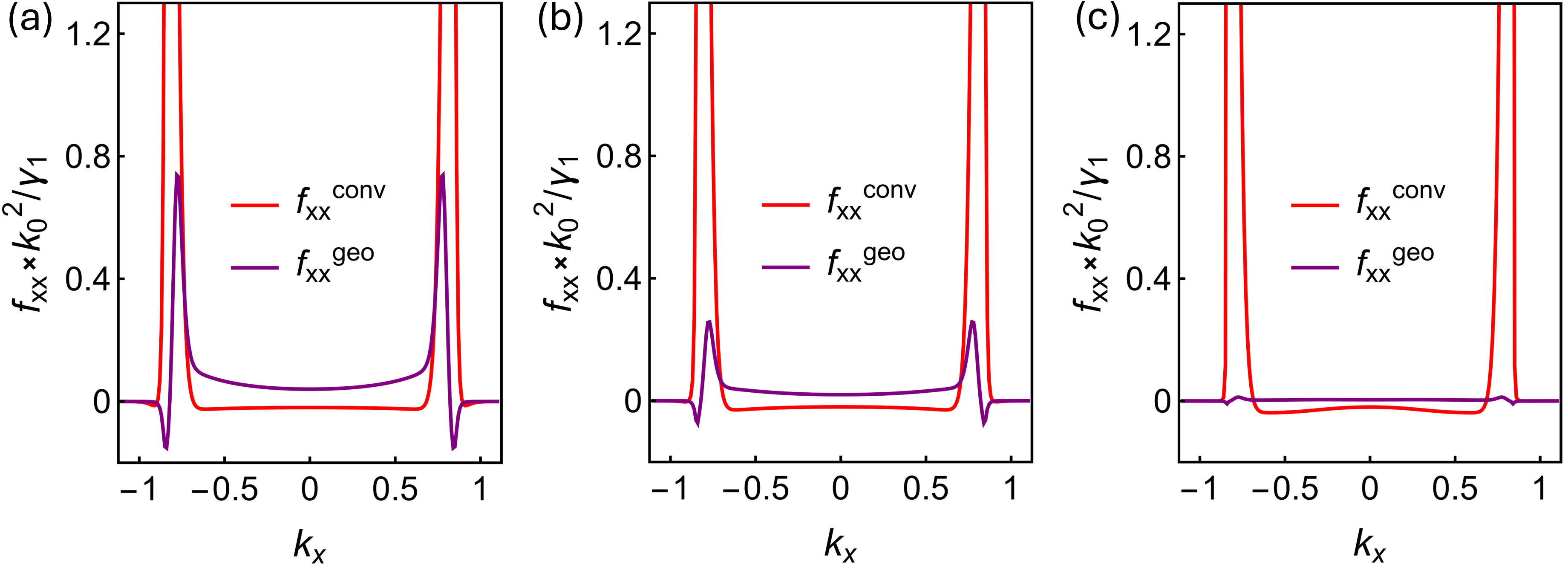}
	\caption{The two integrands vs. $\Delta_{20\text{B}}$, by fixing $m=0.01\gamma_1$, $\mu=-0.01\gamma_1$. (a) $\Delta_{20\text{B}}=0.02\gamma_1$ (same as Fig.~4(a) in the MS); (b) $\Delta_{20\text{B}}=0.01\gamma_1$; (c) $\Delta_{20\text{B}}=0.002\gamma_1$.}
	\label{fig:vsdelta}
\end{figure}

The variations of $f_{\mu\nu}^\text{conv}$ and $f_{\mu\nu}^\text{geo}$ with the superconducting gap, the doping, and the displacement field can be analyzed from their expressions, Eq. (6) and (7) in the MS. We use the parameters chosen in Fig. 4(a) of the MS, $\Delta_{20\text{B}}=0.02\gamma_1$, $m=-\mu=0.01\gamma_1$ ($\mu$ is aligned with the center of the valence band) as a control to study the dependence on each of the three parameters. Here, $\mu$ refers to the chemical potential of the superconducting state (which can be different from that in the normal state due to the strong coupling in flat bands, and the two are related by fixing the same electron density). We note that for the RG model with $\gamma_0,\gamma_1$ hoppings only, the drumhead region has a small bandwidth of the order of $m$ (measured from the center to the rim), and the two surface bands have a $k$-dependent bandgap $2m(1-k^2)$ (see Eq.~\eqref{eq:hmeff}). We choose the parameter $\Delta_{20\text{B}}\leq m$, such that within the interaction scale of $\Delta_{20\text{B}}$ only the valence band is active, which is the condition for Eq. (6) and (7) in the MS to hold.

In Fig.~\ref{fig:vsdelta}, we show the integrands vs. decreasing $\Delta_{20\text{B}}$ values. The geometric integrand $f_{\mu\nu}^\text{geo}$ scales linearly with $\Delta_{20\text{B}}$. The dependence of $f_{\mu\nu}^\text{conv}$ on $\Delta_{20\text{B}}$ is slightly complicated: at the rim, due to the strong dispersion, it falls in the weak coupling or intermediate coupling regime, so the peak does not vary much with $\Delta_{20\text{B}}$; at the center, the order parameter is large or of similar size compared to the drumhead bandwidth. Therefore, $f_{\mu\nu}^\text{conv}$ is negligible there for large $\Delta_{20\text{B}}$, but becomes noticeable when $\Delta_{20\text{B}}$ is small (the sign of $f_{\mu\nu}^\text{conv}$ depends on that of the effective mass $\partial_\mu\partial_\nu\xi_{v,\bd{k}}$).

\begin{figure}[ht]
	\centering	
	\includegraphics[height=4.5cm]{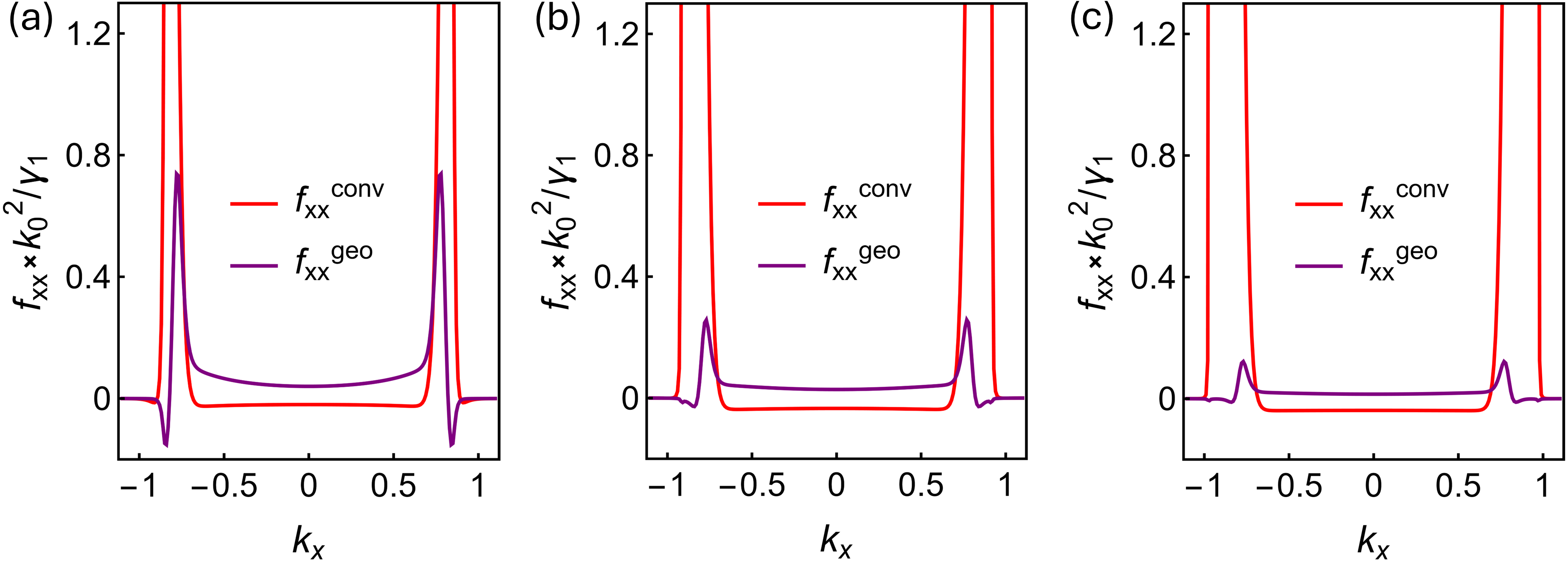}
	\caption{The two integrands vs. $\mu$, by fixing $\Delta_{20\text{B}}=0.02\gamma_1$, $m=0.01\gamma_1$. (a) $\mu=-0.01\gamma_1$ (same as Fig.~4(a) in the MS); (b) $\mu=-0.03\gamma_1$; (c) $\mu=-0.06\gamma_1$.}
	\label{fig:vsmu}
\end{figure}

In Fig.~\ref{fig:vsmu}, we show the integrands vs. decreasing $\mu$ (increasing hole doping). The dependence of $f_{\mu\nu}^\text{geo}$ on $\mu$ (or the filling $\nu$) follows the standard result of flat-band superconductivity~\cite{peotta2015superfluidity}, $f_{\mu\nu}^\text{geo}\propto\nu(1-\nu)$, where $0\leq \nu\leq 1$ in our case is measured with respect to the drumhead flat-band area. Notice that $\mu=-0.01\gamma_1$ [Fig.~\ref{fig:vsmu}(a)] aligns with the center of the valence band, so it is close to the optimal doping, $\nu=1/2$. For the conventional integrand, as the hole doping increases, $\mu$ gets aligned with the more dispersive part of the valence band; therefore, the peak at the rim gets stronger.

\begin{figure}[ht]
	\centering	
	\includegraphics[height=4.5cm]{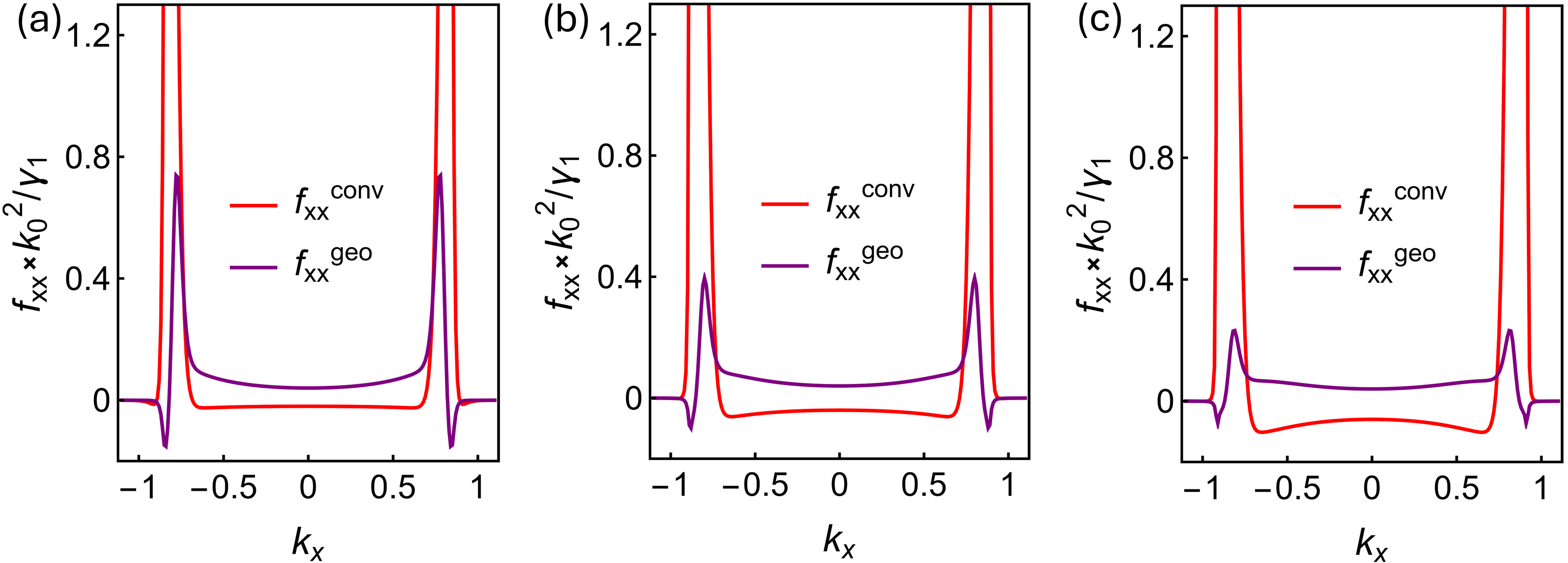}
	\caption{The two integrands vs. $m$, by fixing $\Delta_{20\text{B}}=0.02\gamma_1$. (a) $m=-\mu=0.01\gamma_1$ (same as Fig.~4(a) in the MS); (b) $m=-\mu=0.02\gamma_1$; (c) $m=-\mu=0.03\gamma_3$.}
	\label{fig:vsm}
\end{figure}

In Fig.~\ref{fig:vsmu}, we show the integrands vs. increasing $m$, at about the same doping level, which means we fix $\mu=-m$. As discussed in the MS, increasing $m$ does not affect the eigenstates or IQG (on the contrary, it preserves them); therefore, $f_{\mu\nu}^\text{geo}$ at the center is stable against the change of $m$. At the rim, the nodes of the quasiparticle energy $E_{v,\bd{k}}$ become less singular since we align $\mu=-m$, making the peak of $f_{\mu\nu}^\text{geo}$ less singular. A larger $m$ also leads to stronger dispersion both at the center and the rim, so the pattern of $f_{\mu\nu}^\text{conv}$ in Fig.~\ref{fig:vsm} is also easily expected.

\subsection{4.6 Orbital embedding correction term for the superfluid weight, $D_s^\Delta$}
The geometric contribution $D_s^\text{geo}$ derived from the first term of Eq.~\eqref{eq:dstotal} is generally orbital-embedding-dependent (again, the orbital embedding refers to the gauge freedom of the Bloch Hamiltonian), so an orbital-embedding correction term, $D_s^\Delta$ is needed to make the total geometric superfluid weight $D_s^\text{geo}+D_s^\Delta$ orbital-embedding-independent~\cite{huhtinen2022revisiting,tam2024geometry,jiang2024geometric}. Here, we provide this information for the RG superconductivity, which can be compared with the study of the orbital-embedding-dependence of IQG in Sec.~2.3. We show that when the superconductivity is polarized to one surface of RG, $D_s^\Delta$ vanishes.

After mean-field decoupling, the onsite interaction for the supercurrent state is
\be\label{eq:hintmf}
\hham_\text{MF}^{(int)}(\bd{q})=\sum_{\bd{k},\alpha}(\Delta_{\bd{q},\alpha}c_{\bd{k}+\bd{q},\alpha\uparrow}^\dagger c_{-\bd{k}+\bd{q},\alpha\downarrow}^\dagger+h.c.)+\frac{N_c}{U}\sum_\alpha|\Delta_{\bd{q},\alpha}|^2,
\ee
where the order parameter is
\be\label{eq:gapq1}
\Delta_{\bd{q},\alpha}=-\frac{U}{N_c}\sum_\bd{k}\la c_{-\bd{k}+\bd{q},\alpha\downarrow}c_{\bd{k}+\bd{q},\alpha\uparrow}\ra_\bd{q}=-U\la c_{\bd{R}\alpha\downarrow}c_{\bd{R}\alpha\uparrow}\ra_\bd{q}e^{-2i\bd{q}\cdot(\bd{R}+\bd{r}_\alpha)},
\ee
with $\la\ra_\bd{q}$ the average in the supercurrent state $\Psi_\bd{q}$. Under the orbital embedding transformation $\bd{r}_\alpha\rightarrow \bd{r}_\alpha+\delta\bd{r}_\alpha$, operator $c_{\bd{k}\alpha\sigma}$ transforms as $c_{\bd{k}\alpha\sigma}\rightarrow \wt{c}_{\bd{k}\alpha\sigma}=c_{\bd{k}\alpha\sigma}e^{-i\bd{k}\cdot\delta\bd{r}_\alpha}$. This gauge choice does not affect the state $\Psi_\bd{q}$, so $\la c_{\bd{R}\alpha\downarrow}c_{\bd{R}\alpha\uparrow}\ra_\bd{q}$ is $\bd{r}_\alpha$-independent. Then one can verify that the mean-field interaction Eq.~\eqref{eq:hintmf} is invariant under this $U(1)\times...\times U(1)$ transformation. Comparing with the orbital embedding transformation of IQG, Eq.~\eqref{eq:quadraticform}, here the number of copies of $U(1)$ is restricted to the number of orbitals with nonzero order parameters $\Delta_\alpha$ instead of the number of orbitals with nonzero weight $|\psi_\alpha|^2$ in the band.

To compute $D_s^{\Delta}$, we write down the gap equation for the supercurrent state in the orbital basis:
\be\label{eq:gapq2}
\Delta_{\bd{q},\alpha}=\frac{U}{N_c}\sum_{\bd{k}\in\Lambda_K}\frac{\psi^{(v)*}_{\bd{k}-\bd{q},\alpha}\psi^{(v)}_{\bd{k}+\bd{q},\alpha}\Delta_{v,\bd{k}}(\bd{q})}{\sqrt{(\xi_{v,\bd{k}+\bd{q}}+\xi_{v,\bd{k}-\bd{q}})^2+4|\Delta_{v,\bd{k}}(\bd{q})|^2}}\bigg(\tanh\frac{\beta E_{v+,\bd{k}}(\bd{q})}{2}-\tanh\frac{\beta E_{v-,\bd{k}}(\bd{q})}{2}\bigg),
\ee
where the Bogoliubov spectrum $E_{v\pm,\bd{k}}(\bd{q})$ are given by Eq.~\eqref{eq:bogoliubovq}. Computing the derivative of Eq.~\eqref{eq:gapq2} leads to an equation $\sum_\beta M_{\alpha\beta}\derd_\mu\Delta_\beta=V_{\alpha,\mu}$ (where $M_{\alpha\beta}\equiv\frac{1}{2}\frac{\partial^2\iOmega}{\partial\Delta_{\bd{q},\alpha}^I\partial\Delta_{\bd{q},\beta}^I}\big|_{\bd{q}=0}$, $V_{\alpha,\mu}\equiv-\frac{\partial^2\iOmega}{\partial\Delta_{\bd{q},\alpha}^*\partial q_\mu}\big|_{\bd{q}=0}$, and $\derd_\mu\Delta_\beta\equiv\frac{\derd\Delta_{\bd{q},\beta}}{\derd q_\mu}\big|_{\bd{q}=0}$), with \cite{jiang2024geometric}
\be\label{eq:mmatrix}
M_{\alpha\beta}=\frac{\delta_{\alpha\beta}}{\Delta_\alpha}\sum_{\bd{k}\in\Lambda_K}\frac{\Delta_{v,\bd{k}}(0)}{E_{v,\bd{k}}}\tanh\frac{\beta E_{v,\bd{k}}}{2}|\psi^{(v)}_{\bd{k},\beta}|^2-\sum_{\bd{k}\in\Lambda_K}\frac{1}{E_{v,\bd{k}}}\tanh\frac{\beta E_{v,\bd{k}}}{2}|\psi^{(v)}_{\bd{k},\alpha}|^2|\psi^{(v)}_{\bd{k},\beta}|^2,
\ee
\be\label{eq:vvector}
V_{\alpha,\mu}=\sum_{\bd{k}\in\Lambda_K}\frac{1}{E_{v,\bd{k}}}\tanh\frac{\beta E_{v,\bd{k}}}{2}\big[(\psi^{(v)*}_{\bd{k},\alpha}\partial_\mu \psi^{(v)}_{\bd{k},\alpha}-\partial_\mu \psi^{(v)*}_{\bd{k},\alpha} \psi^{(v)}_{\bd{k},\alpha})\Delta_{v,\bd{k}}(0)-|\psi^{(v)}_{\bd{k},\alpha}|^2(\la \psi^{(v)}_{\bd{k}}|\hat{\Delta}|\partial_\mu \psi^{(v)}_{\bd{k}}\ra-\la \partial_\mu \psi^{(v)}_{\bd{k}}|\hat{\Delta}|\psi^{(v)}_{\bd{k}}\ra)\big].
\ee
Using Eq.~\eqref{eq:mmatrix}, the diagonal entries of $M$ are
\be\label{eq:malphaalpha}
M_{\alpha\alpha}=\sum_{\bd{k}\in\Lambda_K}\sum_{\beta\neq\alpha}|\psi^{(v)}_{\bd{k},\alpha}|^2|\psi^{(v)}_{\bd{k},\beta}|^2\frac{\Delta_\beta}{\Delta_\alpha}\frac{\tanh(\beta E_{v,\bd{k}}/2)}{E_{v,\bd{k}}}.
\ee
Assuming that the only nonzero order parameters are $\Delta_{1\text{A}}$ and $\Delta_{N\text{B}}$, which means that the orbital embedding reduces to a $U(1)\times U(1)$ symmetry of the mean-field Hamiltonian. Then $M_{\alpha\alpha}$ for the other $2N-2$ orbitals, $\alpha\in \{1\text{B}, ..., N\text{A}\}$ diverge. However, $V_{\alpha,\mu}=\sum_\beta M_{\alpha\beta}\derd_\mu\Delta_\beta$ remains finite (as the r.h.s. of Eq.~\eqref{eq:vvector} is finite) due to $\derd_\mu\Delta_\alpha=0$ for these $(2N-2)$ orbitals. As a result, $D_{s,\mu\nu}^{\Delta}=2\sum_\alpha V_{\alpha,\mu}\derd_\nu\Delta_\alpha$ is always finite.

This suggests that one can restrict the summation of orbitals in the second term of Eq.~\eqref{eq:dstotal} to orbitals 1A and $N$B only:
\be\label{eq:ds2}
D_{s,\mu\nu}^{\Delta}=-\sum_{\alpha,\beta=1\text{A},N\text{B}}\frac{\derd\Delta^I_{\bd{q},\alpha}}{\derd q_\mu} \frac{\partial^2\iOmega}{\partial \Delta^I_{\bd{q},\alpha}\partial \Delta^I_{\bd{q},\beta}}\frac{\derd\Delta^I_{\bd{q},\beta}}{\derd q_\nu}\bigg|_{\bd{q}=0}=2\sum_{\alpha,\beta=1\text{A},N\text{B}}\derd_\mu\Delta_\alpha M_{\alpha\beta}\derd_\nu\Delta_\beta.
\ee
Let us denote the restriction of the matrix $M$ to these two orbitals as $M_R$ (a 2 by 2 matrix), and similarly, the vector $V$ to these two orbitals as $V_R$ (a 2-dim vector). The matrix $M_R$ has a kernel eigenvector $\bd{v}_0=(\Delta_{1\text{A}},\Delta_{N\text{B}})$, which enables us to eliminate, e.g., the second row and column of $M_R$ and the second component of $V_R$. Finally, this gives the simple expression
\be\label{eq:dsinmv}
D_{s,\mu\nu}^{\Delta}=2\frac{V_{1\text{A},\mu}V_{1\text{A},\nu}}{M_{1\text{A},1\text{A}}}.
\ee
with
\be
M_{1\text{A},1\text{A}}=\sum_{\bd{k}\in\Lambda_K}|\psi^{(v)}_{\bd{k},1\text{A}}|^2|\psi^{(v)}_{\bd{k},N\text{B}}|^2\frac{\Delta_{N\text{B}}}{\Delta_{1\text{A}}}\frac{\tanh(\beta E_{v,\bd{k}}/2)}{E_{v,\bd{k}}},
\ee
\be
V_{1\text{A},\mu}=\sum_{\bd{k}\in\Lambda_K}\frac{\Delta_{N\text{B}}}{E_{v,\bd{k}}}\tanh\frac{\beta E_{v,\bd{k}}}{2}\big[|\psi^{(v)}_{\bd{k},N\text{B}}|^2(\psi^{(v)*}_{\bd{k},1\text{A}}\partial_\mu \psi^{(v)}_{\bd{k},1\text{A}}-\partial_\mu \psi^{(v)*}_{\bd{k},1\text{A}} \psi^{(v)}_{\bd{k},1\text{A}})-|\psi^{(v)}_{\bd{k},1\text{A}}|^2(\psi^{(v)*}_{\bd{k},N\text{B}}\partial_\mu \psi^{(v)}_{\bd{k},N\text{B}}-\partial_\mu \psi^{(v)*}_{\bd{k},N\text{B}} \psi^{(v)}_{\bd{k},N\text{B}})\big].
\ee
Inserting these into Eq.~\eqref{eq:dsinmv}, we find $D_{s,\mu\nu}^{\Delta}\propto\Delta_{1\text{A}}\Delta_{N\text{B}}$, which vanishes if either one of the two orders vanishes.

To conclude, if the superconducting order parameters of RG are nonzero on both surfaces, then a nonzero $D_{s,\mu\nu}^{\Delta}$ makes the total geometric superfluid weight, $D_s^\text{geo}+D_{s,\mu\nu}^{\Delta}$ orbital-embedding-independent; otherwise, if the order parameter is polarized to only one surface, then $D_{s,\mu\nu}^{\Delta}$ vanishes and the term $D_s^\text{geo}$ itself is orbital-embedding-independent.

\section{5. Rhombohedral trilayer graphene (RTG) and superconducting phase SC1}
\subsection{5.1 SC1 in the phase diagram}
We explain how to use the onsite attractive interaction model to understand some features of the superconducting phase SC1 in the phase diagram of RTG reported in Ref.~\cite{zhou2021superconductivity}.

For RTG, we also include the additional potential parameters $\Delta_2=-0.0023$ eV and $\delta=-0.0105$ eV, which describe the asymmetry between A and B sublattices on the surfaces and the nonlinear potential drop along $z$ direction~\cite{zhou2021half}. Introducing dimensionless parameters, $\wt\Delta_2=\Delta_2/\gamma_1=-0.006$ and $\wt\delta=\delta/\gamma_1=-0.028$, the dimensionless noninteracting continuum model Hamiltonian of RTG, including the onsite potentials, reads
\be\label{eq:htrilayer}
\ham^\text{RTG}(\bd{k})=\begin{pmatrix}
\wt m+\wt\Delta_2+\wt\delta & \pi^* & \wt{v}_4\pi^* & \wt{v}_3\pi & 0 & \wt{\gamma}_2/2\\
\pi & \wt m+\wt\Delta_2 & 1 & \wt{v}_4\pi^* & 0 & 0\\
\wt{v}_4\pi & 1 & -2\wt\Delta_2 & \pi^* & \wt{v}_4\pi^* & \wt{v}_3\pi\\
\wt{v}_3\pi^* & \wt{v}_4\pi & \pi & -2\wt\Delta_2 & 1 & \wt{v}_4\pi^*\\
0 & 0 & \wt{v}_4\pi & 1 & -\wt m+\wt\Delta_2 & \pi^* \\
\wt{\gamma}_2/2 & 0 & \wt{v}_3\pi^* & \wt{v}_4\pi & \pi & -\wt m+\wt\Delta_2+\wt\delta
\end{pmatrix},
\ee
where $\pi=\xi k_x+ik_y$ with $\xi=\pm$ for valley $K$ and $K'$.

\begin{figure}[b]
	\centering	
	\includegraphics[height=6cm]{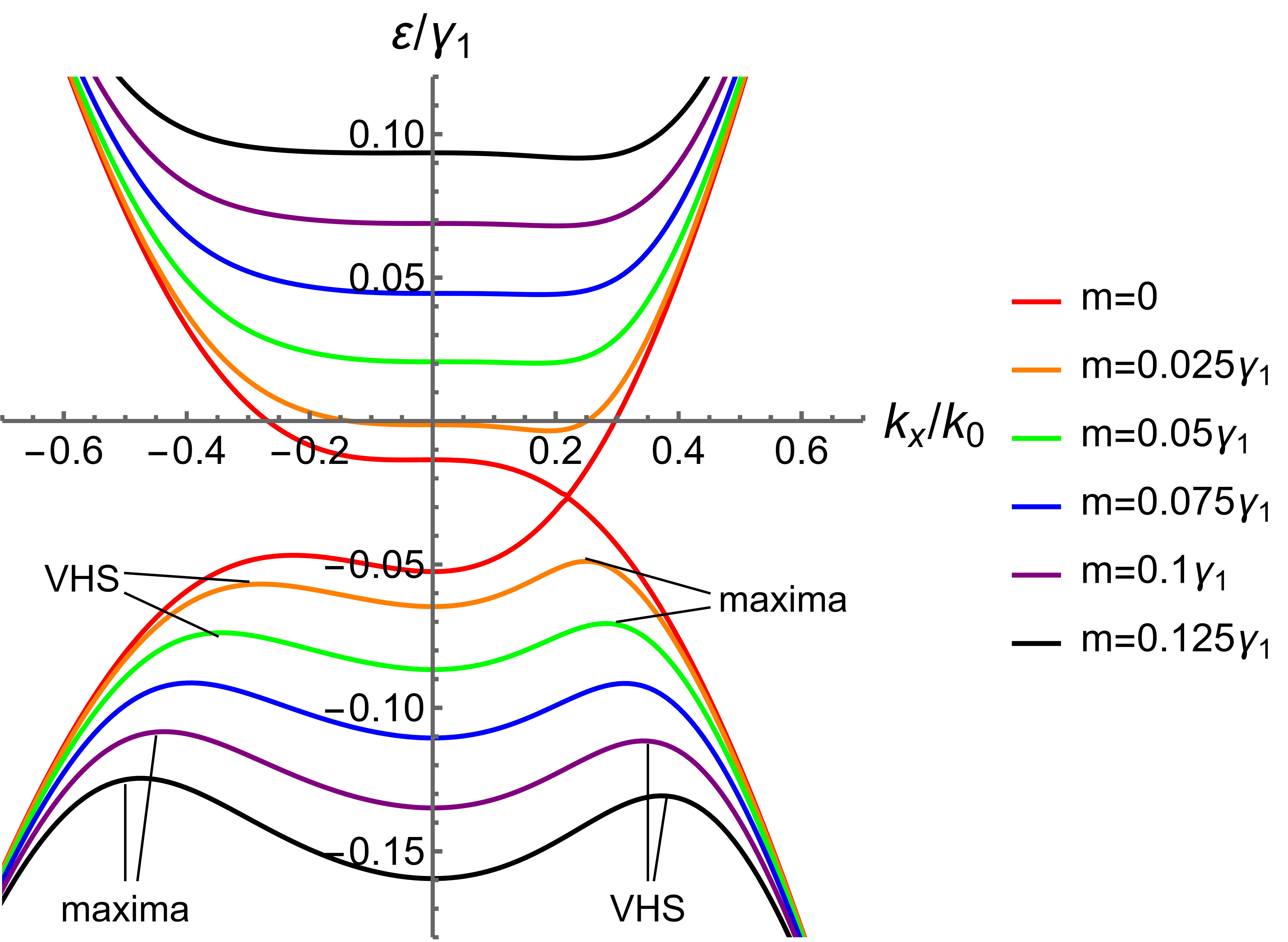}
	\caption{RTG surface bands at valley $K$ with different surface potentials $m$.}
	\label{fig:trilayerdisp}
\end{figure}

One can interpret the superconducting phase diagram of the hole-doped RTG~\cite{zhou2021superconductivity} in two different ways. The first interpretation (interpretation 1) is: the narrow SC1 phase with width $0.1-0.2\times10^{12}$ cm$^{-2}$ in electron density is located at the singular point of the saddle-point van Hove singularity (VHS) \cite{cea2022superconductivity}, and triggered by the enhancement of the density of states there as the displacement field $D$ is applied. An alternative interpretation (interpretation 2) is that SC1 is at the edge of the VHS, while the majority of the VHS region is occupied with other correlated phases, e.g., the inter-valley coherent state \cite{you2022kohn,chatterjee2022inter,chau2024origin}. Both interpretations can be qualitatively fit using the onsite interaction model. The differences are with the interaction strength $U$ and the onset surface potential $m_o$ (the value of $m$ that corresponds to the endpoint of SC1 at $n_e=-10^{12}$ cm$^{-2}$ in Ref. \cite{zhou2021superconductivity}). For interpretation 1, we need to fit $U=30$ meV ($0.08\gamma_1$) and $m_o=50$ meV (this onset $m_o$ agrees with Ref. \cite{cea2022superconductivity}) while for interpretation 2, $U=42$ meV ($0.11\gamma_1$) and $m_o\approx 20$ meV.

We first discuss interpretation 1. It requires that the hole density of a state in SC1 always coincides with that of the VHS singular point as $m$ varies. In Fig.~\ref{fig:trilayerdisp}, we plot the dispersion of the two surface bands with different surface potentials $m$ (band structure is calculated using Eq.~\eqref{eq:htrilayer} with parameters from Sec.~1). Focusing on the hole-doped side, we find the band maximum and the saddle-point VHS switch places at $m=0.075\gamma_1$ ($\sim 29$ meV, blue curve), which means the VHS corresponds to the valence band edge at $m=0.075\gamma_1$. As a result, at $m<0.075\gamma_1$, the VHS singular point shifts to lower hole densities as $m$ increases. In contrast, the experimentally observed SC1 shifts to higher hole densities as $m$ increases; therefore, for interpretation 1 to work, SC1 corresponds to $m$ values larger than $m=0.075\gamma_1$.

\begin{figure}[h]
\centering	
\includegraphics[height=5cm]{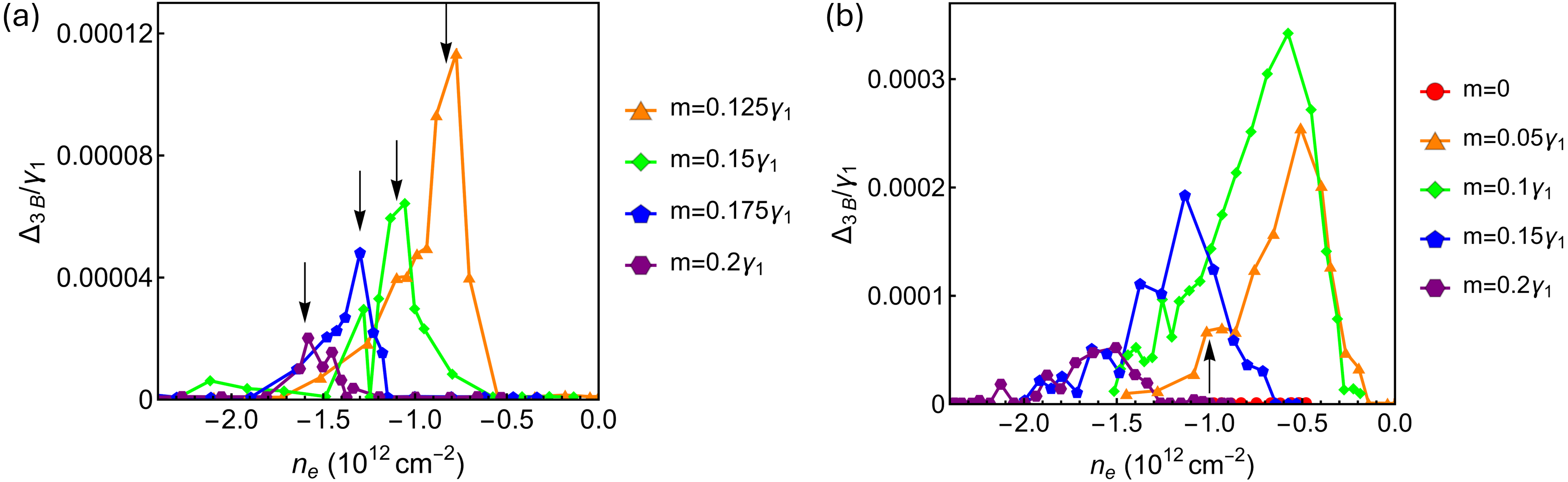}
\caption{Two different interpretations of SC1 of the superconducting phase diagram of RTG. (a) SC1 corresponds to the peaks of the superconducting domes from an attractive Hubbard model (indicated by arrows), with $U=30$ meV. The onset surface potential (i.e., when the peak has electron density $-1\times10^{12}$ cm$^{-2}$) is $m_o\sim0.125\gamma_1\approx50$ meV (orange curve). These peaks have a width of $0.1-0.2\times10^{12}$ cm$^{-2}$. (b) SC1 corresponds to the left edge of the superconducting domes (indicated by an arrow for the orange curve only), with $U=42$ meV and $m_o\sim 0.05\gamma_1\approx 20$ meV. In both interpretations (a) and (b), the observed SC1 has the order parameter $\Delta_{3\text{B}}$ of the order of $0.0001\gamma_1$, giving a mean-field $T_{MF}\sim 100$ mK.}
\label{fig:domes}
\end{figure}

We choose $\Lambda_K$ to be the square area of $2.4k_0\times2.4k_0$ at $K$ point, with a $100\times100$ $\bd{k}$-mesh to solve Eq.~\eqref{eq:gapnumerical} and \eqref{eq:enumbernumerical} numerically for the superconducting dome (i.e., the plot of the superconductor order parameter vs. electron density with fixed $U$). For attractive Hubbard interaction, the peak of the superconducting dome is exactly pinned at the singular point of the VHS since the density of states drives the superconducting transition. Then we obtain the superconducting dome plot for various surface potentials $m$ in Fig.~\ref{fig:domes}(a). We find when the dome is peaked at $n_e=-1\times10^{12}$ cm$^{-2}$, $m\approx50$ meV$\equiv m_o$. This $m_o$ is determined by the band structure only, therefore, it agrees with Ref. \cite{cea2022superconductivity} as expected (see Fig.~3 therein). To fit the transition temperature or $\Delta_{3\text{B}}$ to the order of magnitude $100$ mK, we must fine-tune the attractive interaction $U\sim0.08\gamma_1=30$ meV.

Next, we discuss interpretation 2 above. Since the attractive Hubbard model cannot describe other correlated states, one must assume that the numerically calculated superconducting dome has a much larger span in electron density than the experimentally observed one. In other words, phase SC1 only refers to the left edge of the dome calculated from the attractive Hubbard model, whereas the majority of the dome is replaced by other correlated states \cite{you2022kohn}. In Fig.~\ref{fig:domes}(b), we show such a calculation that qualitatively reproduces the SC1 phase, for which we have to fine-tune $U\sim0.11\gamma_1=42$ meV.

We note that the main difference between the two interpretations above lies in the conversion of the displacement field $D$ to the surface potential $m$. Using data from the bilayer graphene experiment~\cite{zhang2009direct} and taking into account the different thicknesses of the two materials, we estimate $m_o$ for SC1 to be 25-30 meV, which is closer to interpretation 2 above. Moreover, this estimated $m_o$ is close to the $m_s=0.06\gamma_1$ for the saturation of surface polarization in RTG, which suggests that the surface polarization effect plays an important role in the transition to SC1 as the displacement field increases.

\subsection{5.2 Surface polarization effect on the gap equation}
We explain the surface polarization effect on the gap equation of RTG SC1 phase. We show that the surface polarization by the displacement field contributes to an additional factor of 2 in the coupling constant; therefore, it aids the superconducting transition. In the expression of the superconducting gap, the dependence on the displacement field has two sources---the DOS and the surface orbital weight in the band.

Due to the weak-coupling nature, RTG SC1 is a multi-orbital superconductor where only the valence band is active within the attractive interaction scale. The only relevant order parameters are those of the two surface orbitals, $\Delta_{1\text{A}}$ and $\Delta_{3\text{B}}$, which can be solved from the following coupled gap equations:
\be\label{eq:gapcouple}
\begin{split}
&\Delta_{1\text{A}}=\frac{U}{N_c}\sum_{\bd{k}\in\Lambda_K}|\psi^{(v)}_{\bd{k},1\text{A}}|^2\frac{\Delta_{v,\bd{k}}}{\sqrt{\xi_{v,\bd{k}}^2+\Delta_{v,\bd{k}}^2}},\\
&\Delta_{3\text{B}}=\frac{U}{N_c}\sum_{\bd{k}\in\Lambda_K}|\psi^{(v)}_{\bd{k},3\text{B}}|^2\frac{\Delta_{v,\bd{k}}}{\sqrt{\xi_{v,\bd{k}}^2+\Delta_{v,\bd{k}}^2}},
\end{split}
\ee
where
\be
\Delta_{v,\bd{k}}=|\psi^{(v)}_{\bd{k},1\text{A}}|^2\Delta_{1\text{A}}+|\psi^{(v)}_{\bd{k},3\text{B}}|^2\Delta_{3\text{B}}=\rho^{(v)}_{1\text{A}}(\bd{k})\Delta_{1\text{A}}+\rho^{(v)}_{3\text{B}}(\bd{k})\Delta_{3\text{B}}
\ee
is the order parameter in the band basis, which couples $\Delta_{1\text{A}}$ and $\Delta_{3\text{B}}$ through the weight of orbital 1A ($\rho^{(v)}_{1\text{A}}(\bd{k})$) and 3B ($\rho^{(v)}_{3\text{B}}(\bd{k})$) in the valence band.

The solution to Eq.~\eqref{eq:gapcouple} can be accessed if we notice that one order parameter, e.g., $\Delta_{1\text{A}}$ vanishes throughout the metal-superconductor transition. At zero displacement field, $m=0$, it vanishes because SC1 is not activated; when SC1 is activated at a large displacement field, it also vanishes since the orbital weight $\rho^{(v)}_{1\text{A}}(\bd{k})$ is quenched by the surface polarization. As a result, Eq.~\eqref{eq:gapcouple} reduces to a gap equation involving $\Delta_{3\text{B}}$ only:
\be\label{eq:gap3b}
\Delta_{3\text{B}}=\frac{U}{N_c}\sum_{\bd{k}\in\Lambda_K}\frac{(\rho^{(v)}_{3\text{B}})^2\Delta_{3\text{B}}}{\sqrt{\xi_{v,\bd{k}}^2+(\rho^{(v)}_{3\text{B}})^2\Delta_{3\text{B}}^2}},
\ee
where we have made the approximation that $\rho^{(v)}_{3\text{B}}$ is $\bd{k}$-independent near the VHS, but depends on the surface potential $m$ explicitly. Then, Eq.~\eqref{eq:gap3b} can be solved to give the order parameter as a function of $m$,
\be
\Delta_{3\text{B}}(m)=\frac{2\omega_c}{\rho^{(v)}_{3\text{B}}(m)}\exp\bigg\{-\frac{1}{2\rho^{(v)}_{3\text{B}}(m)D_0(m)U}\bigg\}.
\ee
Here, $\omega_c$ is the cutoff energy around the Fermi surface where the interaction is attractive, and $D_0=\frac{1}{N_c}\frac{\derd N_0(\xi)}{\derd\xi}\big|_{\xi=0}$ is the DOS at the Fermi surface. It is assumed that $\omega_c\gg \Delta_{3\text{B}}$ (note: $\Delta_{3\text{B}}\sim 10^{-4}\gamma_1$ for SC1) but $\omega_c$ is smaller than any band structure energy scale of RTG, so $D_0$ can be treated as a constant in the integral. We adopted interpretation 2 from Sec.~5.1 above, where SC1 is located at the edge of the VHS rather than at the VHS singular point, such that $D_0$ is finite, but is enhanced as $m$ increases. We also assumed that $U$ does not change as $m$ varies (in reality, of course, it may renormalize).

\begin{figure}[h!]
	\centering	
	\includegraphics[height=3.5cm]{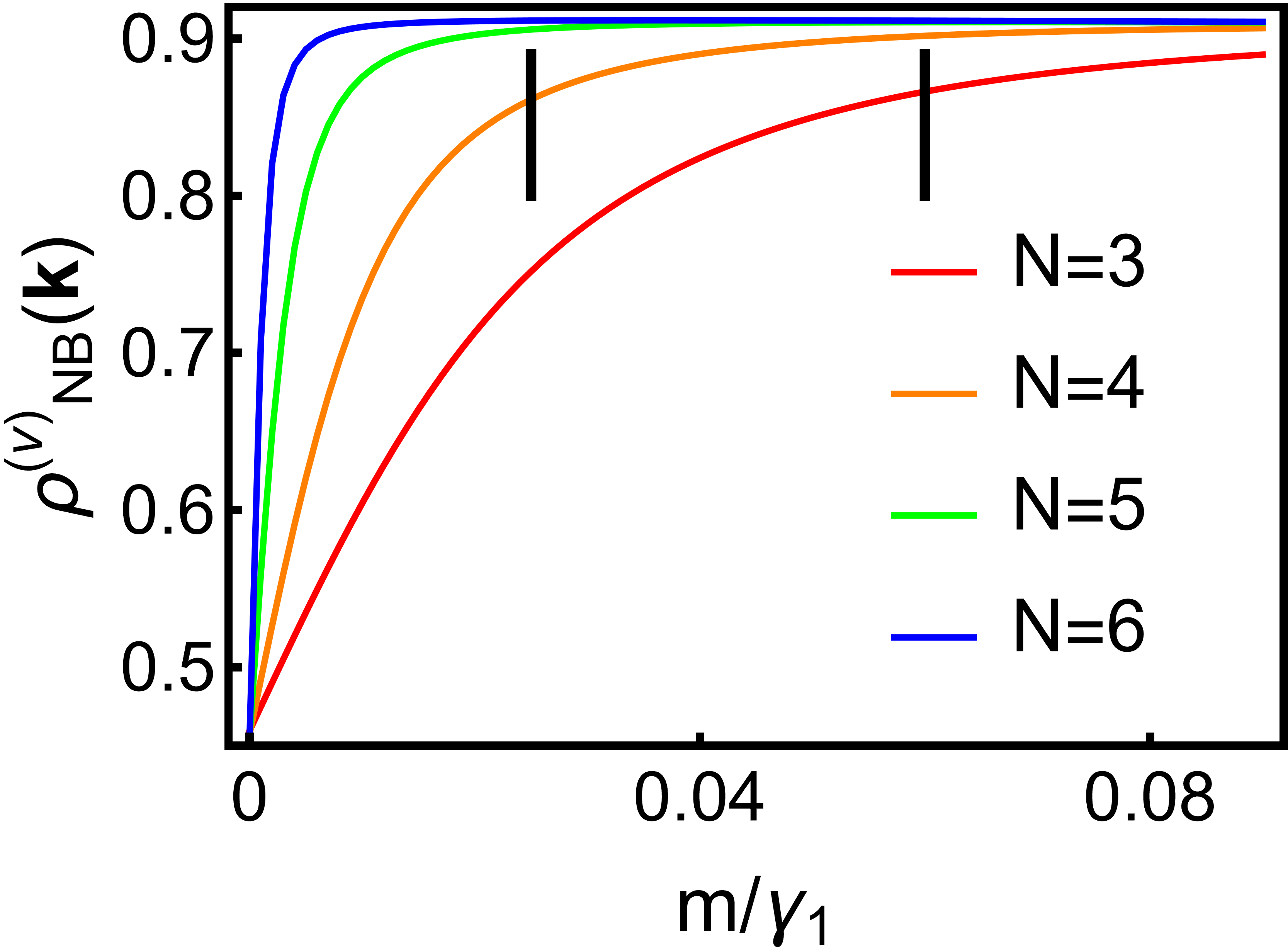}
	\caption{Surface polarization effect on the surface orbital weight $\rho^{(v)}_{N\text{B}}(\bd{k})=|\psi^{(v)}_{\bd{k},N\text{B}}|^2$ in the valence surface band, at $\bd{k}=(-0.3,0)$ (near the saddle-point VHS for RTG, c.f. Fig.~\ref{fig:trilayerdisp}) for various $N$-layer RG. At zero displacement field, $\rho^{(v)}_{N\text{B}}=0.45$ due to the mixing of the two surfaces by the long-range hoppings; it increases to 0.9 and saturates as $m$ increases. The black lines indicate the value of $m$ when $\rho^{(v)}_{N\text{B}}$ roughly saturates, which decreases fast as $N$ increases, as explained in Sec.~3.}
	\label{fig:polarize}
\end{figure}

Comparing with the BCS theory, the coupling constant here, $g=2\rho^{(v)}_{3\text{B}}D_0U$ also takes the orbital weight $\rho^{(v)}_{3\text{B}}$ into account. Its field-dependence comes from both $\rho^{(v)}_{3\text{B}}$ and $D_0$. As the displacement field is turned on until one surface is fully polarized, $\rho^{(v)}_{3\text{B}}$ is doubled (from 0.45 to 0.9 at the VHS of RTG, see Fig.~\ref{fig:polarize}); meanwhile, the DOS $D_0$ at the edge of the VHS is also enhanced by a factor of 2 to 3 (see Ref.~\cite{zhou2021superconductivity} for the $D_0$ vs $m$ plot). Therefore, we conclude that the surface polarization effect on $\rho^{(v)}_{3\text{B}}$ is indispensable for the transition to SC1.

\subsection{5.3 Superfluid weight of SC1}
We briefly discuss the superfluid weight of SC1 of RTG, although SC1 is more like a conventional-type superconductor, which does not align with the superconducting state of many-layer RG as discussed in the MS.

Before calculating the superfluid weight, let us estimate the order of magnitude of different terms contributing to it for SC1. The conventional term $D_s^{\text{conv}}$ does not scale with $\Delta_\alpha$ if the coupling is weak and if $\mu$ measured from the band edge is much larger than $\Delta_\alpha$; the geometric term is always linear in the order parameter, so $D_s^{\text{geo}}\propto\Delta_\alpha$; whereas from Sec.~4.5 we know $D_s^{\Delta}\propto \Delta_{1\text{A}}\Delta_{N\text{B}}$. For SC1 of RTG, $\Delta_{1\text{A}}=0$ due to the polarization, so $D_s^{\Delta}=0$. Meanwhile, $\Delta_{3\text{B}}\sim10^{-4}\gamma_1$ is the smallest energy scale in the system, making $D_s^{\text{geo}}\ll D_s^{\text{conv}}$.

\begin{figure}[h!]
	\centering	
	\includegraphics[height=3.5cm]{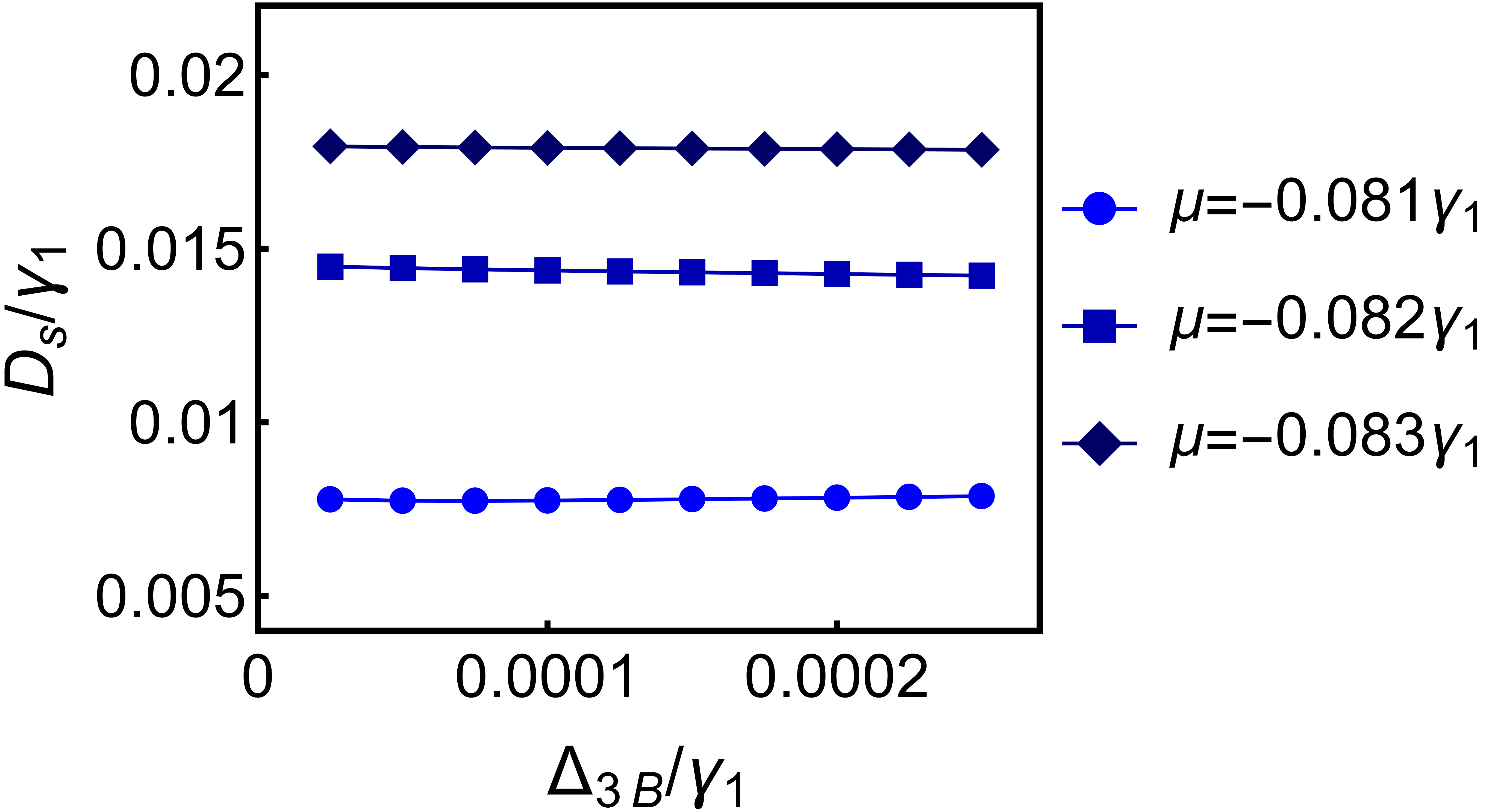}
	\caption{The total superfluid weight $D_s$ vs. $\Delta_{3\text{B}}$ at several chemical potentials $\mu$ close to the VHS, at surface potential $m=0.06\gamma_1$.}
	\label{fig:scaling}
\end{figure}

In Fig.~\ref{fig:scaling}, we present the numerical calculation results for the superfluid weight $D_s$ of SC1 of RTG, at $m=0.06\gamma_1=23$ meV. Since the exact value of the order parameter may not be obtained from the transition temperature $T_c$ directly (there could be phase fluctuations), we calculate $D_s$ for a set of $\Delta_{3\text{B}}$ values of the order of $10^{-4}\gamma_1$. We also allow chemical potential $\mu$ to change around the VHS, including aligned with the singular point and the edge of it. It shows that $D_s$ does not scale with $\Delta_{3\text{B}}$, which means it is purely conventional. Even when $D_s$ depends on $\mu$, the ratio of $D_s/\Delta_{3\text{B}}$ is of the order of 100 (besides an additional factor of 54.9, see Sec.~4.2), regardless of the value of $\Delta_{3\text{B}}$ and $\mu$, indicating almost zero phase fluctuation in SC1 \cite{jiang2025superfluid}. Therefore, the Berezinskii-Kosterlitz-Thouless transition temperature is the same as the mean-field critical temperature.

\begin{thebibliography}{132}%
\makeatletter
\providecommand \@ifxundefined [1]{%
 \@ifx{#1\undefined}
}%
\providecommand \@ifnum [1]{%
 \ifnum #1\expandafter \@firstoftwo
 \else \expandafter \@secondoftwo
 \fi
}%
\providecommand \@ifx [1]{%
 \ifx #1\expandafter \@firstoftwo
 \else \expandafter \@secondoftwo
 \fi
}%
\providecommand \natexlab [1]{#1}%
\providecommand \enquote  [1]{``#1''}%
\providecommand \bibnamefont  [1]{#1}%
\providecommand \bibfnamefont [1]{#1}%
\providecommand \citenamefont [1]{#1}%
\providecommand \href@noop [0]{\@secondoftwo}%
\providecommand \href [0]{\begingroup \@sanitize@url \@href}%
\providecommand \@href[1]{\@@startlink{#1}\@@href}%
\providecommand \@@href[1]{\endgroup#1\@@endlink}%
\providecommand \@sanitize@url [0]{\catcode `\\12\catcode `\$12\catcode
  `\&12\catcode `\#12\catcode `\^12\catcode `\_12\catcode `\%12\relax}%
\providecommand \@@startlink[1]{}%
\providecommand \@@endlink[0]{}%
\providecommand \url  [0]{\begingroup\@sanitize@url \@url }%
\providecommand \@url [1]{\endgroup\@href {#1}{\urlprefix }}%
\providecommand \urlprefix  [0]{URL }%
\providecommand \Eprint [0]{\href }%
\providecommand \doibase [0]{https://doi.org/}%
\providecommand \selectlanguage [0]{\@gobble}%
\providecommand \bibinfo  [0]{\@secondoftwo}%
\providecommand \bibfield  [0]{\@secondoftwo}%
\providecommand \translation [1]{[#1]}%
\providecommand \BibitemOpen [0]{}%
\providecommand \bibitemStop [0]{}%
\providecommand \bibitemNoStop [0]{.\EOS\space}%
\providecommand \EOS [0]{\spacefactor3000\relax}%
\providecommand \BibitemShut  [1]{\csname bibitem#1\endcsname}%
\let\auto@bib@innerbib\@empty
\bibitem [{\citenamefont {Provost}\ and\ \citenamefont
  {Vallee}(1980)}]{provost1980riemannian}%
  \BibitemOpen
  \bibfield  {author} {\bibinfo {author} {\bibfnamefont {J.}~\bibnamefont
  {Provost}}\ and\ \bibinfo {author} {\bibfnamefont {G.}~\bibnamefont
  {Vallee}},\ }\href {https://link.springer.com/article/10.1007/BF02193559}
  {\bibfield  {journal} {\bibinfo  {journal} {Communications in Mathematical
  Physics}\ }\textbf {\bibinfo {volume} {76}},\ \bibinfo {pages} {289}
  (\bibinfo {year} {1980})}\BibitemShut {NoStop}%
\bibitem [{\citenamefont {Resta}(2011)}]{resta2011insulating}%
  \BibitemOpen
  \bibfield  {author} {\bibinfo {author} {\bibfnamefont {R.}~\bibnamefont
  {Resta}},\ }\href {https://doi.org/10.1140/epjb/e2010-10874-4} {\bibfield
  {journal} {\bibinfo  {journal} {The European Physical Journal B}\ }\textbf
  {\bibinfo {volume} {79}},\ \bibinfo {pages} {121} (\bibinfo {year}
  {2011})}\BibitemShut {NoStop}%
\bibitem [{\citenamefont {T\"orm\"a}(2023)}]{paivi2023essay}%
  \BibitemOpen
  \bibfield  {author} {\bibinfo {author} {\bibfnamefont {P.}~\bibnamefont
  {T\"orm\"a}},\ }\href {https://doi.org/10.1103/PhysRevLett.131.240001}
  {\bibfield  {journal} {\bibinfo  {journal} {Phys. Rev. Lett.}\ }\textbf
  {\bibinfo {volume} {131}},\ \bibinfo {pages} {240001} (\bibinfo {year}
  {2023})}\BibitemShut {NoStop}%
\bibitem [{\citenamefont {Yu}\ \emph {et~al.}(2024)\citenamefont {Yu},
  \citenamefont {Bernevig}, \citenamefont {Queiroz}, \citenamefont {Rossi},
  \citenamefont {T{\"o}rm{\"a}},\ and\ \citenamefont {Yang}}]{yu2024quantum}%
  \BibitemOpen
  \bibfield  {author} {\bibinfo {author} {\bibfnamefont {J.}~\bibnamefont
  {Yu}}, \bibinfo {author} {\bibfnamefont {B.~A.}\ \bibnamefont {Bernevig}},
  \bibinfo {author} {\bibfnamefont {R.}~\bibnamefont {Queiroz}}, \bibinfo
  {author} {\bibfnamefont {E.}~\bibnamefont {Rossi}}, \bibinfo {author}
  {\bibfnamefont {P.}~\bibnamefont {T{\"o}rm{\"a}}},\ and\ \bibinfo {author}
  {\bibfnamefont {B.-J.}\ \bibnamefont {Yang}},\ }\href@noop {} {\bibfield
  {journal} {\bibinfo  {journal} {arXiv preprint arXiv:2501.00098}\ } (\bibinfo
  {year} {2024})}\BibitemShut {NoStop}%
\bibitem [{\citenamefont {Liu}\ \emph {et~al.}(2025)\citenamefont {Liu},
  \citenamefont {Qiang}, \citenamefont {Lu},\ and\ \citenamefont
  {Xie}}]{liu2025quantum}%
  \BibitemOpen
  \bibfield  {author} {\bibinfo {author} {\bibfnamefont {T.}~\bibnamefont
  {Liu}}, \bibinfo {author} {\bibfnamefont {X.-B.}\ \bibnamefont {Qiang}},
  \bibinfo {author} {\bibfnamefont {H.-Z.}\ \bibnamefont {Lu}},\ and\ \bibinfo
  {author} {\bibfnamefont {X.}~\bibnamefont {Xie}},\ }\href@noop {} {\bibfield
  {journal} {\bibinfo  {journal} {National Science Review}\ }\textbf {\bibinfo
  {volume} {12}},\ \bibinfo {pages} {nwae334} (\bibinfo {year}
  {2025})}\BibitemShut {NoStop}%
\bibitem [{\citenamefont {Gao}\ \emph {et~al.}(2025)\citenamefont {Gao},
  \citenamefont {Nagaosa}, \citenamefont {Ni},\ and\ \citenamefont
  {Xu}}]{gao2025quantum}%
  \BibitemOpen
  \bibfield  {author} {\bibinfo {author} {\bibfnamefont {A.}~\bibnamefont
  {Gao}}, \bibinfo {author} {\bibfnamefont {N.}~\bibnamefont {Nagaosa}},
  \bibinfo {author} {\bibfnamefont {N.}~\bibnamefont {Ni}},\ and\ \bibinfo
  {author} {\bibfnamefont {S.-Y.}\ \bibnamefont {Xu}},\ }\href@noop {}
  {\bibfield  {journal} {\bibinfo  {journal} {arXiv preprint arXiv:2508.00469}\
  } (\bibinfo {year} {2025})}\BibitemShut {NoStop}%
\bibitem [{\citenamefont {Berry}(1984)}]{berry1984}%
  \BibitemOpen
  \bibfield  {author} {\bibinfo {author} {\bibfnamefont {M.~V.}\ \bibnamefont
  {Berry}},\ }\href
  {https://https://royalsocietypublishing.org/doi/10.1098/rspa.1984.0023}
  {\bibfield  {journal} {\bibinfo  {journal} {Proceedings of the Royal Society
  of London. A. Mathematical and Physical Sciences}\ }\textbf {\bibinfo
  {volume} {392}},\ \bibinfo {pages} {45} (\bibinfo {year} {1984})}\BibitemShut
  {NoStop}%
\bibitem [{\citenamefont {Thouless}\ \emph {et~al.}(1982)\citenamefont
  {Thouless}, \citenamefont {Kohmoto}, \citenamefont {Nightingale},\ and\
  \citenamefont {den Nijs}}]{tknn1982}%
  \BibitemOpen
  \bibfield  {author} {\bibinfo {author} {\bibfnamefont {D.~J.}\ \bibnamefont
  {Thouless}}, \bibinfo {author} {\bibfnamefont {M.}~\bibnamefont {Kohmoto}},
  \bibinfo {author} {\bibfnamefont {M.~P.}\ \bibnamefont {Nightingale}},\ and\
  \bibinfo {author} {\bibfnamefont {M.}~\bibnamefont {den Nijs}},\ }\href
  {https://doi.org/10.1103/PhysRevLett.49.405} {\bibfield  {journal} {\bibinfo
  {journal} {Phys. Rev. Lett.}\ }\textbf {\bibinfo {volume} {49}},\ \bibinfo
  {pages} {405} (\bibinfo {year} {1982})}\BibitemShut {NoStop}%
\bibitem [{\citenamefont {Haldane}(1988)}]{haldane1988}%
  \BibitemOpen
  \bibfield  {author} {\bibinfo {author} {\bibfnamefont {F.~D.~M.}\
  \bibnamefont {Haldane}},\ }\href
  {https://doi.org/10.1103/PhysRevLett.61.2015} {\bibfield  {journal} {\bibinfo
   {journal} {Phys. Rev. Lett.}\ }\textbf {\bibinfo {volume} {61}},\ \bibinfo
  {pages} {2015} (\bibinfo {year} {1988})}\BibitemShut {NoStop}%
\bibitem [{\citenamefont {Peotta}\ and\ \citenamefont
  {T{\"o}rm{\"a}}(2015)}]{peotta2015superfluidity}%
  \BibitemOpen
  \bibfield  {author} {\bibinfo {author} {\bibfnamefont {S.}~\bibnamefont
  {Peotta}}\ and\ \bibinfo {author} {\bibfnamefont {P.}~\bibnamefont
  {T{\"o}rm{\"a}}},\ }\href {https://doi.org/10.1038/ncomms9944} {\bibfield
  {journal} {\bibinfo  {journal} {Nature Communications}\ }\textbf {\bibinfo
  {volume} {6}},\ \bibinfo {pages} {8944} (\bibinfo {year} {2015})}\BibitemShut
  {NoStop}%
\bibitem [{\citenamefont {T{\"o}rm{\"a}}\ \emph {et~al.}(2022)\citenamefont
  {T{\"o}rm{\"a}}, \citenamefont {Peotta},\ and\ \citenamefont
  {Bernevig}}]{torma2022tbgreview}%
  \BibitemOpen
  \bibfield  {author} {\bibinfo {author} {\bibfnamefont {P.}~\bibnamefont
  {T{\"o}rm{\"a}}}, \bibinfo {author} {\bibfnamefont {S.}~\bibnamefont
  {Peotta}},\ and\ \bibinfo {author} {\bibfnamefont {B.~A.}\ \bibnamefont
  {Bernevig}},\ }\href {https://doi.org/10.1038/s42254-022-00466-y} {\bibfield
  {journal} {\bibinfo  {journal} {Nature Reviews Physics}\ }\textbf {\bibinfo
  {volume} {4}},\ \bibinfo {pages} {528} (\bibinfo {year} {2022})}\BibitemShut
  {NoStop}%
\bibitem [{\citenamefont {Roy}(2014)}]{roy2014band}%
  \BibitemOpen
  \bibfield  {author} {\bibinfo {author} {\bibfnamefont {R.}~\bibnamefont
  {Roy}},\ }\href {https://doi.org/10.1103/PhysRevB.90.165139} {\bibfield
  {journal} {\bibinfo  {journal} {Phys. Rev. B}\ }\textbf {\bibinfo {volume}
  {90}},\ \bibinfo {pages} {165139} (\bibinfo {year} {2014})}\BibitemShut
  {NoStop}%
\bibitem [{\citenamefont {Ozawa}\ and\ \citenamefont
  {Mera}(2021)}]{ozawa2021relations}%
  \BibitemOpen
  \bibfield  {author} {\bibinfo {author} {\bibfnamefont {T.}~\bibnamefont
  {Ozawa}}\ and\ \bibinfo {author} {\bibfnamefont {B.}~\bibnamefont {Mera}},\
  }\href {https://doi.org/10.1103/PhysRevB.104.045103} {\bibfield  {journal}
  {\bibinfo  {journal} {Phys. Rev. B}\ }\textbf {\bibinfo {volume} {104}},\
  \bibinfo {pages} {045103} (\bibinfo {year} {2021})}\BibitemShut {NoStop}%
\bibitem [{\citenamefont {Girvin}\ \emph {et~al.}(1986)\citenamefont {Girvin},
  \citenamefont {MacDonald},\ and\ \citenamefont
  {Platzman}}]{girvin1986magneto}%
  \BibitemOpen
  \bibfield  {author} {\bibinfo {author} {\bibfnamefont {S.~M.}\ \bibnamefont
  {Girvin}}, \bibinfo {author} {\bibfnamefont {A.~H.}\ \bibnamefont
  {MacDonald}},\ and\ \bibinfo {author} {\bibfnamefont {P.~M.}\ \bibnamefont
  {Platzman}},\ }\href {https://doi.org/10.1103/PhysRevB.33.2481} {\bibfield
  {journal} {\bibinfo  {journal} {Phys. Rev. B}\ }\textbf {\bibinfo {volume}
  {33}},\ \bibinfo {pages} {2481} (\bibinfo {year} {1986})}\BibitemShut
  {NoStop}%
\bibitem [{\citenamefont {Claassen}\ \emph {et~al.}(2015)\citenamefont
  {Claassen}, \citenamefont {Lee}, \citenamefont {Thomale}, \citenamefont
  {Qi},\ and\ \citenamefont {Devereaux}}]{claassen2015position}%
  \BibitemOpen
  \bibfield  {author} {\bibinfo {author} {\bibfnamefont {M.}~\bibnamefont
  {Claassen}}, \bibinfo {author} {\bibfnamefont {C.~H.}\ \bibnamefont {Lee}},
  \bibinfo {author} {\bibfnamefont {R.}~\bibnamefont {Thomale}}, \bibinfo
  {author} {\bibfnamefont {X.-L.}\ \bibnamefont {Qi}},\ and\ \bibinfo {author}
  {\bibfnamefont {T.~P.}\ \bibnamefont {Devereaux}},\ }\href
  {https://doi.org/10.1103/PhysRevLett.114.236802} {\bibfield  {journal}
  {\bibinfo  {journal} {Phys. Rev. Lett.}\ }\textbf {\bibinfo {volume} {114}},\
  \bibinfo {pages} {236802} (\bibinfo {year} {2015})}\BibitemShut {NoStop}%
\bibitem [{\citenamefont {Wang}\ \emph {et~al.}(2021)\citenamefont {Wang},
  \citenamefont {Cano}, \citenamefont {Millis}, \citenamefont {Liu},\ and\
  \citenamefont {Yang}}]{wang2021exact}%
  \BibitemOpen
  \bibfield  {author} {\bibinfo {author} {\bibfnamefont {J.}~\bibnamefont
  {Wang}}, \bibinfo {author} {\bibfnamefont {J.}~\bibnamefont {Cano}}, \bibinfo
  {author} {\bibfnamefont {A.~J.}\ \bibnamefont {Millis}}, \bibinfo {author}
  {\bibfnamefont {Z.}~\bibnamefont {Liu}},\ and\ \bibinfo {author}
  {\bibfnamefont {B.}~\bibnamefont {Yang}},\ }\href
  {https://doi.org/10.1103/PhysRevLett.127.246403} {\bibfield  {journal}
  {\bibinfo  {journal} {Phys. Rev. Lett.}\ }\textbf {\bibinfo {volume} {127}},\
  \bibinfo {pages} {246403} (\bibinfo {year} {2021})}\BibitemShut {NoStop}%
\bibitem [{\citenamefont {Wang}\ and\ \citenamefont
  {Liu}(2022)}]{wang2022hierachy}%
  \BibitemOpen
  \bibfield  {author} {\bibinfo {author} {\bibfnamefont {J.}~\bibnamefont
  {Wang}}\ and\ \bibinfo {author} {\bibfnamefont {Z.}~\bibnamefont {Liu}},\
  }\href {https://doi.org/10.1103/PhysRevLett.128.176403} {\bibfield  {journal}
  {\bibinfo  {journal} {Phys. Rev. Lett.}\ }\textbf {\bibinfo {volume} {128}},\
  \bibinfo {pages} {176403} (\bibinfo {year} {2022})}\BibitemShut {NoStop}%
\bibitem [{\citenamefont {Parameswaran}\ \emph {et~al.}(2012)\citenamefont
  {Parameswaran}, \citenamefont {Roy},\ and\ \citenamefont
  {Sondhi}}]{parameswaran2012fractional}%
  \BibitemOpen
  \bibfield  {author} {\bibinfo {author} {\bibfnamefont {S.~A.}\ \bibnamefont
  {Parameswaran}}, \bibinfo {author} {\bibfnamefont {R.}~\bibnamefont {Roy}},\
  and\ \bibinfo {author} {\bibfnamefont {S.~L.}\ \bibnamefont {Sondhi}},\
  }\href {https://doi.org/10.1103/PhysRevB.85.241308} {\bibfield  {journal}
  {\bibinfo  {journal} {Phys. Rev. B}\ }\textbf {\bibinfo {volume} {85}},\
  \bibinfo {pages} {241308} (\bibinfo {year} {2012})}\BibitemShut {NoStop}%
\bibitem [{\citenamefont {Jackson}\ \emph {et~al.}(2015)\citenamefont
  {Jackson}, \citenamefont {M{\"o}ller},\ and\ \citenamefont
  {Roy}}]{jackson2015geometric}%
  \BibitemOpen
  \bibfield  {author} {\bibinfo {author} {\bibfnamefont {T.~S.}\ \bibnamefont
  {Jackson}}, \bibinfo {author} {\bibfnamefont {G.}~\bibnamefont
  {M{\"o}ller}},\ and\ \bibinfo {author} {\bibfnamefont {R.}~\bibnamefont
  {Roy}},\ }\href {https://doi.org/10.1038/ncomms9629} {\bibfield  {journal}
  {\bibinfo  {journal} {Nature Communications}\ }\textbf {\bibinfo {volume}
  {6}},\ \bibinfo {pages} {8629} (\bibinfo {year} {2015})}\BibitemShut
  {NoStop}%
\bibitem [{\citenamefont {Neupert}\ \emph {et~al.}(2011)\citenamefont
  {Neupert}, \citenamefont {Santos}, \citenamefont {Chamon},\ and\
  \citenamefont {Mudry}}]{neupert2011fractional}%
  \BibitemOpen
  \bibfield  {author} {\bibinfo {author} {\bibfnamefont {T.}~\bibnamefont
  {Neupert}}, \bibinfo {author} {\bibfnamefont {L.}~\bibnamefont {Santos}},
  \bibinfo {author} {\bibfnamefont {C.}~\bibnamefont {Chamon}},\ and\ \bibinfo
  {author} {\bibfnamefont {C.}~\bibnamefont {Mudry}},\ }\href
  {https://doi.org/10.1103/PhysRevLett.106.236804} {\bibfield  {journal}
  {\bibinfo  {journal} {Phys. Rev. Lett.}\ }\textbf {\bibinfo {volume} {106}},\
  \bibinfo {pages} {236804} (\bibinfo {year} {2011})}\BibitemShut {NoStop}%
\bibitem [{\citenamefont {Regnault}\ and\ \citenamefont
  {Bernevig}(2011)}]{regnault2011fractional}%
  \BibitemOpen
  \bibfield  {author} {\bibinfo {author} {\bibfnamefont {N.}~\bibnamefont
  {Regnault}}\ and\ \bibinfo {author} {\bibfnamefont {B.~A.}\ \bibnamefont
  {Bernevig}},\ }\href {https://doi.org/10.1103/PhysRevX.1.021014} {\bibfield
  {journal} {\bibinfo  {journal} {Phys. Rev. X}\ }\textbf {\bibinfo {volume}
  {1}},\ \bibinfo {pages} {021014} (\bibinfo {year} {2011})}\BibitemShut
  {NoStop}%
\bibitem [{\citenamefont {Sheng}\ \emph {et~al.}(2011)\citenamefont {Sheng},
  \citenamefont {Gu}, \citenamefont {Sun},\ and\ \citenamefont
  {Sheng}}]{sheng2011fractional}%
  \BibitemOpen
  \bibfield  {author} {\bibinfo {author} {\bibfnamefont {D.~N.}\ \bibnamefont
  {Sheng}}, \bibinfo {author} {\bibfnamefont {Z.-C.}\ \bibnamefont {Gu}},
  \bibinfo {author} {\bibfnamefont {K.}~\bibnamefont {Sun}},\ and\ \bibinfo
  {author} {\bibfnamefont {L.}~\bibnamefont {Sheng}},\ }\href
  {https://doi.org/10.1038/ncomms1380} {\bibfield  {journal} {\bibinfo
  {journal} {Nature Communications}\ }\textbf {\bibinfo {volume} {2}},\
  \bibinfo {pages} {389} (\bibinfo {year} {2011})}\BibitemShut {NoStop}%
\bibitem [{\citenamefont {Cai}\ \emph {et~al.}(2023)\citenamefont {Cai},
  \citenamefont {Anderson}, \citenamefont {Wang}, \citenamefont {Zhang},
  \citenamefont {Liu}, \citenamefont {Holtzmann}, \citenamefont {Zhang},
  \citenamefont {Fan}, \citenamefont {Taniguchi}, \citenamefont {Watanabe},
  \citenamefont {Ran}, \citenamefont {Cao}, \citenamefont {Fu}, \citenamefont
  {Xiao}, \citenamefont {Yao},\ and\ \citenamefont {Xu}}]{cai2023signatures}%
  \BibitemOpen
  \bibfield  {author} {\bibinfo {author} {\bibfnamefont {J.}~\bibnamefont
  {Cai}}, \bibinfo {author} {\bibfnamefont {E.}~\bibnamefont {Anderson}},
  \bibinfo {author} {\bibfnamefont {C.}~\bibnamefont {Wang}}, \bibinfo {author}
  {\bibfnamefont {X.}~\bibnamefont {Zhang}}, \bibinfo {author} {\bibfnamefont
  {X.}~\bibnamefont {Liu}}, \bibinfo {author} {\bibfnamefont {W.}~\bibnamefont
  {Holtzmann}}, \bibinfo {author} {\bibfnamefont {Y.}~\bibnamefont {Zhang}},
  \bibinfo {author} {\bibfnamefont {F.}~\bibnamefont {Fan}}, \bibinfo {author}
  {\bibfnamefont {T.}~\bibnamefont {Taniguchi}}, \bibinfo {author}
  {\bibfnamefont {K.}~\bibnamefont {Watanabe}}, \bibinfo {author}
  {\bibfnamefont {Y.}~\bibnamefont {Ran}}, \bibinfo {author} {\bibfnamefont
  {T.}~\bibnamefont {Cao}}, \bibinfo {author} {\bibfnamefont {L.}~\bibnamefont
  {Fu}}, \bibinfo {author} {\bibfnamefont {D.}~\bibnamefont {Xiao}}, \bibinfo
  {author} {\bibfnamefont {W.}~\bibnamefont {Yao}},\ and\ \bibinfo {author}
  {\bibfnamefont {X.}~\bibnamefont {Xu}},\ }\href
  {https://doi.org/10.1038/s41586-023-06289-w} {\bibfield  {journal} {\bibinfo
  {journal} {Nature}\ }\textbf {\bibinfo {volume} {622}},\ \bibinfo {pages}
  {63} (\bibinfo {year} {2023})}\BibitemShut {NoStop}%
\bibitem [{\citenamefont {Zeng}\ \emph {et~al.}(2023)\citenamefont {Zeng},
  \citenamefont {Xia}, \citenamefont {Kang}, \citenamefont {Zhu}, \citenamefont
  {Kn{\"u}ppel}, \citenamefont {Vaswani}, \citenamefont {Watanabe},
  \citenamefont {Taniguchi}, \citenamefont {Mak},\ and\ \citenamefont
  {Shan}}]{zeng2023thermodynamic}%
  \BibitemOpen
  \bibfield  {author} {\bibinfo {author} {\bibfnamefont {Y.}~\bibnamefont
  {Zeng}}, \bibinfo {author} {\bibfnamefont {Z.}~\bibnamefont {Xia}}, \bibinfo
  {author} {\bibfnamefont {K.}~\bibnamefont {Kang}}, \bibinfo {author}
  {\bibfnamefont {J.}~\bibnamefont {Zhu}}, \bibinfo {author} {\bibfnamefont
  {P.}~\bibnamefont {Kn{\"u}ppel}}, \bibinfo {author} {\bibfnamefont
  {C.}~\bibnamefont {Vaswani}}, \bibinfo {author} {\bibfnamefont
  {K.}~\bibnamefont {Watanabe}}, \bibinfo {author} {\bibfnamefont
  {T.}~\bibnamefont {Taniguchi}}, \bibinfo {author} {\bibfnamefont {K.~F.}\
  \bibnamefont {Mak}},\ and\ \bibinfo {author} {\bibfnamefont {J.}~\bibnamefont
  {Shan}},\ }\href {https://doi.org/10.1038/s41586-023-06452-3} {\bibfield
  {journal} {\bibinfo  {journal} {Nature}\ }\textbf {\bibinfo {volume} {622}},\
  \bibinfo {pages} {69} (\bibinfo {year} {2023})}\BibitemShut {NoStop}%
\bibitem [{\citenamefont {Park}\ \emph {et~al.}(2023)\citenamefont {Park},
  \citenamefont {Cai}, \citenamefont {Anderson}, \citenamefont {Zhang},
  \citenamefont {Zhu}, \citenamefont {Liu}, \citenamefont {Wang}, \citenamefont
  {Holtzmann}, \citenamefont {Hu}, \citenamefont {Liu}, \citenamefont
  {Taniguchi}, \citenamefont {Watanabe}, \citenamefont {Chu}, \citenamefont
  {Cao}, \citenamefont {Fu}, \citenamefont {Yao}, \citenamefont {Chang},
  \citenamefont {Cobden}, \citenamefont {Xiao},\ and\ \citenamefont
  {Xu}}]{park2023observation}%
  \BibitemOpen
  \bibfield  {author} {\bibinfo {author} {\bibfnamefont {H.}~\bibnamefont
  {Park}}, \bibinfo {author} {\bibfnamefont {J.}~\bibnamefont {Cai}}, \bibinfo
  {author} {\bibfnamefont {E.}~\bibnamefont {Anderson}}, \bibinfo {author}
  {\bibfnamefont {Y.}~\bibnamefont {Zhang}}, \bibinfo {author} {\bibfnamefont
  {J.}~\bibnamefont {Zhu}}, \bibinfo {author} {\bibfnamefont {X.}~\bibnamefont
  {Liu}}, \bibinfo {author} {\bibfnamefont {C.}~\bibnamefont {Wang}}, \bibinfo
  {author} {\bibfnamefont {W.}~\bibnamefont {Holtzmann}}, \bibinfo {author}
  {\bibfnamefont {C.}~\bibnamefont {Hu}}, \bibinfo {author} {\bibfnamefont
  {Z.}~\bibnamefont {Liu}}, \bibinfo {author} {\bibfnamefont {T.}~\bibnamefont
  {Taniguchi}}, \bibinfo {author} {\bibfnamefont {K.}~\bibnamefont {Watanabe}},
  \bibinfo {author} {\bibfnamefont {J.-H.}\ \bibnamefont {Chu}}, \bibinfo
  {author} {\bibfnamefont {T.}~\bibnamefont {Cao}}, \bibinfo {author}
  {\bibfnamefont {L.}~\bibnamefont {Fu}}, \bibinfo {author} {\bibfnamefont
  {W.}~\bibnamefont {Yao}}, \bibinfo {author} {\bibfnamefont {C.-Z.}\
  \bibnamefont {Chang}}, \bibinfo {author} {\bibfnamefont {D.}~\bibnamefont
  {Cobden}}, \bibinfo {author} {\bibfnamefont {D.}~\bibnamefont {Xiao}},\ and\
  \bibinfo {author} {\bibfnamefont {X.}~\bibnamefont {Xu}},\ }\href
  {https://doi.org/10.1038/s41586-023-06536-0} {\bibfield  {journal} {\bibinfo
  {journal} {Nature}\ }\textbf {\bibinfo {volume} {622}},\ \bibinfo {pages}
  {74} (\bibinfo {year} {2023})}\BibitemShut {NoStop}%
\bibitem [{\citenamefont {Xu}\ \emph {et~al.}(2023)\citenamefont {Xu},
  \citenamefont {Sun}, \citenamefont {Jia}, \citenamefont {Liu}, \citenamefont
  {Xu}, \citenamefont {Li}, \citenamefont {Gu}, \citenamefont {Watanabe},
  \citenamefont {Taniguchi}, \citenamefont {Tong}, \citenamefont {Jia},
  \citenamefont {Shi}, \citenamefont {Jiang}, \citenamefont {Zhang},
  \citenamefont {Liu},\ and\ \citenamefont {Li}}]{xu2023observation}%
  \BibitemOpen
  \bibfield  {author} {\bibinfo {author} {\bibfnamefont {F.}~\bibnamefont
  {Xu}}, \bibinfo {author} {\bibfnamefont {Z.}~\bibnamefont {Sun}}, \bibinfo
  {author} {\bibfnamefont {T.}~\bibnamefont {Jia}}, \bibinfo {author}
  {\bibfnamefont {C.}~\bibnamefont {Liu}}, \bibinfo {author} {\bibfnamefont
  {C.}~\bibnamefont {Xu}}, \bibinfo {author} {\bibfnamefont {C.}~\bibnamefont
  {Li}}, \bibinfo {author} {\bibfnamefont {Y.}~\bibnamefont {Gu}}, \bibinfo
  {author} {\bibfnamefont {K.}~\bibnamefont {Watanabe}}, \bibinfo {author}
  {\bibfnamefont {T.}~\bibnamefont {Taniguchi}}, \bibinfo {author}
  {\bibfnamefont {B.}~\bibnamefont {Tong}}, \bibinfo {author} {\bibfnamefont
  {J.}~\bibnamefont {Jia}}, \bibinfo {author} {\bibfnamefont {Z.}~\bibnamefont
  {Shi}}, \bibinfo {author} {\bibfnamefont {S.}~\bibnamefont {Jiang}}, \bibinfo
  {author} {\bibfnamefont {Y.}~\bibnamefont {Zhang}}, \bibinfo {author}
  {\bibfnamefont {X.}~\bibnamefont {Liu}},\ and\ \bibinfo {author}
  {\bibfnamefont {T.}~\bibnamefont {Li}},\ }\href
  {https://doi.org/10.1103/PhysRevX.13.031037} {\bibfield  {journal} {\bibinfo
  {journal} {Phys. Rev. X}\ }\textbf {\bibinfo {volume} {13}},\ \bibinfo
  {pages} {031037} (\bibinfo {year} {2023})}\BibitemShut {NoStop}%
\bibitem [{\citenamefont {Lu}\ \emph {et~al.}(2024)\citenamefont {Lu},
  \citenamefont {Han}, \citenamefont {Yao}, \citenamefont {Reddy},
  \citenamefont {Yang}, \citenamefont {Seo}, \citenamefont {Watanabe},
  \citenamefont {Taniguchi}, \citenamefont {Fu},\ and\ \citenamefont
  {Ju}}]{lu2024fractional}%
  \BibitemOpen
  \bibfield  {author} {\bibinfo {author} {\bibfnamefont {Z.}~\bibnamefont
  {Lu}}, \bibinfo {author} {\bibfnamefont {T.}~\bibnamefont {Han}}, \bibinfo
  {author} {\bibfnamefont {Y.}~\bibnamefont {Yao}}, \bibinfo {author}
  {\bibfnamefont {A.~P.}\ \bibnamefont {Reddy}}, \bibinfo {author}
  {\bibfnamefont {J.}~\bibnamefont {Yang}}, \bibinfo {author} {\bibfnamefont
  {J.}~\bibnamefont {Seo}}, \bibinfo {author} {\bibfnamefont {K.}~\bibnamefont
  {Watanabe}}, \bibinfo {author} {\bibfnamefont {T.}~\bibnamefont {Taniguchi}},
  \bibinfo {author} {\bibfnamefont {L.}~\bibnamefont {Fu}},\ and\ \bibinfo
  {author} {\bibfnamefont {L.}~\bibnamefont {Ju}},\ }\href
  {https://doi.org/10.1038/s41586-023-07010-7} {\bibfield  {journal} {\bibinfo
  {journal} {Nature}\ }\textbf {\bibinfo {volume} {626}},\ \bibinfo {pages}
  {759} (\bibinfo {year} {2024})}\BibitemShut {NoStop}%
\bibitem [{\citenamefont {Xie}\ \emph {et~al.}(2025)\citenamefont {Xie},
  \citenamefont {Huo}, \citenamefont {Lu}, \citenamefont {Feng}, \citenamefont
  {Zhang}, \citenamefont {Wang}, \citenamefont {Yang}, \citenamefont
  {Watanabe}, \citenamefont {Taniguchi}, \citenamefont {Liu}, \citenamefont
  {Song}, \citenamefont {Xie}, \citenamefont {Liu},\ and\ \citenamefont
  {Lu}}]{xie2025tunable}%
  \BibitemOpen
  \bibfield  {author} {\bibinfo {author} {\bibfnamefont {J.}~\bibnamefont
  {Xie}}, \bibinfo {author} {\bibfnamefont {Z.}~\bibnamefont {Huo}}, \bibinfo
  {author} {\bibfnamefont {X.}~\bibnamefont {Lu}}, \bibinfo {author}
  {\bibfnamefont {Z.}~\bibnamefont {Feng}}, \bibinfo {author} {\bibfnamefont
  {Z.}~\bibnamefont {Zhang}}, \bibinfo {author} {\bibfnamefont
  {W.}~\bibnamefont {Wang}}, \bibinfo {author} {\bibfnamefont {Q.}~\bibnamefont
  {Yang}}, \bibinfo {author} {\bibfnamefont {K.}~\bibnamefont {Watanabe}},
  \bibinfo {author} {\bibfnamefont {T.}~\bibnamefont {Taniguchi}}, \bibinfo
  {author} {\bibfnamefont {K.}~\bibnamefont {Liu}}, \bibinfo {author}
  {\bibfnamefont {Z.}~\bibnamefont {Song}}, \bibinfo {author} {\bibfnamefont
  {X.~C.}\ \bibnamefont {Xie}}, \bibinfo {author} {\bibfnamefont
  {J.}~\bibnamefont {Liu}},\ and\ \bibinfo {author} {\bibfnamefont
  {X.}~\bibnamefont {Lu}},\ }\href {https://doi.org/10.1038/s41563-025-02225-7}
  {\bibfield  {journal} {\bibinfo  {journal} {Nature Materials}\ }\textbf
  {\bibinfo {volume} {24}},\ \bibinfo {pages} {1042} (\bibinfo {year}
  {2025})}\BibitemShut {NoStop}%
\bibitem [{\citenamefont {Lu}\ \emph {et~al.}(2025)\citenamefont {Lu},
  \citenamefont {Han}, \citenamefont {Yao}, \citenamefont {Hadjri},
  \citenamefont {Yang}, \citenamefont {Seo}, \citenamefont {Shi}, \citenamefont
  {Ye}, \citenamefont {Watanabe}, \citenamefont {Taniguchi},\ and\
  \citenamefont {Ju}}]{lu2025extended}%
  \BibitemOpen
  \bibfield  {author} {\bibinfo {author} {\bibfnamefont {Z.}~\bibnamefont
  {Lu}}, \bibinfo {author} {\bibfnamefont {T.}~\bibnamefont {Han}}, \bibinfo
  {author} {\bibfnamefont {Y.}~\bibnamefont {Yao}}, \bibinfo {author}
  {\bibfnamefont {Z.}~\bibnamefont {Hadjri}}, \bibinfo {author} {\bibfnamefont
  {J.}~\bibnamefont {Yang}}, \bibinfo {author} {\bibfnamefont {J.}~\bibnamefont
  {Seo}}, \bibinfo {author} {\bibfnamefont {L.}~\bibnamefont {Shi}}, \bibinfo
  {author} {\bibfnamefont {S.}~\bibnamefont {Ye}}, \bibinfo {author}
  {\bibfnamefont {K.}~\bibnamefont {Watanabe}}, \bibinfo {author}
  {\bibfnamefont {T.}~\bibnamefont {Taniguchi}},\ and\ \bibinfo {author}
  {\bibfnamefont {L.}~\bibnamefont {Ju}},\ }\href
  {https://doi.org/10.1038/s41586-024-08470-1} {\bibfield  {journal} {\bibinfo
  {journal} {Nature}\ }\textbf {\bibinfo {volume} {637}},\ \bibinfo {pages}
  {1090} (\bibinfo {year} {2025})}\BibitemShut {NoStop}%
\bibitem [{\citenamefont {Waters}\ \emph {et~al.}(2025)\citenamefont {Waters},
  \citenamefont {Okounkova}, \citenamefont {Su}, \citenamefont {Zhou},
  \citenamefont {Yao}, \citenamefont {Watanabe}, \citenamefont {Taniguchi},
  \citenamefont {Xu}, \citenamefont {Zhang}, \citenamefont {Folk},\ and\
  \citenamefont {Yankowitz}}]{waters2025chern}%
  \BibitemOpen
  \bibfield  {author} {\bibinfo {author} {\bibfnamefont {D.}~\bibnamefont
  {Waters}}, \bibinfo {author} {\bibfnamefont {A.}~\bibnamefont {Okounkova}},
  \bibinfo {author} {\bibfnamefont {R.}~\bibnamefont {Su}}, \bibinfo {author}
  {\bibfnamefont {B.}~\bibnamefont {Zhou}}, \bibinfo {author} {\bibfnamefont
  {J.}~\bibnamefont {Yao}}, \bibinfo {author} {\bibfnamefont {K.}~\bibnamefont
  {Watanabe}}, \bibinfo {author} {\bibfnamefont {T.}~\bibnamefont {Taniguchi}},
  \bibinfo {author} {\bibfnamefont {X.}~\bibnamefont {Xu}}, \bibinfo {author}
  {\bibfnamefont {Y.-H.}\ \bibnamefont {Zhang}}, \bibinfo {author}
  {\bibfnamefont {J.}~\bibnamefont {Folk}},\ and\ \bibinfo {author}
  {\bibfnamefont {M.}~\bibnamefont {Yankowitz}},\ }\href
  {https://doi.org/10.1103/PhysRevX.15.011045} {\bibfield  {journal} {\bibinfo
  {journal} {Phys. Rev. X}\ }\textbf {\bibinfo {volume} {15}},\ \bibinfo
  {pages} {011045} (\bibinfo {year} {2025})}\BibitemShut {NoStop}%
\bibitem [{\citenamefont {Parameswaran}\ \emph {et~al.}(2013)\citenamefont
  {Parameswaran}, \citenamefont {Roy},\ and\ \citenamefont
  {Sondhi}}]{parameswaran2013fractional}%
  \BibitemOpen
  \bibfield  {author} {\bibinfo {author} {\bibfnamefont {S.~A.}\ \bibnamefont
  {Parameswaran}}, \bibinfo {author} {\bibfnamefont {R.}~\bibnamefont {Roy}},\
  and\ \bibinfo {author} {\bibfnamefont {S.~L.}\ \bibnamefont {Sondhi}},\
  }\href {https://doi.org/https://doi.org/10.1016/j.crhy.2013.04.003}
  {\bibfield  {journal} {\bibinfo  {journal} {Comptes Rendus Physique}\
  }\textbf {\bibinfo {volume} {14}},\ \bibinfo {pages} {816} (\bibinfo {year}
  {2013})},\ \bibinfo {note} {topological insulators / Isolants
  topologiques}\BibitemShut {NoStop}%
\bibitem [{\citenamefont {Liu}\ and\ \citenamefont
  {Bergholtz}(2022)}]{liu2022recent}%
  \BibitemOpen
  \bibfield  {author} {\bibinfo {author} {\bibfnamefont {Z.}~\bibnamefont
  {Liu}}\ and\ \bibinfo {author} {\bibfnamefont {E.~J.}\ \bibnamefont
  {Bergholtz}},\ }\href@noop {} {\bibfield  {journal} {\bibinfo  {journal}
  {arXiv preprint arXiv:2208.08449}\ } (\bibinfo {year} {2022})}\BibitemShut
  {NoStop}%
\bibitem [{\citenamefont {Ledwith}\ \emph {et~al.}(2022)\citenamefont
  {Ledwith}, \citenamefont {Vishwanath},\ and\ \citenamefont
  {Khalaf}}]{ledwith2022family}%
  \BibitemOpen
  \bibfield  {author} {\bibinfo {author} {\bibfnamefont {P.~J.}\ \bibnamefont
  {Ledwith}}, \bibinfo {author} {\bibfnamefont {A.}~\bibnamefont
  {Vishwanath}},\ and\ \bibinfo {author} {\bibfnamefont {E.}~\bibnamefont
  {Khalaf}},\ }\href {https://doi.org/10.1103/PhysRevLett.128.176404}
  {\bibfield  {journal} {\bibinfo  {journal} {Phys. Rev. Lett.}\ }\textbf
  {\bibinfo {volume} {128}},\ \bibinfo {pages} {176404} (\bibinfo {year}
  {2022})}\BibitemShut {NoStop}%
\bibitem [{\citenamefont {Ledwith}\ \emph {et~al.}(2023)\citenamefont
  {Ledwith}, \citenamefont {Vishwanath},\ and\ \citenamefont
  {Parker}}]{ledwith2023vortexability}%
  \BibitemOpen
  \bibfield  {author} {\bibinfo {author} {\bibfnamefont {P.~J.}\ \bibnamefont
  {Ledwith}}, \bibinfo {author} {\bibfnamefont {A.}~\bibnamefont
  {Vishwanath}},\ and\ \bibinfo {author} {\bibfnamefont {D.~E.}\ \bibnamefont
  {Parker}},\ }\href {https://doi.org/10.1103/PhysRevB.108.205144} {\bibfield
  {journal} {\bibinfo  {journal} {Phys. Rev. B}\ }\textbf {\bibinfo {volume}
  {108}},\ \bibinfo {pages} {205144} (\bibinfo {year} {2023})}\BibitemShut
  {NoStop}%
\bibitem [{\citenamefont {Bernevig}\ and\ \citenamefont
  {Kwan}(2025)}]{bernevig2025berry}%
  \BibitemOpen
  \bibfield  {author} {\bibinfo {author} {\bibfnamefont {B.~A.}\ \bibnamefont
  {Bernevig}}\ and\ \bibinfo {author} {\bibfnamefont {Y.~H.}\ \bibnamefont
  {Kwan}},\ }\href@noop {} {\bibfield  {journal} {\bibinfo  {journal} {arXiv
  preprint arXiv:2503.09692}\ } (\bibinfo {year} {2025})}\BibitemShut {NoStop}%
\bibitem [{\citenamefont {Heikkil{\"a}}\ and\ \citenamefont
  {Volovik}(2011)}]{heikkila2011dimensional}%
  \BibitemOpen
  \bibfield  {author} {\bibinfo {author} {\bibfnamefont {T.~T.}\ \bibnamefont
  {Heikkil{\"a}}}\ and\ \bibinfo {author} {\bibfnamefont {G.~E.}\ \bibnamefont
  {Volovik}},\ }\href
  {https://link.springer.com/article/10.1134/S002136401102007X} {\bibfield
  {journal} {\bibinfo  {journal} {JETP letters}\ }\textbf {\bibinfo {volume}
  {93}},\ \bibinfo {pages} {59} (\bibinfo {year} {2011})}\BibitemShut {NoStop}%
\bibitem [{\citenamefont {Heikkil{\"a}}\ \emph {et~al.}(2011)\citenamefont
  {Heikkil{\"a}}, \citenamefont {Kopnin},\ and\ \citenamefont
  {Volovik}}]{heikkila2011flat}%
  \BibitemOpen
  \bibfield  {author} {\bibinfo {author} {\bibfnamefont {T.~T.}\ \bibnamefont
  {Heikkil{\"a}}}, \bibinfo {author} {\bibfnamefont {N.~B.}\ \bibnamefont
  {Kopnin}},\ and\ \bibinfo {author} {\bibfnamefont {G.~E.}\ \bibnamefont
  {Volovik}},\ }\href
  {https://link.springer.com/article/10.1134/S0021364011150045} {\bibfield
  {journal} {\bibinfo  {journal} {JETP letters}\ }\textbf {\bibinfo {volume}
  {94}},\ \bibinfo {pages} {233} (\bibinfo {year} {2011})}\BibitemShut
  {NoStop}%
\bibitem [{\citenamefont {Burkov}\ \emph {et~al.}(2011)\citenamefont {Burkov},
  \citenamefont {Hook},\ and\ \citenamefont {Balents}}]{burkov2011topological}%
  \BibitemOpen
  \bibfield  {author} {\bibinfo {author} {\bibfnamefont {A.~A.}\ \bibnamefont
  {Burkov}}, \bibinfo {author} {\bibfnamefont {M.~D.}\ \bibnamefont {Hook}},\
  and\ \bibinfo {author} {\bibfnamefont {L.}~\bibnamefont {Balents}},\ }\href
  {https://doi.org/10.1103/PhysRevB.84.235126} {\bibfield  {journal} {\bibinfo
  {journal} {Phys. Rev. B}\ }\textbf {\bibinfo {volume} {84}},\ \bibinfo
  {pages} {235126} (\bibinfo {year} {2011})}\BibitemShut {NoStop}%
\bibitem [{\citenamefont {Volovik}(2013)}]{volovik2013topology}%
  \BibitemOpen
  \bibfield  {author} {\bibinfo {author} {\bibfnamefont {G.~E.}\ \bibnamefont
  {Volovik}},\ }in\ \href
  {https://link.springer.com/chapter/10.1007/978-3-319-00266-8_14} {\emph
  {\bibinfo {booktitle} {Analogue Gravity Phenomenology: Analogue Spacetimes
  and Horizons, from Theory to Experiment}}}\ (\bibinfo  {publisher}
  {Springer},\ \bibinfo {year} {2013})\ pp.\ \bibinfo {pages}
  {343--383}\BibitemShut {NoStop}%
\bibitem [{\citenamefont {Weng}\ \emph {et~al.}(2015)\citenamefont {Weng},
  \citenamefont {Liang}, \citenamefont {Xu}, \citenamefont {Yu}, \citenamefont
  {Fang}, \citenamefont {Dai},\ and\ \citenamefont
  {Kawazoe}}]{weng2015topological}%
  \BibitemOpen
  \bibfield  {author} {\bibinfo {author} {\bibfnamefont {H.}~\bibnamefont
  {Weng}}, \bibinfo {author} {\bibfnamefont {Y.}~\bibnamefont {Liang}},
  \bibinfo {author} {\bibfnamefont {Q.}~\bibnamefont {Xu}}, \bibinfo {author}
  {\bibfnamefont {R.}~\bibnamefont {Yu}}, \bibinfo {author} {\bibfnamefont
  {Z.}~\bibnamefont {Fang}}, \bibinfo {author} {\bibfnamefont {X.}~\bibnamefont
  {Dai}},\ and\ \bibinfo {author} {\bibfnamefont {Y.}~\bibnamefont {Kawazoe}},\
  }\href {https://doi.org/10.1103/PhysRevB.92.045108} {\bibfield  {journal}
  {\bibinfo  {journal} {Phys. Rev. B}\ }\textbf {\bibinfo {volume} {92}},\
  \bibinfo {pages} {045108} (\bibinfo {year} {2015})}\BibitemShut {NoStop}%
\bibitem [{\citenamefont {Yu}\ \emph {et~al.}(2015)\citenamefont {Yu},
  \citenamefont {Weng}, \citenamefont {Fang}, \citenamefont {Dai},\ and\
  \citenamefont {Hu}}]{yu2015topological}%
  \BibitemOpen
  \bibfield  {author} {\bibinfo {author} {\bibfnamefont {R.}~\bibnamefont
  {Yu}}, \bibinfo {author} {\bibfnamefont {H.}~\bibnamefont {Weng}}, \bibinfo
  {author} {\bibfnamefont {Z.}~\bibnamefont {Fang}}, \bibinfo {author}
  {\bibfnamefont {X.}~\bibnamefont {Dai}},\ and\ \bibinfo {author}
  {\bibfnamefont {X.}~\bibnamefont {Hu}},\ }\href
  {https://doi.org/10.1103/PhysRevLett.115.036807} {\bibfield  {journal}
  {\bibinfo  {journal} {Phys. Rev. Lett.}\ }\textbf {\bibinfo {volume} {115}},\
  \bibinfo {pages} {036807} (\bibinfo {year} {2015})}\BibitemShut {NoStop}%
\bibitem [{\citenamefont {Kim}\ \emph {et~al.}(2015)\citenamefont {Kim},
  \citenamefont {Wieder}, \citenamefont {Kane},\ and\ \citenamefont
  {Rappe}}]{kim2015dirac}%
  \BibitemOpen
  \bibfield  {author} {\bibinfo {author} {\bibfnamefont {Y.}~\bibnamefont
  {Kim}}, \bibinfo {author} {\bibfnamefont {B.~J.}\ \bibnamefont {Wieder}},
  \bibinfo {author} {\bibfnamefont {C.~L.}\ \bibnamefont {Kane}},\ and\
  \bibinfo {author} {\bibfnamefont {A.~M.}\ \bibnamefont {Rappe}},\ }\href
  {https://doi.org/10.1103/PhysRevLett.115.036806} {\bibfield  {journal}
  {\bibinfo  {journal} {Phys. Rev. Lett.}\ }\textbf {\bibinfo {volume} {115}},\
  \bibinfo {pages} {036806} (\bibinfo {year} {2015})}\BibitemShut {NoStop}%
\bibitem [{\citenamefont {Mullen}\ \emph {et~al.}(2015)\citenamefont {Mullen},
  \citenamefont {Uchoa},\ and\ \citenamefont {Glatzhofer}}]{mullen2015line}%
  \BibitemOpen
  \bibfield  {author} {\bibinfo {author} {\bibfnamefont {K.}~\bibnamefont
  {Mullen}}, \bibinfo {author} {\bibfnamefont {B.}~\bibnamefont {Uchoa}},\ and\
  \bibinfo {author} {\bibfnamefont {D.~T.}\ \bibnamefont {Glatzhofer}},\ }\href
  {https://doi.org/10.1103/PhysRevLett.115.026403} {\bibfield  {journal}
  {\bibinfo  {journal} {Phys. Rev. Lett.}\ }\textbf {\bibinfo {volume} {115}},\
  \bibinfo {pages} {026403} (\bibinfo {year} {2015})}\BibitemShut {NoStop}%
\bibitem [{\citenamefont {Xie}\ \emph {et~al.}(2015)\citenamefont {Xie},
  \citenamefont {Schoop}, \citenamefont {Seibel}, \citenamefont {Gibson},
  \citenamefont {Xie},\ and\ \citenamefont {Cava}}]{xie2015new}%
  \BibitemOpen
  \bibfield  {author} {\bibinfo {author} {\bibfnamefont {L.~S.}\ \bibnamefont
  {Xie}}, \bibinfo {author} {\bibfnamefont {L.~M.}\ \bibnamefont {Schoop}},
  \bibinfo {author} {\bibfnamefont {E.~M.}\ \bibnamefont {Seibel}}, \bibinfo
  {author} {\bibfnamefont {Q.~D.}\ \bibnamefont {Gibson}}, \bibinfo {author}
  {\bibfnamefont {W.}~\bibnamefont {Xie}},\ and\ \bibinfo {author}
  {\bibfnamefont {R.~J.}\ \bibnamefont {Cava}},\ }\href
  {https://pubs.aip.org/aip/apm/article/3/8/083602/120641/A-new-form-of-Ca3P2-with-a-ring-of-Dirac-nodes}
  {\bibfield  {journal} {\bibinfo  {journal} {Apl Materials}\ }\textbf
  {\bibinfo {volume} {3}} (\bibinfo {year} {2015})}\BibitemShut {NoStop}%
\bibitem [{\citenamefont {Hyart}\ \emph {et~al.}(2018)\citenamefont {Hyart},
  \citenamefont {Ojaj{\"a}rvi},\ and\ \citenamefont
  {Heikkil{\"a}}}]{hyart2018two}%
  \BibitemOpen
  \bibfield  {author} {\bibinfo {author} {\bibfnamefont {T.}~\bibnamefont
  {Hyart}}, \bibinfo {author} {\bibfnamefont {R.}~\bibnamefont
  {Ojaj{\"a}rvi}},\ and\ \bibinfo {author} {\bibfnamefont {T.}~\bibnamefont
  {Heikkil{\"a}}},\ }\href
  {https://link.springer.com/article/10.1007/s10909-017-1846-3} {\bibfield
  {journal} {\bibinfo  {journal} {J Low Temp Phys}\ }\textbf {\bibinfo {volume}
  {191}},\ \bibinfo {pages} {35} (\bibinfo {year} {2018})}\BibitemShut
  {NoStop}%
\bibitem [{\citenamefont {Kopnin}\ \emph {et~al.}(2011)\citenamefont {Kopnin},
  \citenamefont {Heikkil\"a},\ and\ \citenamefont {Volovik}}]{kopnin2011high}%
  \BibitemOpen
  \bibfield  {author} {\bibinfo {author} {\bibfnamefont {N.~B.}\ \bibnamefont
  {Kopnin}}, \bibinfo {author} {\bibfnamefont {T.~T.}\ \bibnamefont
  {Heikkil\"a}},\ and\ \bibinfo {author} {\bibfnamefont {G.~E.}\ \bibnamefont
  {Volovik}},\ }\href {https://doi.org/10.1103/PhysRevB.83.220503} {\bibfield
  {journal} {\bibinfo  {journal} {Phys. Rev. B}\ }\textbf {\bibinfo {volume}
  {83}},\ \bibinfo {pages} {220503} (\bibinfo {year} {2011})}\BibitemShut
  {NoStop}%
\bibitem [{\citenamefont {Kopnin}(2011)}]{kopnin2011surface}%
  \BibitemOpen
  \bibfield  {author} {\bibinfo {author} {\bibfnamefont {N.~B.}\ \bibnamefont
  {Kopnin}},\ }\href {https://doi.org/10.1134/S002136401113011X} {\bibfield
  {journal} {\bibinfo  {journal} {JETP Letters}\ }\textbf {\bibinfo {volume}
  {94}},\ \bibinfo {pages} {81} (\bibinfo {year} {2011})}\BibitemShut {NoStop}%
\bibitem [{\citenamefont {Kopnin}\ \emph {et~al.}(2013)\citenamefont {Kopnin},
  \citenamefont {Ij\"as}, \citenamefont {Harju},\ and\ \citenamefont
  {Heikkil\"a}}]{kopnin2013high}%
  \BibitemOpen
  \bibfield  {author} {\bibinfo {author} {\bibfnamefont {N.~B.}\ \bibnamefont
  {Kopnin}}, \bibinfo {author} {\bibfnamefont {M.}~\bibnamefont {Ij\"as}},
  \bibinfo {author} {\bibfnamefont {A.}~\bibnamefont {Harju}},\ and\ \bibinfo
  {author} {\bibfnamefont {T.~T.}\ \bibnamefont {Heikkil\"a}},\ }\href
  {https://doi.org/10.1103/PhysRevB.87.140503} {\bibfield  {journal} {\bibinfo
  {journal} {Phys. Rev. B}\ }\textbf {\bibinfo {volume} {87}},\ \bibinfo
  {pages} {140503} (\bibinfo {year} {2013})}\BibitemShut {NoStop}%
\bibitem [{\citenamefont {Mu\~noz}\ \emph {et~al.}(2013)\citenamefont
  {Mu\~noz}, \citenamefont {Covaci},\ and\ \citenamefont
  {Peeters}}]{munoz2013tight}%
  \BibitemOpen
  \bibfield  {author} {\bibinfo {author} {\bibfnamefont {W.~A.}\ \bibnamefont
  {Mu\~noz}}, \bibinfo {author} {\bibfnamefont {L.}~\bibnamefont {Covaci}},\
  and\ \bibinfo {author} {\bibfnamefont {F.~M.}\ \bibnamefont {Peeters}},\
  }\href {https://doi.org/10.1103/PhysRevB.87.134509} {\bibfield  {journal}
  {\bibinfo  {journal} {Phys. Rev. B}\ }\textbf {\bibinfo {volume} {87}},\
  \bibinfo {pages} {134509} (\bibinfo {year} {2013})}\BibitemShut {NoStop}%
\bibitem [{\citenamefont {L\"othman}\ and\ \citenamefont
  {Black-Schaffer}(2017)}]{lothman2017universal}%
  \BibitemOpen
  \bibfield  {author} {\bibinfo {author} {\bibfnamefont {T.}~\bibnamefont
  {L\"othman}}\ and\ \bibinfo {author} {\bibfnamefont {A.~M.}\ \bibnamefont
  {Black-Schaffer}},\ }\href {https://doi.org/10.1103/PhysRevB.96.064505}
  {\bibfield  {journal} {\bibinfo  {journal} {Phys. Rev. B}\ }\textbf {\bibinfo
  {volume} {96}},\ \bibinfo {pages} {064505} (\bibinfo {year}
  {2017})}\BibitemShut {NoStop}%
\bibitem [{\citenamefont {Zhou}\ \emph
  {et~al.}(2021{\natexlab{a}})\citenamefont {Zhou}, \citenamefont {Xie},
  \citenamefont {Taniguchi}, \citenamefont {Watanabe},\ and\ \citenamefont
  {Young}}]{zhou2021superconductivity}%
  \BibitemOpen
  \bibfield  {author} {\bibinfo {author} {\bibfnamefont {H.}~\bibnamefont
  {Zhou}}, \bibinfo {author} {\bibfnamefont {T.}~\bibnamefont {Xie}}, \bibinfo
  {author} {\bibfnamefont {T.}~\bibnamefont {Taniguchi}}, \bibinfo {author}
  {\bibfnamefont {K.}~\bibnamefont {Watanabe}},\ and\ \bibinfo {author}
  {\bibfnamefont {A.~F.}\ \bibnamefont {Young}},\ }\href
  {https://doi.org/10.1038/s41586-021-03926-0} {\bibfield  {journal} {\bibinfo
  {journal} {Nature}\ }\textbf {\bibinfo {volume} {598}},\ \bibinfo {pages}
  {434} (\bibinfo {year} {2021}{\natexlab{a}})}\BibitemShut {NoStop}%
\bibitem [{\citenamefont {Han}\ \emph {et~al.}(2025)\citenamefont {Han},
  \citenamefont {Lu}, \citenamefont {Hadjri}, \citenamefont {Shi},
  \citenamefont {Wu}, \citenamefont {Xu}, \citenamefont {Yao}, \citenamefont
  {Cotten}, \citenamefont {Sharifi~Sedeh}, \citenamefont {Weldeyesus},
  \citenamefont {Yang}, \citenamefont {Seo}, \citenamefont {Ye}, \citenamefont
  {Zhou}, \citenamefont {Liu}, \citenamefont {Shi}, \citenamefont {Hua},
  \citenamefont {Watanabe}, \citenamefont {Taniguchi}, \citenamefont {Xiong},
  \citenamefont {Zumb{\"u}hl}, \citenamefont {Fu},\ and\ \citenamefont
  {Ju}}]{han2025chiral}%
  \BibitemOpen
  \bibfield  {author} {\bibinfo {author} {\bibfnamefont {T.}~\bibnamefont
  {Han}}, \bibinfo {author} {\bibfnamefont {Z.}~\bibnamefont {Lu}}, \bibinfo
  {author} {\bibfnamefont {Z.}~\bibnamefont {Hadjri}}, \bibinfo {author}
  {\bibfnamefont {L.}~\bibnamefont {Shi}}, \bibinfo {author} {\bibfnamefont
  {Z.}~\bibnamefont {Wu}}, \bibinfo {author} {\bibfnamefont {W.}~\bibnamefont
  {Xu}}, \bibinfo {author} {\bibfnamefont {Y.}~\bibnamefont {Yao}}, \bibinfo
  {author} {\bibfnamefont {A.~A.}\ \bibnamefont {Cotten}}, \bibinfo {author}
  {\bibfnamefont {O.}~\bibnamefont {Sharifi~Sedeh}}, \bibinfo {author}
  {\bibfnamefont {H.}~\bibnamefont {Weldeyesus}}, \bibinfo {author}
  {\bibfnamefont {J.}~\bibnamefont {Yang}}, \bibinfo {author} {\bibfnamefont
  {J.}~\bibnamefont {Seo}}, \bibinfo {author} {\bibfnamefont {S.}~\bibnamefont
  {Ye}}, \bibinfo {author} {\bibfnamefont {M.}~\bibnamefont {Zhou}}, \bibinfo
  {author} {\bibfnamefont {H.}~\bibnamefont {Liu}}, \bibinfo {author}
  {\bibfnamefont {G.}~\bibnamefont {Shi}}, \bibinfo {author} {\bibfnamefont
  {Z.}~\bibnamefont {Hua}}, \bibinfo {author} {\bibfnamefont {K.}~\bibnamefont
  {Watanabe}}, \bibinfo {author} {\bibfnamefont {T.}~\bibnamefont {Taniguchi}},
  \bibinfo {author} {\bibfnamefont {P.}~\bibnamefont {Xiong}}, \bibinfo
  {author} {\bibfnamefont {D.~M.}\ \bibnamefont {Zumb{\"u}hl}}, \bibinfo
  {author} {\bibfnamefont {L.}~\bibnamefont {Fu}},\ and\ \bibinfo {author}
  {\bibfnamefont {L.}~\bibnamefont {Ju}},\ }\href
  {https://doi.org/10.1038/s41586-025-09169-7} {\bibfield  {journal} {\bibinfo
  {journal} {Nature}\ }\textbf {\bibinfo {volume} {643}},\ \bibinfo {pages}
  {654} (\bibinfo {year} {2025})}\BibitemShut {NoStop}%
\bibitem [{\citenamefont {Choi}\ \emph {et~al.}(2024)\citenamefont {Choi},
  \citenamefont {Choi}, \citenamefont {Valentini}, \citenamefont {Patterson},
  \citenamefont {Holleis}, \citenamefont {Sheekey}, \citenamefont {Stoyanov},
  \citenamefont {Cheng}, \citenamefont {Taniguchi}, \citenamefont {Watanabe}
  \emph {et~al.}}]{choi2024electric}%
  \BibitemOpen
  \bibfield  {author} {\bibinfo {author} {\bibfnamefont {Y.}~\bibnamefont
  {Choi}}, \bibinfo {author} {\bibfnamefont {Y.}~\bibnamefont {Choi}}, \bibinfo
  {author} {\bibfnamefont {M.}~\bibnamefont {Valentini}}, \bibinfo {author}
  {\bibfnamefont {C.~L.}\ \bibnamefont {Patterson}}, \bibinfo {author}
  {\bibfnamefont {L.~F.}\ \bibnamefont {Holleis}}, \bibinfo {author}
  {\bibfnamefont {O.~I.}\ \bibnamefont {Sheekey}}, \bibinfo {author}
  {\bibfnamefont {H.}~\bibnamefont {Stoyanov}}, \bibinfo {author}
  {\bibfnamefont {X.}~\bibnamefont {Cheng}}, \bibinfo {author} {\bibfnamefont
  {T.}~\bibnamefont {Taniguchi}}, \bibinfo {author} {\bibfnamefont
  {K.}~\bibnamefont {Watanabe}}, \emph {et~al.},\ }\href
  {https://arxiv.org/abs/2408.12584} {\bibfield  {journal} {\bibinfo  {journal}
  {arXiv preprint arXiv:2408.12584}\ } (\bibinfo {year} {2024})}\BibitemShut
  {NoStop}%
\bibitem [{\citenamefont {Morissette}\ \emph {et~al.}(2025)\citenamefont
  {Morissette}, \citenamefont {Qin}, \citenamefont {Wu}, \citenamefont {Zhang},
  \citenamefont {Watanabe}, \citenamefont {Taniguchi},\ and\ \citenamefont
  {Li}}]{morissette2025superconductivity}%
  \BibitemOpen
  \bibfield  {author} {\bibinfo {author} {\bibfnamefont {E.}~\bibnamefont
  {Morissette}}, \bibinfo {author} {\bibfnamefont {P.}~\bibnamefont {Qin}},
  \bibinfo {author} {\bibfnamefont {H.-T.}\ \bibnamefont {Wu}}, \bibinfo
  {author} {\bibfnamefont {N.~J.}\ \bibnamefont {Zhang}}, \bibinfo {author}
  {\bibfnamefont {K.}~\bibnamefont {Watanabe}}, \bibinfo {author}
  {\bibfnamefont {T.}~\bibnamefont {Taniguchi}},\ and\ \bibinfo {author}
  {\bibfnamefont {J.}~\bibnamefont {Li}},\ }\href@noop {} {\bibfield  {journal}
  {\bibinfo  {journal} {arXiv preprint arXiv:2504.05129}\ } (\bibinfo {year}
  {2025})}\BibitemShut {NoStop}%
\bibitem [{\citenamefont {Kumar}\ \emph {et~al.}(2025)\citenamefont {Kumar},
  \citenamefont {Waleffe}, \citenamefont {Okounkova}, \citenamefont {Tejani},
  \citenamefont {Phong}, \citenamefont {Watanabe}, \citenamefont {Taniguchi},
  \citenamefont {Lewandowski}, \citenamefont {Folk},\ and\ \citenamefont
  {Yankowitz}}]{kumar2025superconductivity}%
  \BibitemOpen
  \bibfield  {author} {\bibinfo {author} {\bibfnamefont {M.}~\bibnamefont
  {Kumar}}, \bibinfo {author} {\bibfnamefont {D.}~\bibnamefont {Waleffe}},
  \bibinfo {author} {\bibfnamefont {A.}~\bibnamefont {Okounkova}}, \bibinfo
  {author} {\bibfnamefont {R.}~\bibnamefont {Tejani}}, \bibinfo {author}
  {\bibfnamefont {V.~T.}\ \bibnamefont {Phong}}, \bibinfo {author}
  {\bibfnamefont {K.}~\bibnamefont {Watanabe}}, \bibinfo {author}
  {\bibfnamefont {T.}~\bibnamefont {Taniguchi}}, \bibinfo {author}
  {\bibfnamefont {C.}~\bibnamefont {Lewandowski}}, \bibinfo {author}
  {\bibfnamefont {J.}~\bibnamefont {Folk}},\ and\ \bibinfo {author}
  {\bibfnamefont {M.}~\bibnamefont {Yankowitz}},\ }\href@noop {} {\bibfield
  {journal} {\bibinfo  {journal} {arXiv preprint arXiv:2507.18598}\ } (\bibinfo
  {year} {2025})}\BibitemShut {NoStop}%
\bibitem [{\citenamefont {Nguyen}\ \emph {et~al.}(2025)\citenamefont {Nguyen},
  \citenamefont {Wu}, \citenamefont {Morissette}, \citenamefont {Zhang},
  \citenamefont {Qin}, \citenamefont {Watanabe}, \citenamefont {Taniguchi},
  \citenamefont {Hui}, \citenamefont {Feldman},\ and\ \citenamefont
  {Li}}]{nguyen2025hierarchy}%
  \BibitemOpen
  \bibfield  {author} {\bibinfo {author} {\bibfnamefont {R.~Q.}\ \bibnamefont
  {Nguyen}}, \bibinfo {author} {\bibfnamefont {H.-T.}\ \bibnamefont {Wu}},
  \bibinfo {author} {\bibfnamefont {E.}~\bibnamefont {Morissette}}, \bibinfo
  {author} {\bibfnamefont {N.~J.}\ \bibnamefont {Zhang}}, \bibinfo {author}
  {\bibfnamefont {P.}~\bibnamefont {Qin}}, \bibinfo {author} {\bibfnamefont
  {K.}~\bibnamefont {Watanabe}}, \bibinfo {author} {\bibfnamefont
  {T.}~\bibnamefont {Taniguchi}}, \bibinfo {author} {\bibfnamefont {A.~W.}\
  \bibnamefont {Hui}}, \bibinfo {author} {\bibfnamefont {D.~E.}\ \bibnamefont
  {Feldman}},\ and\ \bibinfo {author} {\bibfnamefont {J.}~\bibnamefont {Li}},\
  }\href@noop {} {\bibfield  {journal} {\bibinfo  {journal} {arXiv preprint
  arXiv:2507.22026}\ } (\bibinfo {year} {2025})}\BibitemShut {NoStop}%
\bibitem [{\citenamefont {Yang}\ \emph {et~al.}(2025)\citenamefont {Yang},
  \citenamefont {Shi}, \citenamefont {Ye}, \citenamefont {Yoon}, \citenamefont
  {Lu}, \citenamefont {Kakani}, \citenamefont {Han}, \citenamefont {Seo},
  \citenamefont {Shi}, \citenamefont {Watanabe}, \citenamefont {Taniguchi},
  \citenamefont {Zhang},\ and\ \citenamefont {Ju}}]{yang2025}%
  \BibitemOpen
  \bibfield  {author} {\bibinfo {author} {\bibfnamefont {J.}~\bibnamefont
  {Yang}}, \bibinfo {author} {\bibfnamefont {X.}~\bibnamefont {Shi}}, \bibinfo
  {author} {\bibfnamefont {S.}~\bibnamefont {Ye}}, \bibinfo {author}
  {\bibfnamefont {C.}~\bibnamefont {Yoon}}, \bibinfo {author} {\bibfnamefont
  {Z.}~\bibnamefont {Lu}}, \bibinfo {author} {\bibfnamefont {V.}~\bibnamefont
  {Kakani}}, \bibinfo {author} {\bibfnamefont {T.}~\bibnamefont {Han}},
  \bibinfo {author} {\bibfnamefont {J.}~\bibnamefont {Seo}}, \bibinfo {author}
  {\bibfnamefont {L.}~\bibnamefont {Shi}}, \bibinfo {author} {\bibfnamefont
  {K.}~\bibnamefont {Watanabe}}, \bibinfo {author} {\bibfnamefont
  {T.}~\bibnamefont {Taniguchi}}, \bibinfo {author} {\bibfnamefont
  {F.}~\bibnamefont {Zhang}},\ and\ \bibinfo {author} {\bibfnamefont
  {L.}~\bibnamefont {Ju}},\ }\href {https://doi.org/10.1038/s41563-025-02156-3}
  {\bibfield  {journal} {\bibinfo  {journal} {Nature Materials}\ }\textbf
  {\bibinfo {volume} {24}},\ \bibinfo {pages} {1058} (\bibinfo {year}
  {2025})}\BibitemShut {NoStop}%
\bibitem [{\citenamefont {Patterson}\ \emph {et~al.}()\citenamefont
  {Patterson}, \citenamefont {Sheekey}, \citenamefont {Arp}, \citenamefont
  {Holleis}, \citenamefont {Koh}, \citenamefont {Choi}, \citenamefont {Xie},
  \citenamefont {Xu}, \citenamefont {Redekop}, \citenamefont {Babikyan} \emph
  {et~al.}}]{patterson2024}%
  \BibitemOpen
  \bibfield  {author} {\bibinfo {author} {\bibfnamefont {C.}~\bibnamefont
  {Patterson}}, \bibinfo {author} {\bibfnamefont {O.}~\bibnamefont {Sheekey}},
  \bibinfo {author} {\bibfnamefont {T.}~\bibnamefont {Arp}}, \bibinfo {author}
  {\bibfnamefont {L.}~\bibnamefont {Holleis}}, \bibinfo {author} {\bibfnamefont
  {J.}~\bibnamefont {Koh}}, \bibinfo {author} {\bibfnamefont {Y.}~\bibnamefont
  {Choi}}, \bibinfo {author} {\bibfnamefont {T.}~\bibnamefont {Xie}}, \bibinfo
  {author} {\bibfnamefont {S.}~\bibnamefont {Xu}}, \bibinfo {author}
  {\bibfnamefont {E.}~\bibnamefont {Redekop}}, \bibinfo {author} {\bibfnamefont
  {G.}~\bibnamefont {Babikyan}}, \emph {et~al.},\ }\href@noop {} {\bibinfo
  {journal} {arXiv preprint arXiv:2408.10190}\ }\BibitemShut {NoStop}%
\bibitem [{\citenamefont {Lee}\ \emph {et~al.}(2013)\citenamefont {Lee},
  \citenamefont {Myhro}, \citenamefont {Tran}, \citenamefont {Gilgren},
  \citenamefont {Jr.}, \citenamefont {Bao}, \citenamefont {Deo},\ and\
  \citenamefont {Lau}}]{lee2013band}%
  \BibitemOpen
\bibfield  {journal} {  }\bibfield  {author} {\bibinfo {author} {\bibfnamefont
  {Y.}~\bibnamefont {Lee}}, \bibinfo {author} {\bibfnamefont {K.}~\bibnamefont
  {Myhro}}, \bibinfo {author} {\bibfnamefont {D.}~\bibnamefont {Tran}},
  \bibinfo {author} {\bibfnamefont {N.}~\bibnamefont {Gilgren}}, \bibinfo
  {author} {\bibfnamefont {J.~V.}\ \bibnamefont {Jr.}}, \bibinfo {author}
  {\bibfnamefont {W.}~\bibnamefont {Bao}}, \bibinfo {author} {\bibfnamefont
  {M.}~\bibnamefont {Deo}},\ and\ \bibinfo {author} {\bibfnamefont {C.~N.}\
  \bibnamefont {Lau}},\ }in\ \href {https://doi.org/10.1117/12.2016521} {\emph
  {\bibinfo {booktitle} {Micro- and Nanotechnology Sensors, Systems, and
  Applications V}}},\ Vol.\ \bibinfo {volume} {8725},\ \bibinfo {editor}
  {edited by\ \bibinfo {editor} {\bibfnamefont {T.}~\bibnamefont {George}},
  \bibinfo {editor} {\bibfnamefont {M.~S.}\ \bibnamefont {Islam}},\ and\
  \bibinfo {editor} {\bibfnamefont {A.~K.}\ \bibnamefont {Dutta}}},\ \bibinfo
  {organization} {International Society for Optics and Photonics}\ (\bibinfo
  {publisher} {SPIE},\ \bibinfo {year} {2013})\ p.\ \bibinfo {pages}
  {872506}\BibitemShut {NoStop}%
\bibitem [{\citenamefont {Myhro}\ \emph {et~al.}(2018)\citenamefont {Myhro},
  \citenamefont {Che}, \citenamefont {Shi}, \citenamefont {Lee}, \citenamefont
  {Thilahar}, \citenamefont {Bleich}, \citenamefont {Smirnov},\ and\
  \citenamefont {Lau}}]{myhro2018large}%
  \BibitemOpen
  \bibfield  {author} {\bibinfo {author} {\bibfnamefont {K.}~\bibnamefont
  {Myhro}}, \bibinfo {author} {\bibfnamefont {S.}~\bibnamefont {Che}}, \bibinfo
  {author} {\bibfnamefont {Y.}~\bibnamefont {Shi}}, \bibinfo {author}
  {\bibfnamefont {Y.}~\bibnamefont {Lee}}, \bibinfo {author} {\bibfnamefont
  {K.}~\bibnamefont {Thilahar}}, \bibinfo {author} {\bibfnamefont
  {K.}~\bibnamefont {Bleich}}, \bibinfo {author} {\bibfnamefont
  {D.}~\bibnamefont {Smirnov}},\ and\ \bibinfo {author} {\bibfnamefont {C.~N.}\
  \bibnamefont {Lau}},\ }\href {https://doi.org/10.1088/2053-1583/aad2f2}
  {\bibfield  {journal} {\bibinfo  {journal} {2D Materials}\ }\textbf {\bibinfo
  {volume} {5}},\ \bibinfo {pages} {045013} (\bibinfo {year}
  {2018})}\BibitemShut {NoStop}%
\bibitem [{\citenamefont {Shi}\ \emph {et~al.}(2020)\citenamefont {Shi},
  \citenamefont {Xu}, \citenamefont {Yang}, \citenamefont {Slizovskiy},
  \citenamefont {Morozov}, \citenamefont {Son}, \citenamefont {Ozdemir},
  \citenamefont {Mullan}, \citenamefont {Barrier}, \citenamefont {Yin},
  \citenamefont {Berdyugin}, \citenamefont {Piot}, \citenamefont {Taniguchi},
  \citenamefont {Watanabe}, \citenamefont {Fal'ko}, \citenamefont {Novoselov},
  \citenamefont {Geim},\ and\ \citenamefont {Mishchenko}}]{shi2020electronic}%
  \BibitemOpen
  \bibfield  {author} {\bibinfo {author} {\bibfnamefont {Y.}~\bibnamefont
  {Shi}}, \bibinfo {author} {\bibfnamefont {S.}~\bibnamefont {Xu}}, \bibinfo
  {author} {\bibfnamefont {Y.}~\bibnamefont {Yang}}, \bibinfo {author}
  {\bibfnamefont {S.}~\bibnamefont {Slizovskiy}}, \bibinfo {author}
  {\bibfnamefont {S.~V.}\ \bibnamefont {Morozov}}, \bibinfo {author}
  {\bibfnamefont {S.-K.}\ \bibnamefont {Son}}, \bibinfo {author} {\bibfnamefont
  {S.}~\bibnamefont {Ozdemir}}, \bibinfo {author} {\bibfnamefont
  {C.}~\bibnamefont {Mullan}}, \bibinfo {author} {\bibfnamefont
  {J.}~\bibnamefont {Barrier}}, \bibinfo {author} {\bibfnamefont
  {J.}~\bibnamefont {Yin}}, \bibinfo {author} {\bibfnamefont {A.~I.}\
  \bibnamefont {Berdyugin}}, \bibinfo {author} {\bibfnamefont {B.~A.}\
  \bibnamefont {Piot}}, \bibinfo {author} {\bibfnamefont {T.}~\bibnamefont
  {Taniguchi}}, \bibinfo {author} {\bibfnamefont {K.}~\bibnamefont {Watanabe}},
  \bibinfo {author} {\bibfnamefont {V.~I.}\ \bibnamefont {Fal'ko}}, \bibinfo
  {author} {\bibfnamefont {K.~S.}\ \bibnamefont {Novoselov}}, \bibinfo {author}
  {\bibfnamefont {A.~K.}\ \bibnamefont {Geim}},\ and\ \bibinfo {author}
  {\bibfnamefont {A.}~\bibnamefont {Mishchenko}},\ }\href
  {https://doi.org/10.1038/s41586-020-2568-2} {\bibfield  {journal} {\bibinfo
  {journal} {Nature}\ }\textbf {\bibinfo {volume} {584}},\ \bibinfo {pages}
  {210} (\bibinfo {year} {2020})}\BibitemShut {NoStop}%
\bibitem [{\citenamefont {Hagymási}\ \emph {et~al.}(2022)\citenamefont
  {Hagymási}, \citenamefont {Isa}, \citenamefont {Tajkov}, \citenamefont
  {Márity}, \citenamefont {Oroszlány}, \citenamefont {Koltai}, \citenamefont
  {Alassaf}, \citenamefont {Kun}, \citenamefont {Kandrai}, \citenamefont
  {Pálinkás}, \citenamefont {Vancsó}, \citenamefont {Tapasztó},\ and\
  \citenamefont {Nemes-Incze}}]{hagymasi2022observation}%
  \BibitemOpen
  \bibfield  {author} {\bibinfo {author} {\bibfnamefont {I.}~\bibnamefont
  {Hagymási}}, \bibinfo {author} {\bibfnamefont {M.~S.~M.}\ \bibnamefont
  {Isa}}, \bibinfo {author} {\bibfnamefont {Z.}~\bibnamefont {Tajkov}},
  \bibinfo {author} {\bibfnamefont {K.}~\bibnamefont {Márity}}, \bibinfo
  {author} {\bibfnamefont {L.}~\bibnamefont {Oroszlány}}, \bibinfo {author}
  {\bibfnamefont {J.}~\bibnamefont {Koltai}}, \bibinfo {author} {\bibfnamefont
  {A.}~\bibnamefont {Alassaf}}, \bibinfo {author} {\bibfnamefont
  {P.}~\bibnamefont {Kun}}, \bibinfo {author} {\bibfnamefont {K.}~\bibnamefont
  {Kandrai}}, \bibinfo {author} {\bibfnamefont {A.}~\bibnamefont {Pálinkás}},
  \bibinfo {author} {\bibfnamefont {P.}~\bibnamefont {Vancsó}}, \bibinfo
  {author} {\bibfnamefont {L.}~\bibnamefont {Tapasztó}},\ and\ \bibinfo
  {author} {\bibfnamefont {P.}~\bibnamefont {Nemes-Incze}},\ }\href
  {https://doi.org/10.1126/sciadv.abo6879} {\bibfield  {journal} {\bibinfo
  {journal} {Science Advances}\ }\textbf {\bibinfo {volume} {8}},\ \bibinfo
  {pages} {eabo6879} (\bibinfo {year} {2022})}\BibitemShut {NoStop}%
\bibitem [{\citenamefont {Zhou}\ \emph
  {et~al.}(2021{\natexlab{b}})\citenamefont {Zhou}, \citenamefont {Xie},
  \citenamefont {Ghazaryan}, \citenamefont {Holder}, \citenamefont {Ehrets},
  \citenamefont {Spanton}, \citenamefont {Taniguchi}, \citenamefont {Watanabe},
  \citenamefont {Berg}, \citenamefont {Serbyn},\ and\ \citenamefont
  {Young}}]{zhou2021half}%
  \BibitemOpen
  \bibfield  {author} {\bibinfo {author} {\bibfnamefont {H.}~\bibnamefont
  {Zhou}}, \bibinfo {author} {\bibfnamefont {T.}~\bibnamefont {Xie}}, \bibinfo
  {author} {\bibfnamefont {A.}~\bibnamefont {Ghazaryan}}, \bibinfo {author}
  {\bibfnamefont {T.}~\bibnamefont {Holder}}, \bibinfo {author} {\bibfnamefont
  {J.~R.}\ \bibnamefont {Ehrets}}, \bibinfo {author} {\bibfnamefont {E.~M.}\
  \bibnamefont {Spanton}}, \bibinfo {author} {\bibfnamefont {T.}~\bibnamefont
  {Taniguchi}}, \bibinfo {author} {\bibfnamefont {K.}~\bibnamefont {Watanabe}},
  \bibinfo {author} {\bibfnamefont {E.}~\bibnamefont {Berg}}, \bibinfo {author}
  {\bibfnamefont {M.}~\bibnamefont {Serbyn}},\ and\ \bibinfo {author}
  {\bibfnamefont {A.~F.}\ \bibnamefont {Young}},\ }\href
  {https://doi.org/10.1038/s41586-021-03938-w} {\bibfield  {journal} {\bibinfo
  {journal} {Nature}\ }\textbf {\bibinfo {volume} {598}},\ \bibinfo {pages}
  {429} (\bibinfo {year} {2021}{\natexlab{b}})}\BibitemShut {NoStop}%
\bibitem [{\citenamefont {Lee}\ \emph {et~al.}(2022)\citenamefont {Lee},
  \citenamefont {Che}, \citenamefont {Velasco~Jr.}, \citenamefont {Gao},
  \citenamefont {Shi}, \citenamefont {Tran}, \citenamefont {Baima},
  \citenamefont {Mauri}, \citenamefont {Calandra}, \citenamefont {Bockrath},\
  and\ \citenamefont {Lau}}]{lee2022gate}%
  \BibitemOpen
  \bibfield  {author} {\bibinfo {author} {\bibfnamefont {Y.}~\bibnamefont
  {Lee}}, \bibinfo {author} {\bibfnamefont {S.}~\bibnamefont {Che}}, \bibinfo
  {author} {\bibfnamefont {J.}~\bibnamefont {Velasco~Jr.}}, \bibinfo {author}
  {\bibfnamefont {X.}~\bibnamefont {Gao}}, \bibinfo {author} {\bibfnamefont
  {Y.}~\bibnamefont {Shi}}, \bibinfo {author} {\bibfnamefont {D.}~\bibnamefont
  {Tran}}, \bibinfo {author} {\bibfnamefont {J.}~\bibnamefont {Baima}},
  \bibinfo {author} {\bibfnamefont {F.}~\bibnamefont {Mauri}}, \bibinfo
  {author} {\bibfnamefont {M.}~\bibnamefont {Calandra}}, \bibinfo {author}
  {\bibfnamefont {M.}~\bibnamefont {Bockrath}},\ and\ \bibinfo {author}
  {\bibfnamefont {C.~N.}\ \bibnamefont {Lau}},\ }\href
  {https://doi.org/10.1021/acs.nanolett.2c00466} {\bibfield  {journal}
  {\bibinfo  {journal} {Nano Letters}\ }\textbf {\bibinfo {volume} {22}},\
  \bibinfo {pages} {5094} (\bibinfo {year} {2022})}\BibitemShut {NoStop}%
\bibitem [{\citenamefont {Zhou}\ \emph
  {et~al.}(2024{\natexlab{a}})\citenamefont {Zhou}, \citenamefont {Ding},
  \citenamefont {Hua}, \citenamefont {Zhang}, \citenamefont {Watanabe},
  \citenamefont {Taniguchi}, \citenamefont {Zhu},\ and\ \citenamefont
  {Xu}}]{zhou2024layer}%
  \BibitemOpen
  \bibfield  {author} {\bibinfo {author} {\bibfnamefont {W.}~\bibnamefont
  {Zhou}}, \bibinfo {author} {\bibfnamefont {J.}~\bibnamefont {Ding}}, \bibinfo
  {author} {\bibfnamefont {J.}~\bibnamefont {Hua}}, \bibinfo {author}
  {\bibfnamefont {L.}~\bibnamefont {Zhang}}, \bibinfo {author} {\bibfnamefont
  {K.}~\bibnamefont {Watanabe}}, \bibinfo {author} {\bibfnamefont
  {T.}~\bibnamefont {Taniguchi}}, \bibinfo {author} {\bibfnamefont
  {W.}~\bibnamefont {Zhu}},\ and\ \bibinfo {author} {\bibfnamefont
  {S.}~\bibnamefont {Xu}},\ }\href {https://doi.org/10.1038/s41467-024-46913-5}
  {\bibfield  {journal} {\bibinfo  {journal} {Nature Communications}\ }\textbf
  {\bibinfo {volume} {15}},\ \bibinfo {pages} {2597} (\bibinfo {year}
  {2024}{\natexlab{a}})}\BibitemShut {NoStop}%
\bibitem [{\citenamefont {Bao}\ \emph {et~al.}(2011)\citenamefont {Bao},
  \citenamefont {Jing}, \citenamefont {Velasco}, \citenamefont {Lee},
  \citenamefont {Liu}, \citenamefont {Tran}, \citenamefont {Standley},
  \citenamefont {Aykol}, \citenamefont {Cronin}, \citenamefont {Smirnov},
  \citenamefont {Koshino}, \citenamefont {McCann}, \citenamefont {Bockrath},\
  and\ \citenamefont {Lau}}]{bao2011stacking}%
  \BibitemOpen
  \bibfield  {author} {\bibinfo {author} {\bibfnamefont {W.}~\bibnamefont
  {Bao}}, \bibinfo {author} {\bibfnamefont {L.}~\bibnamefont {Jing}}, \bibinfo
  {author} {\bibfnamefont {J.}~\bibnamefont {Velasco}}, \bibinfo {author}
  {\bibfnamefont {Y.}~\bibnamefont {Lee}}, \bibinfo {author} {\bibfnamefont
  {G.}~\bibnamefont {Liu}}, \bibinfo {author} {\bibfnamefont {D.}~\bibnamefont
  {Tran}}, \bibinfo {author} {\bibfnamefont {B.}~\bibnamefont {Standley}},
  \bibinfo {author} {\bibfnamefont {M.}~\bibnamefont {Aykol}}, \bibinfo
  {author} {\bibfnamefont {S.~B.}\ \bibnamefont {Cronin}}, \bibinfo {author}
  {\bibfnamefont {D.}~\bibnamefont {Smirnov}}, \bibinfo {author} {\bibfnamefont
  {M.}~\bibnamefont {Koshino}}, \bibinfo {author} {\bibfnamefont
  {E.}~\bibnamefont {McCann}}, \bibinfo {author} {\bibfnamefont
  {M.}~\bibnamefont {Bockrath}},\ and\ \bibinfo {author} {\bibfnamefont
  {C.~N.}\ \bibnamefont {Lau}},\ }\href {https://doi.org/10.1038/nphys2103}
  {\bibfield  {journal} {\bibinfo  {journal} {Nature Physics}\ }\textbf
  {\bibinfo {volume} {7}},\ \bibinfo {pages} {948} (\bibinfo {year}
  {2011})}\BibitemShut {NoStop}%
\bibitem [{\citenamefont {Lee}\ \emph {et~al.}(2014)\citenamefont {Lee},
  \citenamefont {Tran}, \citenamefont {Myhro}, \citenamefont {Velasco},
  \citenamefont {Gillgren}, \citenamefont {Lau}, \citenamefont {Barlas},
  \citenamefont {Poumirol}, \citenamefont {Smirnov},\ and\ \citenamefont
  {Guinea}}]{lee2014competition}%
  \BibitemOpen
  \bibfield  {author} {\bibinfo {author} {\bibfnamefont {Y.}~\bibnamefont
  {Lee}}, \bibinfo {author} {\bibfnamefont {D.}~\bibnamefont {Tran}}, \bibinfo
  {author} {\bibfnamefont {K.}~\bibnamefont {Myhro}}, \bibinfo {author}
  {\bibfnamefont {J.}~\bibnamefont {Velasco}}, \bibinfo {author} {\bibfnamefont
  {N.}~\bibnamefont {Gillgren}}, \bibinfo {author} {\bibfnamefont {C.~N.}\
  \bibnamefont {Lau}}, \bibinfo {author} {\bibfnamefont {Y.}~\bibnamefont
  {Barlas}}, \bibinfo {author} {\bibfnamefont {J.~M.}\ \bibnamefont
  {Poumirol}}, \bibinfo {author} {\bibfnamefont {D.}~\bibnamefont {Smirnov}},\
  and\ \bibinfo {author} {\bibfnamefont {F.}~\bibnamefont {Guinea}},\ }\href
  {https://doi.org/10.1038/ncomms6656} {\bibfield  {journal} {\bibinfo
  {journal} {Nature Communications}\ }\textbf {\bibinfo {volume} {5}},\
  \bibinfo {pages} {5656} (\bibinfo {year} {2014})}\BibitemShut {NoStop}%
\bibitem [{\citenamefont {Han}\ \emph {et~al.}(2024{\natexlab{a}})\citenamefont
  {Han}, \citenamefont {Lu}, \citenamefont {Scuri}, \citenamefont {Sung},
  \citenamefont {Wang}, \citenamefont {Han}, \citenamefont {Watanabe},
  \citenamefont {Taniguchi}, \citenamefont {Park},\ and\ \citenamefont
  {Ju}}]{han2024correlated}%
  \BibitemOpen
  \bibfield  {author} {\bibinfo {author} {\bibfnamefont {T.}~\bibnamefont
  {Han}}, \bibinfo {author} {\bibfnamefont {Z.}~\bibnamefont {Lu}}, \bibinfo
  {author} {\bibfnamefont {G.}~\bibnamefont {Scuri}}, \bibinfo {author}
  {\bibfnamefont {J.}~\bibnamefont {Sung}}, \bibinfo {author} {\bibfnamefont
  {J.}~\bibnamefont {Wang}}, \bibinfo {author} {\bibfnamefont {T.}~\bibnamefont
  {Han}}, \bibinfo {author} {\bibfnamefont {K.}~\bibnamefont {Watanabe}},
  \bibinfo {author} {\bibfnamefont {T.}~\bibnamefont {Taniguchi}}, \bibinfo
  {author} {\bibfnamefont {H.}~\bibnamefont {Park}},\ and\ \bibinfo {author}
  {\bibfnamefont {L.}~\bibnamefont {Ju}},\ }\href
  {https://doi.org/10.1038/s41565-023-01520-1} {\bibfield  {journal} {\bibinfo
  {journal} {Nature Nanotechnology}\ }\textbf {\bibinfo {volume} {19}},\
  \bibinfo {pages} {181} (\bibinfo {year} {2024}{\natexlab{a}})}\BibitemShut
  {NoStop}%
\bibitem [{\citenamefont {Arp}\ \emph {et~al.}(2024)\citenamefont {Arp},
  \citenamefont {Sheekey}, \citenamefont {Zhou}, \citenamefont {Tschirhart},
  \citenamefont {Patterson}, \citenamefont {Yoo}, \citenamefont {Holleis},
  \citenamefont {Redekop}, \citenamefont {Babikyan}, \citenamefont {Xie},
  \citenamefont {Xiao}, \citenamefont {Vituri}, \citenamefont {Holder},
  \citenamefont {Taniguchi}, \citenamefont {Watanabe}, \citenamefont {Huber},
  \citenamefont {Berg},\ and\ \citenamefont {Young}}]{arp2024intervalley}%
  \BibitemOpen
  \bibfield  {author} {\bibinfo {author} {\bibfnamefont {T.}~\bibnamefont
  {Arp}}, \bibinfo {author} {\bibfnamefont {O.}~\bibnamefont {Sheekey}},
  \bibinfo {author} {\bibfnamefont {H.}~\bibnamefont {Zhou}}, \bibinfo {author}
  {\bibfnamefont {C.~L.}\ \bibnamefont {Tschirhart}}, \bibinfo {author}
  {\bibfnamefont {C.~L.}\ \bibnamefont {Patterson}}, \bibinfo {author}
  {\bibfnamefont {H.~M.}\ \bibnamefont {Yoo}}, \bibinfo {author} {\bibfnamefont
  {L.}~\bibnamefont {Holleis}}, \bibinfo {author} {\bibfnamefont
  {E.}~\bibnamefont {Redekop}}, \bibinfo {author} {\bibfnamefont
  {G.}~\bibnamefont {Babikyan}}, \bibinfo {author} {\bibfnamefont
  {T.}~\bibnamefont {Xie}}, \bibinfo {author} {\bibfnamefont {J.}~\bibnamefont
  {Xiao}}, \bibinfo {author} {\bibfnamefont {Y.}~\bibnamefont {Vituri}},
  \bibinfo {author} {\bibfnamefont {T.}~\bibnamefont {Holder}}, \bibinfo
  {author} {\bibfnamefont {T.}~\bibnamefont {Taniguchi}}, \bibinfo {author}
  {\bibfnamefont {K.}~\bibnamefont {Watanabe}}, \bibinfo {author}
  {\bibfnamefont {M.~E.}\ \bibnamefont {Huber}}, \bibinfo {author}
  {\bibfnamefont {E.}~\bibnamefont {Berg}},\ and\ \bibinfo {author}
  {\bibfnamefont {A.~F.}\ \bibnamefont {Young}},\ }\href
  {https://doi.org/10.1038/s41567-024-02560-7} {\bibfield  {journal} {\bibinfo
  {journal} {Nature Physics}\ }\textbf {\bibinfo {volume} {20}},\ \bibinfo
  {pages} {1413} (\bibinfo {year} {2024})}\BibitemShut {NoStop}%
\bibitem [{\citenamefont {Zhang}\ \emph
  {et~al.}(2024{\natexlab{a}})\citenamefont {Zhang}, \citenamefont {Li},
  \citenamefont {Scheer}, \citenamefont {Wang}, \citenamefont {Tuo},
  \citenamefont {Zou}, \citenamefont {Chen}, \citenamefont {Li}, \citenamefont
  {Cai}, \citenamefont {Bao}, \citenamefont {Li}, \citenamefont {Deng},
  \citenamefont {Watanabe}, \citenamefont {Taniguchi}, \citenamefont {Ye},
  \citenamefont {Tang}, \citenamefont {Xu}, \citenamefont {Yu}, \citenamefont
  {Avila}, \citenamefont {Dudin}, \citenamefont {Denlinger}, \citenamefont
  {Yao}, \citenamefont {Lian}, \citenamefont {Duan},\ and\ \citenamefont
  {Zhou}}]{zhang2024correlated}%
  \BibitemOpen
  \bibfield  {author} {\bibinfo {author} {\bibfnamefont {H.}~\bibnamefont
  {Zhang}}, \bibinfo {author} {\bibfnamefont {Q.}~\bibnamefont {Li}}, \bibinfo
  {author} {\bibfnamefont {M.~G.}\ \bibnamefont {Scheer}}, \bibinfo {author}
  {\bibfnamefont {R.}~\bibnamefont {Wang}}, \bibinfo {author} {\bibfnamefont
  {C.}~\bibnamefont {Tuo}}, \bibinfo {author} {\bibfnamefont {N.}~\bibnamefont
  {Zou}}, \bibinfo {author} {\bibfnamefont {W.}~\bibnamefont {Chen}}, \bibinfo
  {author} {\bibfnamefont {J.}~\bibnamefont {Li}}, \bibinfo {author}
  {\bibfnamefont {X.}~\bibnamefont {Cai}}, \bibinfo {author} {\bibfnamefont
  {C.}~\bibnamefont {Bao}}, \bibinfo {author} {\bibfnamefont {M.-R.}\
  \bibnamefont {Li}}, \bibinfo {author} {\bibfnamefont {K.}~\bibnamefont
  {Deng}}, \bibinfo {author} {\bibfnamefont {K.}~\bibnamefont {Watanabe}},
  \bibinfo {author} {\bibfnamefont {T.}~\bibnamefont {Taniguchi}}, \bibinfo
  {author} {\bibfnamefont {M.}~\bibnamefont {Ye}}, \bibinfo {author}
  {\bibfnamefont {P.}~\bibnamefont {Tang}}, \bibinfo {author} {\bibfnamefont
  {Y.}~\bibnamefont {Xu}}, \bibinfo {author} {\bibfnamefont {P.}~\bibnamefont
  {Yu}}, \bibinfo {author} {\bibfnamefont {J.}~\bibnamefont {Avila}}, \bibinfo
  {author} {\bibfnamefont {P.}~\bibnamefont {Dudin}}, \bibinfo {author}
  {\bibfnamefont {J.~D.}\ \bibnamefont {Denlinger}}, \bibinfo {author}
  {\bibfnamefont {H.}~\bibnamefont {Yao}}, \bibinfo {author} {\bibfnamefont
  {B.}~\bibnamefont {Lian}}, \bibinfo {author} {\bibfnamefont {W.}~\bibnamefont
  {Duan}},\ and\ \bibinfo {author} {\bibfnamefont {S.}~\bibnamefont {Zhou}},\
  }\href {https://doi.org/10.1073/pnas.2410714121} {\bibfield  {journal}
  {\bibinfo  {journal} {Proceedings of the National Academy of Sciences}\
  }\textbf {\bibinfo {volume} {121}},\ \bibinfo {pages} {e2410714121} (\bibinfo
  {year} {2024}{\natexlab{a}})}\BibitemShut {NoStop}%
\bibitem [{\citenamefont {Zhang}\ \emph
  {et~al.}(2024{\natexlab{b}})\citenamefont {Zhang}, \citenamefont {Zhou},
  \citenamefont {Zhang}, \citenamefont {Cai}, \citenamefont {Tong},
  \citenamefont {Liao}, \citenamefont {Zou}, \citenamefont {Xue}, \citenamefont
  {Tian}, \citenamefont {Chen}, \citenamefont {Tian}, \citenamefont {Zhang},
  \citenamefont {Wang}, \citenamefont {Zou}, \citenamefont {Liu}, \citenamefont
  {Hu}, \citenamefont {Ren}, \citenamefont {Zhang}, \citenamefont {Zhang},
  \citenamefont {Wang}, \citenamefont {He}, \citenamefont {Liao}, \citenamefont
  {Qin},\ and\ \citenamefont {Yin}}]{zhang2024layer}%
  \BibitemOpen
  \bibfield  {author} {\bibinfo {author} {\bibfnamefont {Y.}~\bibnamefont
  {Zhang}}, \bibinfo {author} {\bibfnamefont {Y.-Y.}\ \bibnamefont {Zhou}},
  \bibinfo {author} {\bibfnamefont {S.}~\bibnamefont {Zhang}}, \bibinfo
  {author} {\bibfnamefont {H.}~\bibnamefont {Cai}}, \bibinfo {author}
  {\bibfnamefont {L.-H.}\ \bibnamefont {Tong}}, \bibinfo {author}
  {\bibfnamefont {W.-Y.}\ \bibnamefont {Liao}}, \bibinfo {author}
  {\bibfnamefont {R.-J.}\ \bibnamefont {Zou}}, \bibinfo {author} {\bibfnamefont
  {S.-M.}\ \bibnamefont {Xue}}, \bibinfo {author} {\bibfnamefont
  {Y.}~\bibnamefont {Tian}}, \bibinfo {author} {\bibfnamefont {T.}~\bibnamefont
  {Chen}}, \bibinfo {author} {\bibfnamefont {Q.}~\bibnamefont {Tian}}, \bibinfo
  {author} {\bibfnamefont {C.}~\bibnamefont {Zhang}}, \bibinfo {author}
  {\bibfnamefont {Y.}~\bibnamefont {Wang}}, \bibinfo {author} {\bibfnamefont
  {X.}~\bibnamefont {Zou}}, \bibinfo {author} {\bibfnamefont {X.}~\bibnamefont
  {Liu}}, \bibinfo {author} {\bibfnamefont {Y.}~\bibnamefont {Hu}}, \bibinfo
  {author} {\bibfnamefont {Y.-N.}\ \bibnamefont {Ren}}, \bibinfo {author}
  {\bibfnamefont {L.}~\bibnamefont {Zhang}}, \bibinfo {author} {\bibfnamefont
  {L.}~\bibnamefont {Zhang}}, \bibinfo {author} {\bibfnamefont {W.-X.}\
  \bibnamefont {Wang}}, \bibinfo {author} {\bibfnamefont {L.}~\bibnamefont
  {He}}, \bibinfo {author} {\bibfnamefont {L.}~\bibnamefont {Liao}}, \bibinfo
  {author} {\bibfnamefont {Z.}~\bibnamefont {Qin}},\ and\ \bibinfo {author}
  {\bibfnamefont {L.-J.}\ \bibnamefont {Yin}},\ }\bibfield  {journal} {\bibinfo
   {journal} {Nature Nanotechnology}\ }\href
  {https://doi.org/10.1038/s41565-024-01822-y} {10.1038/s41565-024-01822-y}
  (\bibinfo {year} {2024}{\natexlab{b}})\BibitemShut {NoStop}%
\bibitem [{\citenamefont {Koshino}(2010)}]{koshino2010interlayer}%
  \BibitemOpen
  \bibfield  {author} {\bibinfo {author} {\bibfnamefont {M.}~\bibnamefont
  {Koshino}},\ }\href {https://doi.org/10.1103/PhysRevB.81.125304} {\bibfield
  {journal} {\bibinfo  {journal} {Phys. Rev. B}\ }\textbf {\bibinfo {volume}
  {81}},\ \bibinfo {pages} {125304} (\bibinfo {year} {2010})}\BibitemShut
  {NoStop}%
\bibitem [{Note1()}]{Note1}%
  \BibitemOpen
  \bibinfo {note} {Function $\kappa (\protect \textbf {k})$ is understood as in
  the polar representation, with the gradient $\nabla \kappa (\protect \textbf
  {k})$ branch-independent. The singularity of $\kappa (\protect \textbf {k})$
  at $\protect \textbf {k}=0$ becomes removable as we compute the
  QGT.}\BibitemShut {Stop}%
\bibitem [{\citenamefont {Koshino}\ and\ \citenamefont
  {McCann}(2009)}]{koshino2009trigonal}%
  \BibitemOpen
  \bibfield  {author} {\bibinfo {author} {\bibfnamefont {M.}~\bibnamefont
  {Koshino}}\ and\ \bibinfo {author} {\bibfnamefont {E.}~\bibnamefont
  {McCann}},\ }\href {https://doi.org/10.1103/PhysRevB.80.165409} {\bibfield
  {journal} {\bibinfo  {journal} {Phys. Rev. B}\ }\textbf {\bibinfo {volume}
  {80}},\ \bibinfo {pages} {165409} (\bibinfo {year} {2009})}\BibitemShut
  {NoStop}%
\bibitem [{Note2()}]{Note2}%
  \BibitemOpen
  \bibinfo {note} {This relation is obtained by setting the thickness equal to
  $\eta $ times the decay length, $N= \eta \lambda =-\eta /\ln k_h$, which is
  also equivalent to setting the dispersion function of $N$-layer RG equal to
  some small energy scale, $k_h^N=e^{-\eta }$. Since there is an arbitrariness
  in choosing $\eta $, $k_h$ cannot be defined precisely, but has a small span,
  which decreases with increasing $N$.}\BibitemShut {Stop}%
\bibitem [{\citenamefont {Zhu}\ \emph {et~al.}(2018)\citenamefont {Zhu},
  \citenamefont {Liu}, \citenamefont {Yu}, \citenamefont {Wang}, \citenamefont
  {Zhao}, \citenamefont {Feng}, \citenamefont {Sheng},\ and\ \citenamefont
  {Yang}}]{zhu2018quadratic}%
  \BibitemOpen
  \bibfield  {author} {\bibinfo {author} {\bibfnamefont {Z.}~\bibnamefont
  {Zhu}}, \bibinfo {author} {\bibfnamefont {Y.}~\bibnamefont {Liu}}, \bibinfo
  {author} {\bibfnamefont {Z.-M.}\ \bibnamefont {Yu}}, \bibinfo {author}
  {\bibfnamefont {S.-S.}\ \bibnamefont {Wang}}, \bibinfo {author}
  {\bibfnamefont {Y.~X.}\ \bibnamefont {Zhao}}, \bibinfo {author}
  {\bibfnamefont {Y.}~\bibnamefont {Feng}}, \bibinfo {author} {\bibfnamefont
  {X.-L.}\ \bibnamefont {Sheng}},\ and\ \bibinfo {author} {\bibfnamefont
  {S.~A.}\ \bibnamefont {Yang}},\ }\href
  {https://doi.org/10.1103/PhysRevB.98.125104} {\bibfield  {journal} {\bibinfo
  {journal} {Phys. Rev. B}\ }\textbf {\bibinfo {volume} {98}},\ \bibinfo
  {pages} {125104} (\bibinfo {year} {2018})}\BibitemShut {NoStop}%
\bibitem [{\citenamefont {Gao}\ \emph {et~al.}(2023)\citenamefont {Gao},
  \citenamefont {Zhao}, \citenamefont {Wu}, \citenamefont {Feng}, \citenamefont
  {Zhang}, \citenamefont {Qiao}, \citenamefont {Chiu},\ and\ \citenamefont
  {Feng}}]{gao2023topological}%
  \BibitemOpen
  \bibfield  {author} {\bibinfo {author} {\bibfnamefont {Z.}~\bibnamefont
  {Gao}}, \bibinfo {author} {\bibfnamefont {H.}~\bibnamefont {Zhao}}, \bibinfo
  {author} {\bibfnamefont {T.}~\bibnamefont {Wu}}, \bibinfo {author}
  {\bibfnamefont {X.}~\bibnamefont {Feng}}, \bibinfo {author} {\bibfnamefont
  {Z.}~\bibnamefont {Zhang}}, \bibinfo {author} {\bibfnamefont
  {X.}~\bibnamefont {Qiao}}, \bibinfo {author} {\bibfnamefont {C.-K.}\
  \bibnamefont {Chiu}},\ and\ \bibinfo {author} {\bibfnamefont
  {L.}~\bibnamefont {Feng}},\ }\href
  {https://doi.org/10.1038/s41467-023-38861-3} {\bibfield  {journal} {\bibinfo
  {journal} {Nature Communications}\ }\textbf {\bibinfo {volume} {14}},\
  \bibinfo {pages} {3206} (\bibinfo {year} {2023})}\BibitemShut {NoStop}%
\bibitem [{\citenamefont {Mera}\ and\ \citenamefont
  {Ozawa}(2021)}]{mera2021kahler}%
  \BibitemOpen
  \bibfield  {author} {\bibinfo {author} {\bibfnamefont {B.}~\bibnamefont
  {Mera}}\ and\ \bibinfo {author} {\bibfnamefont {T.}~\bibnamefont {Ozawa}},\
  }\href {https://doi.org/10.1103/PhysRevB.104.045104} {\bibfield  {journal}
  {\bibinfo  {journal} {Phys. Rev. B}\ }\textbf {\bibinfo {volume} {104}},\
  \bibinfo {pages} {045104} (\bibinfo {year} {2021})}\BibitemShut {NoStop}%
\bibitem [{\citenamefont {Nissinen}\ \emph {et~al.}(2021)\citenamefont
  {Nissinen}, \citenamefont {Heikkil\"a},\ and\ \citenamefont
  {Volovik}}]{nissinen2021topological}%
  \BibitemOpen
  \bibfield  {author} {\bibinfo {author} {\bibfnamefont {J.}~\bibnamefont
  {Nissinen}}, \bibinfo {author} {\bibfnamefont {T.~T.}\ \bibnamefont
  {Heikkil\"a}},\ and\ \bibinfo {author} {\bibfnamefont {G.~E.}\ \bibnamefont
  {Volovik}},\ }\href {https://doi.org/10.1103/PhysRevB.103.245115} {\bibfield
  {journal} {\bibinfo  {journal} {Phys. Rev. B}\ }\textbf {\bibinfo {volume}
  {103}},\ \bibinfo {pages} {245115} (\bibinfo {year} {2021})}\BibitemShut
  {NoStop}%
\bibitem [{\citenamefont {Cea}\ \emph {et~al.}(2022)\citenamefont {Cea},
  \citenamefont {Pantale\'on}, \citenamefont {Phong},\ and\ \citenamefont
  {Guinea}}]{cea2022superconductivity}%
  \BibitemOpen
  \bibfield  {author} {\bibinfo {author} {\bibfnamefont {T.}~\bibnamefont
  {Cea}}, \bibinfo {author} {\bibfnamefont {P.~A.}\ \bibnamefont
  {Pantale\'on}}, \bibinfo {author} {\bibfnamefont {V.~o.~T.}\ \bibnamefont
  {Phong}},\ and\ \bibinfo {author} {\bibfnamefont {F.}~\bibnamefont
  {Guinea}},\ }\href {https://doi.org/10.1103/PhysRevB.105.075432} {\bibfield
  {journal} {\bibinfo  {journal} {Phys. Rev. B}\ }\textbf {\bibinfo {volume}
  {105}},\ \bibinfo {pages} {075432} (\bibinfo {year} {2022})}\BibitemShut
  {NoStop}%
\bibitem [{\citenamefont {Pantale{\'o}n}\ \emph {et~al.}(2023)\citenamefont
  {Pantale{\'o}n}, \citenamefont {Jimeno-Pozo}, \citenamefont {Sainz-Cruz},
  \citenamefont {Phong}, \citenamefont {Cea},\ and\ \citenamefont
  {Guinea}}]{pantaleon2023review}%
  \BibitemOpen
  \bibfield  {author} {\bibinfo {author} {\bibfnamefont {P.~A.}\ \bibnamefont
  {Pantale{\'o}n}}, \bibinfo {author} {\bibfnamefont {A.}~\bibnamefont
  {Jimeno-Pozo}}, \bibinfo {author} {\bibfnamefont {H.}~\bibnamefont
  {Sainz-Cruz}}, \bibinfo {author} {\bibfnamefont {V.~T.}\ \bibnamefont
  {Phong}}, \bibinfo {author} {\bibfnamefont {T.}~\bibnamefont {Cea}},\ and\
  \bibinfo {author} {\bibfnamefont {F.}~\bibnamefont {Guinea}},\ }\href
  {https://doi.org/10.1038/s42254-023-00575-2} {\bibfield  {journal} {\bibinfo
  {journal} {Nature Reviews Physics}\ }\textbf {\bibinfo {volume} {5}},\
  \bibinfo {pages} {304} (\bibinfo {year} {2023})}\BibitemShut {NoStop}%
\bibitem [{\citenamefont {Scalapino}\ \emph {et~al.}(1992)\citenamefont
  {Scalapino}, \citenamefont {White},\ and\ \citenamefont
  {Zhang}}]{scalapino1992superfluid}%
  \BibitemOpen
  \bibfield  {author} {\bibinfo {author} {\bibfnamefont {D.~J.}\ \bibnamefont
  {Scalapino}}, \bibinfo {author} {\bibfnamefont {S.~R.}\ \bibnamefont
  {White}},\ and\ \bibinfo {author} {\bibfnamefont {S.~C.}\ \bibnamefont
  {Zhang}},\ }\href {https://doi.org/10.1103/PhysRevLett.68.2830} {\bibfield
  {journal} {\bibinfo  {journal} {Phys. Rev. Lett.}\ }\textbf {\bibinfo
  {volume} {68}},\ \bibinfo {pages} {2830} (\bibinfo {year}
  {1992})}\BibitemShut {NoStop}%
\bibitem [{\citenamefont {Scalapino}\ \emph {et~al.}(1993)\citenamefont
  {Scalapino}, \citenamefont {White},\ and\ \citenamefont
  {Zhang}}]{scalapino1993insulator}%
  \BibitemOpen
  \bibfield  {author} {\bibinfo {author} {\bibfnamefont {D.~J.}\ \bibnamefont
  {Scalapino}}, \bibinfo {author} {\bibfnamefont {S.~R.}\ \bibnamefont
  {White}},\ and\ \bibinfo {author} {\bibfnamefont {S.}~\bibnamefont {Zhang}},\
  }\href {https://doi.org/10.1103/PhysRevB.47.7995} {\bibfield  {journal}
  {\bibinfo  {journal} {Phys. Rev. B}\ }\textbf {\bibinfo {volume} {47}},\
  \bibinfo {pages} {7995} (\bibinfo {year} {1993})}\BibitemShut {NoStop}%
\bibitem [{\citenamefont {Julku}\ \emph {et~al.}(2016)\citenamefont {Julku},
  \citenamefont {Peotta}, \citenamefont {Vanhala}, \citenamefont {Kim},\ and\
  \citenamefont {T\"orm\"a}}]{julku2016geometric}%
  \BibitemOpen
  \bibfield  {author} {\bibinfo {author} {\bibfnamefont {A.}~\bibnamefont
  {Julku}}, \bibinfo {author} {\bibfnamefont {S.}~\bibnamefont {Peotta}},
  \bibinfo {author} {\bibfnamefont {T.~I.}\ \bibnamefont {Vanhala}}, \bibinfo
  {author} {\bibfnamefont {D.-H.}\ \bibnamefont {Kim}},\ and\ \bibinfo {author}
  {\bibfnamefont {P.}~\bibnamefont {T\"orm\"a}},\ }\href
  {https://doi.org/10.1103/PhysRevLett.117.045303} {\bibfield  {journal}
  {\bibinfo  {journal} {Phys. Rev. Lett.}\ }\textbf {\bibinfo {volume} {117}},\
  \bibinfo {pages} {045303} (\bibinfo {year} {2016})}\BibitemShut {NoStop}%
\bibitem [{\citenamefont {Liang}\ \emph {et~al.}(2017)\citenamefont {Liang},
  \citenamefont {Vanhala}, \citenamefont {Peotta}, \citenamefont {Siro},
  \citenamefont {Harju},\ and\ \citenamefont {T\"orm\"a}}]{liang2017band}%
  \BibitemOpen
  \bibfield  {author} {\bibinfo {author} {\bibfnamefont {L.}~\bibnamefont
  {Liang}}, \bibinfo {author} {\bibfnamefont {T.~I.}\ \bibnamefont {Vanhala}},
  \bibinfo {author} {\bibfnamefont {S.}~\bibnamefont {Peotta}}, \bibinfo
  {author} {\bibfnamefont {T.}~\bibnamefont {Siro}}, \bibinfo {author}
  {\bibfnamefont {A.}~\bibnamefont {Harju}},\ and\ \bibinfo {author}
  {\bibfnamefont {P.}~\bibnamefont {T\"orm\"a}},\ }\href
  {https://doi.org/10.1103/PhysRevB.95.024515} {\bibfield  {journal} {\bibinfo
  {journal} {Phys. Rev. B}\ }\textbf {\bibinfo {volume} {95}},\ \bibinfo
  {pages} {024515} (\bibinfo {year} {2017})}\BibitemShut {NoStop}%
\bibitem [{\citenamefont {Huhtinen}\ \emph {et~al.}(2022)\citenamefont
  {Huhtinen}, \citenamefont {Herzog-Arbeitman}, \citenamefont {Chew},
  \citenamefont {Bernevig},\ and\ \citenamefont
  {T\"orm\"a}}]{huhtinen2022revisiting}%
  \BibitemOpen
  \bibfield  {author} {\bibinfo {author} {\bibfnamefont {K.-E.}\ \bibnamefont
  {Huhtinen}}, \bibinfo {author} {\bibfnamefont {J.}~\bibnamefont
  {Herzog-Arbeitman}}, \bibinfo {author} {\bibfnamefont {A.}~\bibnamefont
  {Chew}}, \bibinfo {author} {\bibfnamefont {B.~A.}\ \bibnamefont {Bernevig}},\
  and\ \bibinfo {author} {\bibfnamefont {P.}~\bibnamefont {T\"orm\"a}},\ }\href
  {https://doi.org/10.1103/PhysRevB.106.014518} {\bibfield  {journal} {\bibinfo
   {journal} {Phys. Rev. B}\ }\textbf {\bibinfo {volume} {106}},\ \bibinfo
  {pages} {014518} (\bibinfo {year} {2022})}\BibitemShut {NoStop}%
\bibitem [{\citenamefont {Jiang}\ and\ \citenamefont
  {Barlas}(2023)}]{jiang2023pdw}%
  \BibitemOpen
  \bibfield  {author} {\bibinfo {author} {\bibfnamefont {G.}~\bibnamefont
  {Jiang}}\ and\ \bibinfo {author} {\bibfnamefont {Y.}~\bibnamefont {Barlas}},\
  }\href {https://doi.org/10.1103/PhysRevLett.131.016002} {\bibfield  {journal}
  {\bibinfo  {journal} {Phys. Rev. Lett.}\ }\textbf {\bibinfo {volume} {131}},\
  \bibinfo {pages} {016002} (\bibinfo {year} {2023})}\BibitemShut {NoStop}%
\bibitem [{\citenamefont {Jiang}\ and\ \citenamefont
  {Barlas}(2024)}]{jiang2024geometric}%
  \BibitemOpen
  \bibfield  {author} {\bibinfo {author} {\bibfnamefont {G.}~\bibnamefont
  {Jiang}}\ and\ \bibinfo {author} {\bibfnamefont {Y.}~\bibnamefont {Barlas}},\
  }\href {https://doi.org/10.1103/PhysRevB.109.214518} {\bibfield  {journal}
  {\bibinfo  {journal} {Phys. Rev. B}\ }\textbf {\bibinfo {volume} {109}},\
  \bibinfo {pages} {214518} (\bibinfo {year} {2024})}\BibitemShut {NoStop}%
\bibitem [{\citenamefont {Daido}\ \emph {et~al.}(2024)\citenamefont {Daido},
  \citenamefont {Kitamura},\ and\ \citenamefont {Yanase}}]{daido2024quantum}%
  \BibitemOpen
  \bibfield  {author} {\bibinfo {author} {\bibfnamefont {A.}~\bibnamefont
  {Daido}}, \bibinfo {author} {\bibfnamefont {T.}~\bibnamefont {Kitamura}},\
  and\ \bibinfo {author} {\bibfnamefont {Y.}~\bibnamefont {Yanase}},\ }\href
  {https://doi.org/10.1103/PhysRevB.110.094505} {\bibfield  {journal} {\bibinfo
   {journal} {Phys. Rev. B}\ }\textbf {\bibinfo {volume} {110}},\ \bibinfo
  {pages} {094505} (\bibinfo {year} {2024})}\BibitemShut {NoStop}%
\bibitem [{\citenamefont {Han}\ \emph {et~al.}(2024{\natexlab{b}})\citenamefont
  {Han}, \citenamefont {Herzog-Arbeitman}, \citenamefont {Bernevig},\ and\
  \citenamefont {Kivelson}}]{han2024qgn}%
  \BibitemOpen
  \bibfield  {author} {\bibinfo {author} {\bibfnamefont {Z.}~\bibnamefont
  {Han}}, \bibinfo {author} {\bibfnamefont {J.}~\bibnamefont
  {Herzog-Arbeitman}}, \bibinfo {author} {\bibfnamefont {B.~A.}\ \bibnamefont
  {Bernevig}},\ and\ \bibinfo {author} {\bibfnamefont {S.~A.}\ \bibnamefont
  {Kivelson}},\ }\href {https://doi.org/10.1103/PhysRevX.14.041004} {\bibfield
  {journal} {\bibinfo  {journal} {Phys. Rev. X}\ }\textbf {\bibinfo {volume}
  {14}},\ \bibinfo {pages} {041004} (\bibinfo {year}
  {2024}{\natexlab{b}})}\BibitemShut {NoStop}%
\bibitem [{\citenamefont {Zhang}\ \emph {et~al.}(2025)\citenamefont {Zhang},
  \citenamefont {Wang}, \citenamefont {Balents},\ and\ \citenamefont
  {Savary}}]{zhang2025identifying}%
  \BibitemOpen
  \bibfield  {author} {\bibinfo {author} {\bibfnamefont {J.-X.}\ \bibnamefont
  {Zhang}}, \bibinfo {author} {\bibfnamefont {W.~O.}\ \bibnamefont {Wang}},
  \bibinfo {author} {\bibfnamefont {L.}~\bibnamefont {Balents}},\ and\ \bibinfo
  {author} {\bibfnamefont {L.}~\bibnamefont {Savary}},\ }\href@noop {}
  {\bibfield  {journal} {\bibinfo  {journal} {arXiv preprint arXiv:2504.03882}\
  } (\bibinfo {year} {2025})}\BibitemShut {NoStop}%
\bibitem [{Note3()}]{Note3}%
  \BibitemOpen
  \bibinfo {note} {The order parameter used, $\Delta _{20\protect \text
  {B}}=0.02\gamma _1$, is one order-of-magnitude larger than that is estimated
  for 20-layer RG using $U\sim 40$ meV, which is extrapolated from SC1 of
  RTG.}\BibitemShut {Stop}%
\bibitem [{\citenamefont {Steglich}\ \emph {et~al.}(1979)\citenamefont
  {Steglich}, \citenamefont {Aarts}, \citenamefont {Bredl}, \citenamefont
  {Lieke}, \citenamefont {Meschede}, \citenamefont {Franz},\ and\ \citenamefont
  {Sch\"afer}}]{steglich1979sc}%
  \BibitemOpen
  \bibfield  {author} {\bibinfo {author} {\bibfnamefont {F.}~\bibnamefont
  {Steglich}}, \bibinfo {author} {\bibfnamefont {J.}~\bibnamefont {Aarts}},
  \bibinfo {author} {\bibfnamefont {C.~D.}\ \bibnamefont {Bredl}}, \bibinfo
  {author} {\bibfnamefont {W.}~\bibnamefont {Lieke}}, \bibinfo {author}
  {\bibfnamefont {D.}~\bibnamefont {Meschede}}, \bibinfo {author}
  {\bibfnamefont {W.}~\bibnamefont {Franz}},\ and\ \bibinfo {author}
  {\bibfnamefont {H.}~\bibnamefont {Sch\"afer}},\ }\href
  {https://doi.org/10.1103/PhysRevLett.43.1892} {\bibfield  {journal} {\bibinfo
   {journal} {Phys. Rev. Lett.}\ }\textbf {\bibinfo {volume} {43}},\ \bibinfo
  {pages} {1892} (\bibinfo {year} {1979})}\BibitemShut {NoStop}%
\bibitem [{\citenamefont {Sauls}(1994)}]{sauls1994order}%
  \BibitemOpen
  \bibfield  {author} {\bibinfo {author} {\bibfnamefont {J.~A.}\ \bibnamefont
  {Sauls}},\ }\href {https://doi.org/10.1080/00018739400101475} {\bibfield
  {journal} {\bibinfo  {journal} {Advances in Physics}\ }\textbf {\bibinfo
  {volume} {43}},\ \bibinfo {pages} {113} (\bibinfo {year} {1994})}\BibitemShut
  {NoStop}%
\bibitem [{\citenamefont {Heffner}\ and\ \citenamefont
  {Norman}(1995)}]{heffner1995heavy}%
  \BibitemOpen
  \bibfield  {author} {\bibinfo {author} {\bibfnamefont {R.~H.}\ \bibnamefont
  {Heffner}}\ and\ \bibinfo {author} {\bibfnamefont {M.~R.}\ \bibnamefont
  {Norman}},\ }\href {https://arxiv.org/abs/cond-mat/9506043} {\bibfield
  {journal} {\bibinfo  {journal} {arXiv preprint cond-mat/9506043}\ } (\bibinfo
  {year} {1995})}\BibitemShut {NoStop}%
\bibitem [{\citenamefont {Pfleiderer}(2009)}]{pfleiderer2009superconducting}%
  \BibitemOpen
  \bibfield  {author} {\bibinfo {author} {\bibfnamefont {C.}~\bibnamefont
  {Pfleiderer}},\ }\href {https://doi.org/10.1103/RevModPhys.81.1551}
  {\bibfield  {journal} {\bibinfo  {journal} {Rev. Mod. Phys.}\ }\textbf
  {\bibinfo {volume} {81}},\ \bibinfo {pages} {1551} (\bibinfo {year}
  {2009})}\BibitemShut {NoStop}%
\bibitem [{\citenamefont {Song}\ and\ \citenamefont
  {Bernevig}(2022)}]{song2022magic}%
  \BibitemOpen
  \bibfield  {author} {\bibinfo {author} {\bibfnamefont {Z.-D.}\ \bibnamefont
  {Song}}\ and\ \bibinfo {author} {\bibfnamefont {B.~A.}\ \bibnamefont
  {Bernevig}},\ }\href {https://doi.org/10.1103/PhysRevLett.129.047601}
  {\bibfield  {journal} {\bibinfo  {journal} {Phys. Rev. Lett.}\ }\textbf
  {\bibinfo {volume} {129}},\ \bibinfo {pages} {047601} (\bibinfo {year}
  {2022})}\BibitemShut {NoStop}%
\bibitem [{\citenamefont {Shi}\ and\ \citenamefont {Dai}(2022)}]{shi2022heavy}%
  \BibitemOpen
  \bibfield  {author} {\bibinfo {author} {\bibfnamefont {H.}~\bibnamefont
  {Shi}}\ and\ \bibinfo {author} {\bibfnamefont {X.}~\bibnamefont {Dai}},\
  }\href {https://doi.org/10.1103/PhysRevB.106.245129} {\bibfield  {journal}
  {\bibinfo  {journal} {Phys. Rev. B}\ }\textbf {\bibinfo {volume} {106}},\
  \bibinfo {pages} {245129} (\bibinfo {year} {2022})}\BibitemShut {NoStop}%
\bibitem [{\citenamefont {Herzog-Arbeitman}\ \emph
  {et~al.}(2024{\natexlab{a}})\citenamefont {Herzog-Arbeitman}, \citenamefont
  {Yu}, \citenamefont {C{\u{a}}lug{\u{a}}ru}, \citenamefont {Hu}, \citenamefont
  {Regnault}, \citenamefont {Liu}, \citenamefont {Sarma}, \citenamefont
  {Vafek}, \citenamefont {Coleman}, \citenamefont {Tsvelik} \emph
  {et~al.}}]{herzog2024topological}%
  \BibitemOpen
  \bibfield  {author} {\bibinfo {author} {\bibfnamefont {J.}~\bibnamefont
  {Herzog-Arbeitman}}, \bibinfo {author} {\bibfnamefont {J.}~\bibnamefont
  {Yu}}, \bibinfo {author} {\bibfnamefont {D.}~\bibnamefont
  {C{\u{a}}lug{\u{a}}ru}}, \bibinfo {author} {\bibfnamefont {H.}~\bibnamefont
  {Hu}}, \bibinfo {author} {\bibfnamefont {N.}~\bibnamefont {Regnault}},
  \bibinfo {author} {\bibfnamefont {C.}~\bibnamefont {Liu}}, \bibinfo {author}
  {\bibfnamefont {S.~D.}\ \bibnamefont {Sarma}}, \bibinfo {author}
  {\bibfnamefont {O.}~\bibnamefont {Vafek}}, \bibinfo {author} {\bibfnamefont
  {P.}~\bibnamefont {Coleman}}, \bibinfo {author} {\bibfnamefont
  {A.}~\bibnamefont {Tsvelik}}, \emph {et~al.},\ }\href
  {https://arxiv.org/abs/2404.07253} {\bibfield  {journal} {\bibinfo  {journal}
  {arXiv preprint arXiv:2404.07253}\ } (\bibinfo {year}
  {2024}{\natexlab{a}})}\BibitemShut {NoStop}%
\bibitem [{\citenamefont {Bistritzer}\ and\ \citenamefont
  {MacDonald}(2011)}]{bistritzer2011moire}%
  \BibitemOpen
  \bibfield  {author} {\bibinfo {author} {\bibfnamefont {R.}~\bibnamefont
  {Bistritzer}}\ and\ \bibinfo {author} {\bibfnamefont {A.~H.}\ \bibnamefont
  {MacDonald}},\ }\href {https://doi.org/10.1073/pnas.1108174108} {\bibfield
  {journal} {\bibinfo  {journal} {Proceedings of the National Academy of
  Sciences}\ }\textbf {\bibinfo {volume} {108}},\ \bibinfo {pages} {12233}
  (\bibinfo {year} {2011})}\BibitemShut {NoStop}%
\bibitem [{\citenamefont {Bernevig}\ \emph {et~al.}(2021)\citenamefont
  {Bernevig}, \citenamefont {Song}, \citenamefont {Regnault},\ and\
  \citenamefont {Lian}}]{bernevig2021tbg}%
  \BibitemOpen
  \bibfield  {author} {\bibinfo {author} {\bibfnamefont {B.~A.}\ \bibnamefont
  {Bernevig}}, \bibinfo {author} {\bibfnamefont {Z.-D.}\ \bibnamefont {Song}},
  \bibinfo {author} {\bibfnamefont {N.}~\bibnamefont {Regnault}},\ and\
  \bibinfo {author} {\bibfnamefont {B.}~\bibnamefont {Lian}},\ }\href
  {https://doi.org/10.1103/PhysRevB.103.205411} {\bibfield  {journal} {\bibinfo
   {journal} {Phys. Rev. B}\ }\textbf {\bibinfo {volume} {103}},\ \bibinfo
  {pages} {205411} (\bibinfo {year} {2021})}\BibitemShut {NoStop}%
\bibitem [{\citenamefont {Hu}\ \emph {et~al.}(2019)\citenamefont {Hu},
  \citenamefont {Hyart}, \citenamefont {Pikulin},\ and\ \citenamefont
  {Rossi}}]{hu2019geometric}%
  \BibitemOpen
  \bibfield  {author} {\bibinfo {author} {\bibfnamefont {X.}~\bibnamefont
  {Hu}}, \bibinfo {author} {\bibfnamefont {T.}~\bibnamefont {Hyart}}, \bibinfo
  {author} {\bibfnamefont {D.~I.}\ \bibnamefont {Pikulin}},\ and\ \bibinfo
  {author} {\bibfnamefont {E.}~\bibnamefont {Rossi}},\ }\href
  {https://doi.org/10.1103/PhysRevLett.123.237002} {\bibfield  {journal}
  {\bibinfo  {journal} {Phys. Rev. Lett.}\ }\textbf {\bibinfo {volume} {123}},\
  \bibinfo {pages} {237002} (\bibinfo {year} {2019})}\BibitemShut {NoStop}%
\bibitem [{\citenamefont {Julku}\ \emph {et~al.}(2020)\citenamefont {Julku},
  \citenamefont {Peltonen}, \citenamefont {Liang}, \citenamefont {Heikkil\"a},\
  and\ \citenamefont {T\"orm\"a}}]{julku2020superfluid}%
  \BibitemOpen
  \bibfield  {author} {\bibinfo {author} {\bibfnamefont {A.}~\bibnamefont
  {Julku}}, \bibinfo {author} {\bibfnamefont {T.~J.}\ \bibnamefont {Peltonen}},
  \bibinfo {author} {\bibfnamefont {L.}~\bibnamefont {Liang}}, \bibinfo
  {author} {\bibfnamefont {T.~T.}\ \bibnamefont {Heikkil\"a}},\ and\ \bibinfo
  {author} {\bibfnamefont {P.}~\bibnamefont {T\"orm\"a}},\ }\href
  {https://doi.org/10.1103/PhysRevB.101.060505} {\bibfield  {journal} {\bibinfo
   {journal} {Phys. Rev. B}\ }\textbf {\bibinfo {volume} {101}},\ \bibinfo
  {pages} {060505} (\bibinfo {year} {2020})}\BibitemShut {NoStop}%
\bibitem [{\citenamefont {Xie}\ \emph {et~al.}(2020)\citenamefont {Xie},
  \citenamefont {Song}, \citenamefont {Lian},\ and\ \citenamefont
  {Bernevig}}]{xie2020topology}%
  \BibitemOpen
  \bibfield  {author} {\bibinfo {author} {\bibfnamefont {F.}~\bibnamefont
  {Xie}}, \bibinfo {author} {\bibfnamefont {Z.}~\bibnamefont {Song}}, \bibinfo
  {author} {\bibfnamefont {B.}~\bibnamefont {Lian}},\ and\ \bibinfo {author}
  {\bibfnamefont {B.~A.}\ \bibnamefont {Bernevig}},\ }\href
  {https://doi.org/10.1103/PhysRevLett.124.167002} {\bibfield  {journal}
  {\bibinfo  {journal} {Phys. Rev. Lett.}\ }\textbf {\bibinfo {volume} {124}},\
  \bibinfo {pages} {167002} (\bibinfo {year} {2020})}\BibitemShut {NoStop}%
\bibitem [{\citenamefont {Dong}\ \emph
  {et~al.}(2024{\natexlab{a}})\citenamefont {Dong}, \citenamefont {Wang},
  \citenamefont {Wang}, \citenamefont {Soejima}, \citenamefont {Zaletel},
  \citenamefont {Vishwanath},\ and\ \citenamefont
  {Parker}}]{dong2024anomalous}%
  \BibitemOpen
  \bibfield  {author} {\bibinfo {author} {\bibfnamefont {J.}~\bibnamefont
  {Dong}}, \bibinfo {author} {\bibfnamefont {T.}~\bibnamefont {Wang}}, \bibinfo
  {author} {\bibfnamefont {T.}~\bibnamefont {Wang}}, \bibinfo {author}
  {\bibfnamefont {T.}~\bibnamefont {Soejima}}, \bibinfo {author} {\bibfnamefont
  {M.~P.}\ \bibnamefont {Zaletel}}, \bibinfo {author} {\bibfnamefont
  {A.}~\bibnamefont {Vishwanath}},\ and\ \bibinfo {author} {\bibfnamefont
  {D.~E.}\ \bibnamefont {Parker}},\ }\href
  {https://doi.org/10.1103/PhysRevLett.133.206503} {\bibfield  {journal}
  {\bibinfo  {journal} {Phys. Rev. Lett.}\ }\textbf {\bibinfo {volume} {133}},\
  \bibinfo {pages} {206503} (\bibinfo {year} {2024}{\natexlab{a}})}\BibitemShut
  {NoStop}%
\bibitem [{\citenamefont {Herzog-Arbeitman}\ \emph
  {et~al.}(2024{\natexlab{b}})\citenamefont {Herzog-Arbeitman}, \citenamefont
  {Wang}, \citenamefont {Liu}, \citenamefont {Tam}, \citenamefont {Qi},
  \citenamefont {Jia}, \citenamefont {Efetov}, \citenamefont {Vafek},
  \citenamefont {Regnault}, \citenamefont {Weng}, \citenamefont {Wu},
  \citenamefont {Bernevig},\ and\ \citenamefont {Yu}}]{herzog2024moire}%
  \BibitemOpen
  \bibfield  {author} {\bibinfo {author} {\bibfnamefont {J.}~\bibnamefont
  {Herzog-Arbeitman}}, \bibinfo {author} {\bibfnamefont {Y.}~\bibnamefont
  {Wang}}, \bibinfo {author} {\bibfnamefont {J.}~\bibnamefont {Liu}}, \bibinfo
  {author} {\bibfnamefont {P.~M.}\ \bibnamefont {Tam}}, \bibinfo {author}
  {\bibfnamefont {Z.}~\bibnamefont {Qi}}, \bibinfo {author} {\bibfnamefont
  {Y.}~\bibnamefont {Jia}}, \bibinfo {author} {\bibfnamefont {D.~K.}\
  \bibnamefont {Efetov}}, \bibinfo {author} {\bibfnamefont {O.}~\bibnamefont
  {Vafek}}, \bibinfo {author} {\bibfnamefont {N.}~\bibnamefont {Regnault}},
  \bibinfo {author} {\bibfnamefont {H.}~\bibnamefont {Weng}}, \bibinfo {author}
  {\bibfnamefont {Q.}~\bibnamefont {Wu}}, \bibinfo {author} {\bibfnamefont
  {B.~A.}\ \bibnamefont {Bernevig}},\ and\ \bibinfo {author} {\bibfnamefont
  {J.}~\bibnamefont {Yu}},\ }\href
  {https://doi.org/10.1103/PhysRevB.109.205122} {\bibfield  {journal} {\bibinfo
   {journal} {Phys. Rev. B}\ }\textbf {\bibinfo {volume} {109}},\ \bibinfo
  {pages} {205122} (\bibinfo {year} {2024}{\natexlab{b}})}\BibitemShut
  {NoStop}%
\bibitem [{\citenamefont {Dong}\ \emph
  {et~al.}(2024{\natexlab{b}})\citenamefont {Dong}, \citenamefont {Patri},\
  and\ \citenamefont {Senthil}}]{dong2024theory}%
  \BibitemOpen
  \bibfield  {author} {\bibinfo {author} {\bibfnamefont {Z.}~\bibnamefont
  {Dong}}, \bibinfo {author} {\bibfnamefont {A.~S.}\ \bibnamefont {Patri}},\
  and\ \bibinfo {author} {\bibfnamefont {T.}~\bibnamefont {Senthil}},\ }\href
  {https://doi.org/10.1103/PhysRevLett.133.206502} {\bibfield  {journal}
  {\bibinfo  {journal} {Phys. Rev. Lett.}\ }\textbf {\bibinfo {volume} {133}},\
  \bibinfo {pages} {206502} (\bibinfo {year} {2024}{\natexlab{b}})}\BibitemShut
  {NoStop}%
\bibitem [{\citenamefont {Guo}\ \emph {et~al.}(2024)\citenamefont {Guo},
  \citenamefont {Lu}, \citenamefont {Xie},\ and\ \citenamefont
  {Liu}}]{guo2024fractional}%
  \BibitemOpen
  \bibfield  {author} {\bibinfo {author} {\bibfnamefont {Z.}~\bibnamefont
  {Guo}}, \bibinfo {author} {\bibfnamefont {X.}~\bibnamefont {Lu}}, \bibinfo
  {author} {\bibfnamefont {B.}~\bibnamefont {Xie}},\ and\ \bibinfo {author}
  {\bibfnamefont {J.}~\bibnamefont {Liu}},\ }\href
  {https://doi.org/10.1103/PhysRevB.110.075109} {\bibfield  {journal} {\bibinfo
   {journal} {Phys. Rev. B}\ }\textbf {\bibinfo {volume} {110}},\ \bibinfo
  {pages} {075109} (\bibinfo {year} {2024})}\BibitemShut {NoStop}%
\bibitem [{\citenamefont {Zhou}\ \emph
  {et~al.}(2024{\natexlab{b}})\citenamefont {Zhou}, \citenamefont {Yang},\ and\
  \citenamefont {Zhang}}]{zhou2024fractional}%
  \BibitemOpen
  \bibfield  {author} {\bibinfo {author} {\bibfnamefont {B.}~\bibnamefont
  {Zhou}}, \bibinfo {author} {\bibfnamefont {H.}~\bibnamefont {Yang}},\ and\
  \bibinfo {author} {\bibfnamefont {Y.-H.}\ \bibnamefont {Zhang}},\ }\href
  {https://doi.org/10.1103/PhysRevLett.133.206504} {\bibfield  {journal}
  {\bibinfo  {journal} {Phys. Rev. Lett.}\ }\textbf {\bibinfo {volume} {133}},\
  \bibinfo {pages} {206504} (\bibinfo {year} {2024}{\natexlab{b}})}\BibitemShut
  {NoStop}%
\bibitem [{\citenamefont {Huang}\ \emph {et~al.}(2025)\citenamefont {Huang},
  \citenamefont {Das~Sarma},\ and\ \citenamefont {Li}}]{huang2025fractional}%
  \BibitemOpen
  \bibfield  {author} {\bibinfo {author} {\bibfnamefont {K.}~\bibnamefont
  {Huang}}, \bibinfo {author} {\bibfnamefont {S.}~\bibnamefont {Das~Sarma}},\
  and\ \bibinfo {author} {\bibfnamefont {X.}~\bibnamefont {Li}},\ }\href
  {https://doi.org/10.1103/PhysRevB.111.075130} {\bibfield  {journal} {\bibinfo
   {journal} {Phys. Rev. B}\ }\textbf {\bibinfo {volume} {111}},\ \bibinfo
  {pages} {075130} (\bibinfo {year} {2025})}\BibitemShut {NoStop}%
\bibitem [{\citenamefont {Otani}\ \emph {et~al.}(2010)\citenamefont {Otani},
  \citenamefont {Koshino}, \citenamefont {Takagi},\ and\ \citenamefont
  {Okada}}]{otani2010intrinsic}%
  \BibitemOpen
  \bibfield  {author} {\bibinfo {author} {\bibfnamefont {M.}~\bibnamefont
  {Otani}}, \bibinfo {author} {\bibfnamefont {M.}~\bibnamefont {Koshino}},
  \bibinfo {author} {\bibfnamefont {Y.}~\bibnamefont {Takagi}},\ and\ \bibinfo
  {author} {\bibfnamefont {S.}~\bibnamefont {Okada}},\ }\href
  {https://doi.org/10.1103/PhysRevB.81.161403} {\bibfield  {journal} {\bibinfo
  {journal} {Phys. Rev. B}\ }\textbf {\bibinfo {volume} {81}},\ \bibinfo
  {pages} {161403} (\bibinfo {year} {2010})}\BibitemShut {NoStop}%
\bibitem [{\citenamefont {Cuong}\ \emph {et~al.}(2012)\citenamefont {Cuong},
  \citenamefont {Otani},\ and\ \citenamefont {Okada}}]{cuong2012magnetic}%
  \BibitemOpen
  \bibfield  {author} {\bibinfo {author} {\bibfnamefont {N.~T.}\ \bibnamefont
  {Cuong}}, \bibinfo {author} {\bibfnamefont {M.}~\bibnamefont {Otani}},\ and\
  \bibinfo {author} {\bibfnamefont {S.}~\bibnamefont {Okada}},\ }\href
  {https://doi.org/https://doi.org/10.1016/j.susc.2011.10.001} {\bibfield
  {journal} {\bibinfo  {journal} {Surface Science}\ }\textbf {\bibinfo {volume}
  {606}},\ \bibinfo {pages} {253} (\bibinfo {year} {2012})}\BibitemShut
  {NoStop}%
\bibitem [{\citenamefont {Xu}\ \emph {et~al.}(2012)\citenamefont {Xu},
  \citenamefont {Yuan}, \citenamefont {Yao}, \citenamefont {Zhou},
  \citenamefont {Gao},\ and\ \citenamefont {Zhang}}]{xu2012stacking}%
  \BibitemOpen
  \bibfield  {author} {\bibinfo {author} {\bibfnamefont {D.-H.}\ \bibnamefont
  {Xu}}, \bibinfo {author} {\bibfnamefont {J.}~\bibnamefont {Yuan}}, \bibinfo
  {author} {\bibfnamefont {Z.-J.}\ \bibnamefont {Yao}}, \bibinfo {author}
  {\bibfnamefont {Y.}~\bibnamefont {Zhou}}, \bibinfo {author} {\bibfnamefont
  {J.-H.}\ \bibnamefont {Gao}},\ and\ \bibinfo {author} {\bibfnamefont {F.-C.}\
  \bibnamefont {Zhang}},\ }\href {https://doi.org/10.1103/PhysRevB.86.201404}
  {\bibfield  {journal} {\bibinfo  {journal} {Phys. Rev. B}\ }\textbf {\bibinfo
  {volume} {86}},\ \bibinfo {pages} {201404} (\bibinfo {year}
  {2012})}\BibitemShut {NoStop}%
\bibitem [{\citenamefont {Olsen}\ \emph {et~al.}(2013)\citenamefont {Olsen},
  \citenamefont {van Gelderen},\ and\ \citenamefont {Smith}}]{olsen2013ferro}%
  \BibitemOpen
  \bibfield  {author} {\bibinfo {author} {\bibfnamefont {R.}~\bibnamefont
  {Olsen}}, \bibinfo {author} {\bibfnamefont {R.}~\bibnamefont {van
  Gelderen}},\ and\ \bibinfo {author} {\bibfnamefont {C.~M.}\ \bibnamefont
  {Smith}},\ }\href {https://doi.org/10.1103/PhysRevB.87.115414} {\bibfield
  {journal} {\bibinfo  {journal} {Phys. Rev. B}\ }\textbf {\bibinfo {volume}
  {87}},\ \bibinfo {pages} {115414} (\bibinfo {year} {2013})}\BibitemShut
  {NoStop}%
\bibitem [{\citenamefont {Awoga}\ \emph {et~al.}(2023)\citenamefont {Awoga},
  \citenamefont {L\"othman},\ and\ \citenamefont
  {Black-Schaffer}}]{awoga2023superconductivity}%
  \BibitemOpen
  \bibfield  {author} {\bibinfo {author} {\bibfnamefont {O.~A.}\ \bibnamefont
  {Awoga}}, \bibinfo {author} {\bibfnamefont {T.}~\bibnamefont {L\"othman}},\
  and\ \bibinfo {author} {\bibfnamefont {A.~M.}\ \bibnamefont
  {Black-Schaffer}},\ }\href {https://doi.org/10.1103/PhysRevB.108.144504}
  {\bibfield  {journal} {\bibinfo  {journal} {Phys. Rev. B}\ }\textbf {\bibinfo
  {volume} {108}},\ \bibinfo {pages} {144504} (\bibinfo {year}
  {2023})}\BibitemShut {NoStop}%
\bibitem [{\citenamefont {Wu}\ and\ \citenamefont
  {Das~Sarma}(2020)}]{wu2020quantum}%
  \BibitemOpen
  \bibfield  {author} {\bibinfo {author} {\bibfnamefont {F.}~\bibnamefont
  {Wu}}\ and\ \bibinfo {author} {\bibfnamefont {S.}~\bibnamefont {Das~Sarma}},\
  }\href {https://doi.org/10.1103/PhysRevB.102.165118} {\bibfield  {journal}
  {\bibinfo  {journal} {Phys. Rev. B}\ }\textbf {\bibinfo {volume} {102}},\
  \bibinfo {pages} {165118} (\bibinfo {year} {2020})}\BibitemShut {NoStop}%
\bibitem [{\citenamefont {Kang}\ \emph {et~al.}(2024)\citenamefont {Kang},
  \citenamefont {Oh}, \citenamefont {Lee},\ and\ \citenamefont
  {Yang}}]{kang2024quantum}%
  \BibitemOpen
  \bibfield  {author} {\bibinfo {author} {\bibfnamefont {J.}~\bibnamefont
  {Kang}}, \bibinfo {author} {\bibfnamefont {T.}~\bibnamefont {Oh}}, \bibinfo
  {author} {\bibfnamefont {J.}~\bibnamefont {Lee}},\ and\ \bibinfo {author}
  {\bibfnamefont {B.-J.}\ \bibnamefont {Yang}},\ }\href@noop {} {\bibfield
  {journal} {\bibinfo  {journal} {arXiv preprint arXiv:2402.07171}\ } (\bibinfo
  {year} {2024})}\BibitemShut {NoStop}%
\bibitem [{\citenamefont {McCann}\ and\ \citenamefont
  {Fal'ko}(2006)}]{mccann2006landau}%
  \BibitemOpen
  \bibfield  {author} {\bibinfo {author} {\bibfnamefont {E.}~\bibnamefont
  {McCann}}\ and\ \bibinfo {author} {\bibfnamefont {V.~I.}\ \bibnamefont
  {Fal'ko}},\ }\href {https://doi.org/10.1103/PhysRevLett.96.086805} {\bibfield
   {journal} {\bibinfo  {journal} {Phys. Rev. Lett.}\ }\textbf {\bibinfo
  {volume} {96}},\ \bibinfo {pages} {086805} (\bibinfo {year}
  {2006})}\BibitemShut {NoStop}%
\bibitem [{\citenamefont {Min}\ and\ \citenamefont
  {MacDonald}(2008)}]{min2008chiral}%
  \BibitemOpen
  \bibfield  {author} {\bibinfo {author} {\bibfnamefont {H.}~\bibnamefont
  {Min}}\ and\ \bibinfo {author} {\bibfnamefont {A.~H.}\ \bibnamefont
  {MacDonald}},\ }\href {https://doi.org/10.1103/PhysRevB.77.155416} {\bibfield
   {journal} {\bibinfo  {journal} {Phys. Rev. B}\ }\textbf {\bibinfo {volume}
  {77}},\ \bibinfo {pages} {155416} (\bibinfo {year} {2008})}\BibitemShut
  {NoStop}%
\bibitem [{\citenamefont {Guinea}\ \emph {et~al.}(2006)\citenamefont {Guinea},
  \citenamefont {Castro~Neto},\ and\ \citenamefont
  {Peres}}]{guinea2006electronic}%
  \BibitemOpen
  \bibfield  {author} {\bibinfo {author} {\bibfnamefont {F.}~\bibnamefont
  {Guinea}}, \bibinfo {author} {\bibfnamefont {A.~H.}\ \bibnamefont
  {Castro~Neto}},\ and\ \bibinfo {author} {\bibfnamefont {N.~M.~R.}\
  \bibnamefont {Peres}},\ }\href {https://doi.org/10.1103/PhysRevB.73.245426}
  {\bibfield  {journal} {\bibinfo  {journal} {Phys. Rev. B}\ }\textbf {\bibinfo
  {volume} {73}},\ \bibinfo {pages} {245426} (\bibinfo {year}
  {2006})}\BibitemShut {NoStop}%
\bibitem [{\citenamefont {Hu}\ \emph {et~al.}(2023)\citenamefont {Hu},
  \citenamefont {Rossi},\ and\ \citenamefont {Barlas}}]{hu2023effect}%
  \BibitemOpen
  \bibfield  {author} {\bibinfo {author} {\bibfnamefont {X.}~\bibnamefont
  {Hu}}, \bibinfo {author} {\bibfnamefont {E.}~\bibnamefont {Rossi}},\ and\
  \bibinfo {author} {\bibfnamefont {Y.}~\bibnamefont {Barlas}},\ }\href
  {https://arxiv.org/abs/2304.04825} {\bibfield  {journal} {\bibinfo  {journal}
  {arXiv preprint arXiv:2304.04825}\ } (\bibinfo {year} {2023})}\BibitemShut
  {NoStop}%
\bibitem [{\citenamefont {Jiang}\ \emph {et~al.}(2025)\citenamefont {Jiang},
  \citenamefont {T{\"o}rm{\"a}},\ and\ \citenamefont
  {Barlas}}]{jiang2025superfluid}%
  \BibitemOpen
  \bibfield  {author} {\bibinfo {author} {\bibfnamefont {G.}~\bibnamefont
  {Jiang}}, \bibinfo {author} {\bibfnamefont {P.}~\bibnamefont
  {T{\"o}rm{\"a}}},\ and\ \bibinfo {author} {\bibfnamefont {Y.}~\bibnamefont
  {Barlas}},\ }\href {https://doi.org/10.1073/pnas.2416726122} {\bibfield
  {journal} {\bibinfo  {journal} {Proceedings of the National Academy of
  Sciences}\ }\textbf {\bibinfo {volume} {122}},\ \bibinfo {pages}
  {e2416726122} (\bibinfo {year} {2025})}\BibitemShut {NoStop}%
\bibitem [{\citenamefont {Simon}\ and\ \citenamefont
  {Rudner}(2020)}]{simon2020contrasting}%
  \BibitemOpen
  \bibfield  {author} {\bibinfo {author} {\bibfnamefont {S.~H.}\ \bibnamefont
  {Simon}}\ and\ \bibinfo {author} {\bibfnamefont {M.~S.}\ \bibnamefont
  {Rudner}},\ }\href {https://doi.org/10.1103/PhysRevB.102.165148} {\bibfield
  {journal} {\bibinfo  {journal} {Phys. Rev. B}\ }\textbf {\bibinfo {volume}
  {102}},\ \bibinfo {pages} {165148} (\bibinfo {year} {2020})}\BibitemShut
  {NoStop}%
\bibitem [{\citenamefont {Bouadim}\ \emph {et~al.}(2011)\citenamefont
  {Bouadim}, \citenamefont {Loh}, \citenamefont {Randeria},\ and\ \citenamefont
  {Trivedi}}]{bouadim2011}%
  \BibitemOpen
  \bibfield  {author} {\bibinfo {author} {\bibfnamefont {K.}~\bibnamefont
  {Bouadim}}, \bibinfo {author} {\bibfnamefont {Y.~L.}\ \bibnamefont {Loh}},
  \bibinfo {author} {\bibfnamefont {M.}~\bibnamefont {Randeria}},\ and\
  \bibinfo {author} {\bibfnamefont {N.}~\bibnamefont {Trivedi}},\ }\href
  {https://doi.org/10.1038/nphys2037} {\bibfield  {journal} {\bibinfo
  {journal} {Nature Physics}\ }\textbf {\bibinfo {volume} {7}},\ \bibinfo
  {pages} {884} (\bibinfo {year} {2011})}\BibitemShut {NoStop}%
\bibitem [{\citenamefont {Seibold}\ \emph {et~al.}(2012)\citenamefont
  {Seibold}, \citenamefont {Benfatto}, \citenamefont {Castellani},\ and\
  \citenamefont {Lorenzana}}]{seibold2012}%
  \BibitemOpen
  \bibfield  {author} {\bibinfo {author} {\bibfnamefont {G.}~\bibnamefont
  {Seibold}}, \bibinfo {author} {\bibfnamefont {L.}~\bibnamefont {Benfatto}},
  \bibinfo {author} {\bibfnamefont {C.}~\bibnamefont {Castellani}},\ and\
  \bibinfo {author} {\bibfnamefont {J.}~\bibnamefont {Lorenzana}},\ }\href
  {https://doi.org/10.1103/PhysRevLett.108.207004} {\bibfield  {journal}
  {\bibinfo  {journal} {Phys. Rev. Lett.}\ }\textbf {\bibinfo {volume} {108}},\
  \bibinfo {pages} {207004} (\bibinfo {year} {2012})}\BibitemShut {NoStop}%
\bibitem [{\citenamefont {Samanta}\ \emph {et~al.}(2022)\citenamefont
  {Samanta}, \citenamefont {Das}, \citenamefont {Trivedi},\ and\ \citenamefont
  {Sensarma}}]{samanta2022}%
  \BibitemOpen
  \bibfield  {author} {\bibinfo {author} {\bibfnamefont {A.}~\bibnamefont
  {Samanta}}, \bibinfo {author} {\bibfnamefont {A.}~\bibnamefont {Das}},
  \bibinfo {author} {\bibfnamefont {N.}~\bibnamefont {Trivedi}},\ and\ \bibinfo
  {author} {\bibfnamefont {R.}~\bibnamefont {Sensarma}},\ }\href
  {https://doi.org/10.1103/PhysRevB.105.104503} {\bibfield  {journal} {\bibinfo
   {journal} {Phys. Rev. B}\ }\textbf {\bibinfo {volume} {105}},\ \bibinfo
  {pages} {104503} (\bibinfo {year} {2022})}\BibitemShut {NoStop}%
\bibitem [{\citenamefont {Kopnin}\ and\ \citenamefont
  {Sonin}(2008)}]{kopnin2008bcs}%
  \BibitemOpen
  \bibfield  {author} {\bibinfo {author} {\bibfnamefont {N.~B.}\ \bibnamefont
  {Kopnin}}\ and\ \bibinfo {author} {\bibfnamefont {E.~B.}\ \bibnamefont
  {Sonin}},\ }\href {https://doi.org/10.1103/PhysRevLett.100.246808} {\bibfield
   {journal} {\bibinfo  {journal} {Phys. Rev. Lett.}\ }\textbf {\bibinfo
  {volume} {100}},\ \bibinfo {pages} {246808} (\bibinfo {year}
  {2008})}\BibitemShut {NoStop}%
\bibitem [{\citenamefont {Tam}\ and\ \citenamefont
  {Peotta}(2024)}]{tam2024geometry}%
  \BibitemOpen
  \bibfield  {author} {\bibinfo {author} {\bibfnamefont {M.}~\bibnamefont
  {Tam}}\ and\ \bibinfo {author} {\bibfnamefont {S.}~\bibnamefont {Peotta}},\
  }\href {https://doi.org/10.1103/PhysRevResearch.6.013256} {\bibfield
  {journal} {\bibinfo  {journal} {Phys. Rev. Res.}\ }\textbf {\bibinfo {volume}
  {6}},\ \bibinfo {pages} {013256} (\bibinfo {year} {2024})}\BibitemShut
  {NoStop}%
\bibitem [{\citenamefont {You}\ and\ \citenamefont
  {Vishwanath}(2022)}]{you2022kohn}%
  \BibitemOpen
  \bibfield  {author} {\bibinfo {author} {\bibfnamefont {Y.-Z.}\ \bibnamefont
  {You}}\ and\ \bibinfo {author} {\bibfnamefont {A.}~\bibnamefont
  {Vishwanath}},\ }\href {https://doi.org/10.1103/PhysRevB.105.134524}
  {\bibfield  {journal} {\bibinfo  {journal} {Phys. Rev. B}\ }\textbf {\bibinfo
  {volume} {105}},\ \bibinfo {pages} {134524} (\bibinfo {year}
  {2022})}\BibitemShut {NoStop}%
\bibitem [{\citenamefont {Chatterjee}\ \emph {et~al.}(2022)\citenamefont
  {Chatterjee}, \citenamefont {Wang}, \citenamefont {Berg},\ and\ \citenamefont
  {Zaletel}}]{chatterjee2022inter}%
  \BibitemOpen
  \bibfield  {author} {\bibinfo {author} {\bibfnamefont {S.}~\bibnamefont
  {Chatterjee}}, \bibinfo {author} {\bibfnamefont {T.}~\bibnamefont {Wang}},
  \bibinfo {author} {\bibfnamefont {E.}~\bibnamefont {Berg}},\ and\ \bibinfo
  {author} {\bibfnamefont {M.~P.}\ \bibnamefont {Zaletel}},\ }\href
  {https://doi.org/10.1038/s41467-022-33561-w} {\bibfield  {journal} {\bibinfo
  {journal} {Nature Communications}\ }\textbf {\bibinfo {volume} {13}},\
  \bibinfo {pages} {6013} (\bibinfo {year} {2022})}\BibitemShut {NoStop}%
\bibitem [{\citenamefont {Chau}\ \emph {et~al.}(2024)\citenamefont {Chau},
  \citenamefont {Chen},\ and\ \citenamefont {Law}}]{chau2024origin}%
  \BibitemOpen
  \bibfield  {author} {\bibinfo {author} {\bibfnamefont {C.~W.}\ \bibnamefont
  {Chau}}, \bibinfo {author} {\bibfnamefont {S.~A.}\ \bibnamefont {Chen}},\
  and\ \bibinfo {author} {\bibfnamefont {K.}~\bibnamefont {Law}},\ }\href
  {https://arxiv.org/abs/2404.19237} {\bibfield  {journal} {\bibinfo  {journal}
  {arXiv preprint arXiv:2404.19237}\ } (\bibinfo {year} {2024})}\BibitemShut
  {NoStop}%
\bibitem [{\citenamefont {Zhang}\ \emph {et~al.}(2009)\citenamefont {Zhang},
  \citenamefont {Tang}, \citenamefont {Girit}, \citenamefont {Hao},
  \citenamefont {Martin}, \citenamefont {Zettl}, \citenamefont {Crommie},
  \citenamefont {Shen},\ and\ \citenamefont {Wang}}]{zhang2009direct}%
  \BibitemOpen
  \bibfield  {author} {\bibinfo {author} {\bibfnamefont {Y.}~\bibnamefont
  {Zhang}}, \bibinfo {author} {\bibfnamefont {T.-T.}\ \bibnamefont {Tang}},
  \bibinfo {author} {\bibfnamefont {C.}~\bibnamefont {Girit}}, \bibinfo
  {author} {\bibfnamefont {Z.}~\bibnamefont {Hao}}, \bibinfo {author}
  {\bibfnamefont {M.~C.}\ \bibnamefont {Martin}}, \bibinfo {author}
  {\bibfnamefont {A.}~\bibnamefont {Zettl}}, \bibinfo {author} {\bibfnamefont
  {M.~F.}\ \bibnamefont {Crommie}}, \bibinfo {author} {\bibfnamefont {Y.~R.}\
  \bibnamefont {Shen}},\ and\ \bibinfo {author} {\bibfnamefont
  {F.}~\bibnamefont {Wang}},\ }\href {https://doi.org/10.1038/nature08105}
  {\bibfield  {journal} {\bibinfo  {journal} {Nature}\ }\textbf {\bibinfo
  {volume} {459}},\ \bibinfo {pages} {820} (\bibinfo {year}
  {2009})}\BibitemShut {NoStop}%
\end{thebibliography}
\end{document}